\documentclass{appolb}
\usepackage{epsfig}
\usepackage{wrapfig,rotating}
\usepackage{amssymb}
\usepackage{amsmath}
\usepackage{graphicx}

\def\figurename{Fig.}

\usepackage{graphicx}

\newcommand\epsfigure[4][width=\hsize]{%
\begin{figure}%
  \begin{center}%
     \IfFileExists{#2.eps.bb}%
       {\includegraphics[draft,#1]{#2}}%
       {\includegraphics[#1]{#2}}%
  \end{center}%
\caption{ #3}\label{fig:#4}%
\end{figure}%
}

\def \q2{\ensuremath{Q^2}}
\def \r0{\ensuremath{\rho^0}}

\newcommand {\pom} {I\!\!P}
\newcommand {\pomsub} {{\scriptscriptstyle \pom}}

\newcommand {\xpom} {x_{\pomsub}}
\newcommand{\pomeron}{\mathbb{P}}
\newcommand{\invpb}{\units{pb}^{-1}}
\newcommand{\invfb}{\units{fb}^{-1}}
\newcommand{\lumi}{$\mathcal{L}$}

\newlength{\picwidth}
\newcommand\units{\,\mathrm}

\newcommand{\gevtwo}{\units{GeV^2}}

\newcommand{\dkap}{\Delta\kappa^{\gamma}}

\newcommand{\lam}{\lambda^{\gamma}}
\newcommand{\wwgamma}{WW\gamma}
\newcommand{\be}{\begin{equation}}
\newcommand{\ee}{\end{equation}}

\begin{document}
\title{Understanding the structure of the proton: From HERA and Tevatron to LHC
}
\author{M. Boonekamp, F. Chevallier, C. Royon, L. Schoeffel
\address{CEA Saclay/IRFU-SPP, 91191 Gif-sur-Yvette, France}
}
\maketitle
%%%%%%%%%%%%%%%%%%%%%%%%%%%%%%%
\begin{abstract}
%%%%%%%%%%%%%%%%%%%%%%%%%%%%%%%
In this review, we first discuss the perspectives concerning a better determination
of the proton structure in terms of quarks and gluons at LHC after
describing the results coming from HERA and Tevatron. In a second part of
the review, we describe the diffractive phenomena at HERA and Tevatron and the
consequences for LHC.
\end{abstract}
%%%\PACS{PACS numbers come here}

%%%%%%%%%%%%%%%%%%%%%%%%%%%%%%%
%%%%%%%%%%%%%%%%%%%%%%%%%%%%%%%
%%%%%%%%%%%%%%%%%%%%%%%%%%%%%%%
%\section{Introduction }
Understanding the fundamental structure of matter requires an understanding of how
quarks and gluons are assembled to form hadrons and of the structure of the
protons which are the colliding particles at LHC.
The arrangement of quarks and gluons
inside nucleons can be probed by accelerating electrons, hadrons or nuclei to precisely 
controlled energies, smashing them into a target nucleus
and examining in detail the final products.

In this review, we first discuss the structure of the proton in terms of
quarks and gluons. We first present briefly the results from HERA and the
Tevatron and then discuss two aspects at LHC: how can we improve our
knowledge on parton distribution functions (PDFs) of the proton, and is it
possible to find some observables less sensitive to PDF uncertainties to probe
new physics beyond the standard model. In a second part of the review, we will
discuss diffraction at HERA, Tevatron and the prospects for LHC.

%%%%%%%%%%%%%%%%%%% LHC SECTION

%\input{lhcsection}
\section{The parton distribution functions at the LHC}

\subsection{The challenges of LHC physics}

The LHC physics program is rich and has been widely
described~\cite{atlastdr,atlascsc,cmstdr}. It encompasses the searches for new
particles up to masses of several TeV, including the elucidation of
electroweak symmetry breaking and the possible observation of new
symmetries at higher scales, and precision measurements of fundamental
parameters in the electroweak and strong gauge sectors.

A common requirement for this program to succeed is a good control of
the proton parton densities. To be more specific, the discovery of the
Higgs boson, and subsequent measurements of its couplings relies on
precise predictions of the gluon density in the range $x \sim  
10^{-2}-10^{-1}$, and at corresponding scales $Q^2 = m_H^2 \sim
10^4 - 10^6 \,\,\, \mathrm{GeV}^2$. The high-$x$, high-$Q^2$ gluon
density also determines the production rate of high-$E_T$ jets, and
affects e.g. the measurement of the running of $\alpha_S$ and the
search for extra dimensions through this final state.

\begin{figure}[htbp]
\begin{center}
  \includegraphics[width=\textwidth]{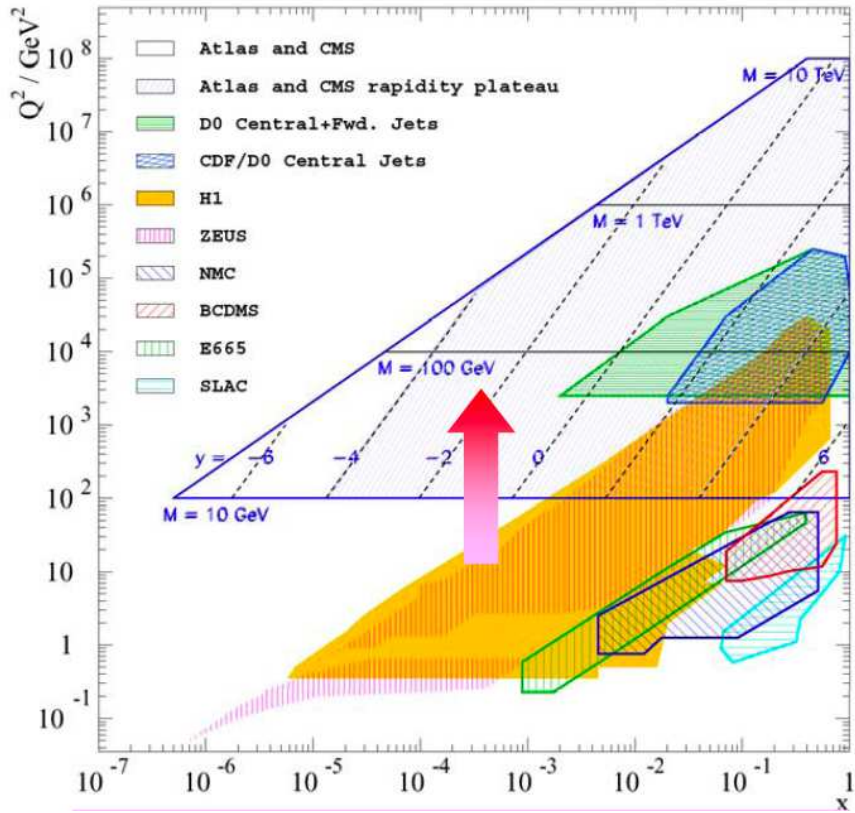}
\end{center}
\caption{\label{XQ2} ($x$, $Q^2$) domain probed by the fixed target,
HERA, Tevatron and LHC experiments.}
\label{kinplane}
\end{figure}

The LHC also allows to reach very low values of $x$ as it is indicated in
Fig.~\ref{kinplane}. Dedicated processes at LHC will allow to study the low
$x$ region in detail as we will see in the following, for instance using
Mueller-Navelet jets. In addition, the saturation region where the gluon density
is large and gluons overlap in the proton might be accessible at LHC as we will
discuss it further. 

The on-shell production of electroweak gauge bosons is mainly controlled by
the sea quarks. In this case $Q^2$ is essentially fixed, but the
vector boson rapidity distribution probes the range $x \sim
10^{-5}-10^{-1}$. Precise measurements of the electroweak parameters,
notably the $W$ boson mass $m_W$ and the weak mixing angle,
$\sin^2\theta_W$, require tightly constrained parton densities in this
range. The interpretation of high-mass Drell-Yan events or 
gauge boson pairs in terms of electroweak interactions again requires a
good control of these densities at scales up to $10^6 \,\,\,
\mathrm{GeV}^2$. 

In the process of quantifying the PDF-induced uncertainties on the above
observables, and improving them where needed, one has to remember the
hypotheses under which the PDFs have been determined in the first
place. In particular, the dependence of the results on the choice of
functional form of the parton densities at $Q_0 \sim m_p$ can often
not be ignored. Hypotheses concerning the initial flavor composition
(u,d,s and possibly c quarks) need to be accounted for as
well. In addition, the QCD evolution of these densities is performed
assuming different schemes (Dokshitzer Gribov Lipatov Altarelli Parisi, 
DGLAP~\cite{dglap}, Balitski Fadin Kuraev Lipatov, BFKL~\cite{bfkl}), 
perturbative orders (leading order (LO),
next-to-leading order (NLO), next-to-next-to-leading order (NNLO)), 
including saturation effects or not. When quoting a
PDF-induced uncertainty, one needs to ascertain whether these
underlying hypotheses affect the result or can be ignored.

In the following, we review the most prominent examples of the
influence of parton densities on the LHC physics program. After an
introduction about the status of PDF determination and their uncertainties,
coming mainly from fixed target experiments and from HERA, we discuss 
briefly the input from the Tevatron.  
The third section describes how
LHC measurements are sensitive to gluon PDF and how to reduce its
uncertainty. The fourth section is devoted to quark PDF. The last
section deals with observables less sensitive to PDF uncertainties to look
for beyond standard model effects; we can quote for instance ratios
of cross sections with the goal
to measure separately a subset of parton densities or to reduce the impact of
the PDF uncertainties.

\subsection{Status of PDF uncertainties}

The understanding of the proton structure has made great progress since the
observation of the broken scale invariance in the early 70's. The measurement
of the proton structure function at HERA in the H1 and ZEUS experiment allowed
to make considerable progress on the knowledge of the proton
structure~\cite{reviewhera}. The HERA data allowed to constrain strongly the PDF
uncertainties at medium $x \sim 10^{-2}$, and allowed to access
a completely new kinematical region in $x$ down to 10$^{-5}$. These data are
fundamental to get precise cross sections at LHC for beyond standard model
effects and the background. In this section, we will only summarize the status
on the PDF uncertainties since many reviews described already the impact of HERA
on PDF determination~\cite{reviewhera}.

According to the last version of global QCD fits, the uncertainties on the quark and
gluon densities in the proton reach few per-cent in most of the kinematic plane
in ($x$, $Q^2$)~\cite{cteq66,mstw09}. However, at LHC, the uncertainty
on parton distributions leads to one of the most important uncertainties
for many measurements, greater than the expected statistical errors, and thus reduce
the sensitivity of these measurements to new physics effects. Several
reasons explain why PDF-induced uncertainties on observables can be so
large and why they must be quoted with critical thinking.

First, the uncertainties on PDFs come from the uncertainties on parameters
of the functional form at $Q_0^2\sim m_p^2$. These parameters are the output
of global QCD fits on data (fixed-target, HERA, and Tevatron experiments).
Valence quark PDFs can be essentially measured in the high-$x$ region, so the
largest uncertainties ($>20\%$) on valence quark distributions are found at
low $x$ ($x<10^{-3}$) and very high $x$ ($x>0.8$). On the contrary, the
uncertainty on sea quark and gluon distributions reaches 20\% at high $x$ ($x>0.2$).
This is due to the lack of data on processes using sea quarks and gluon in
this region and to the rapid fall off of the parton distributions at high $x$. 
At hadron colliders, many measurements are sensitive to a large
$x$ or $Q^2$ range, and not only to the intermediate region $10^{-3}<x<10^{-1}$
in which PDFs are best known. Large uncertainties on PDFs are found at high $x$
($x>0.2$) or low-$x$ ($x<10^{-4}$), and the error can be greater than 20\%.
Thus, precise measurements at LHC could help to reduce the errors on
PDF and improve our knowledge of the proton substructure.
 
Second, one has to remember the hypotheses under which the PDFs have been
determined in the first place. In particular, the dependence of the results
on the choice of the functional form of the parton densities at $Q_0^2\sim m_p^2$
can often not be ignored. These hypotheses are needed to decrease the numbers
of degrees of freedom in QCD fits, but the systematic uncertainty
induced by such approximations cannot be evaluated any longer with one
single PDF set. Among these assumptions, 
the ratio of the $\bar{d}, \bar{u}$ sea quark PDFs in the asymptotic
low $x$ region are often set to one, the sea quark and antiquark
densities at $Q_0^2$ often have the same parametrization,
etc. Hypotheses concerning the initial flavor composition need to be
accounted for as well. The description of the strange quark and
antiquark distributions at $Q_0^2$ may require additional degrees of
freedom~\cite{cteq66}. The proton could have an intrinsic charm quark
component at $Q_0^2\sim m_p^2$, thus enhancing the rate of charm
quark-induced processes~\cite{CharmComponent}. More generally, the extraction of heavy
flavour PDFs and the comparison to data is still difficult and suffers
from large statistical and systematic uncertainties. 

\begin{figure}[htbp]
    \begin{center}
      \hspace{-0.7cm}
      \includegraphics[width=.53\textwidth]{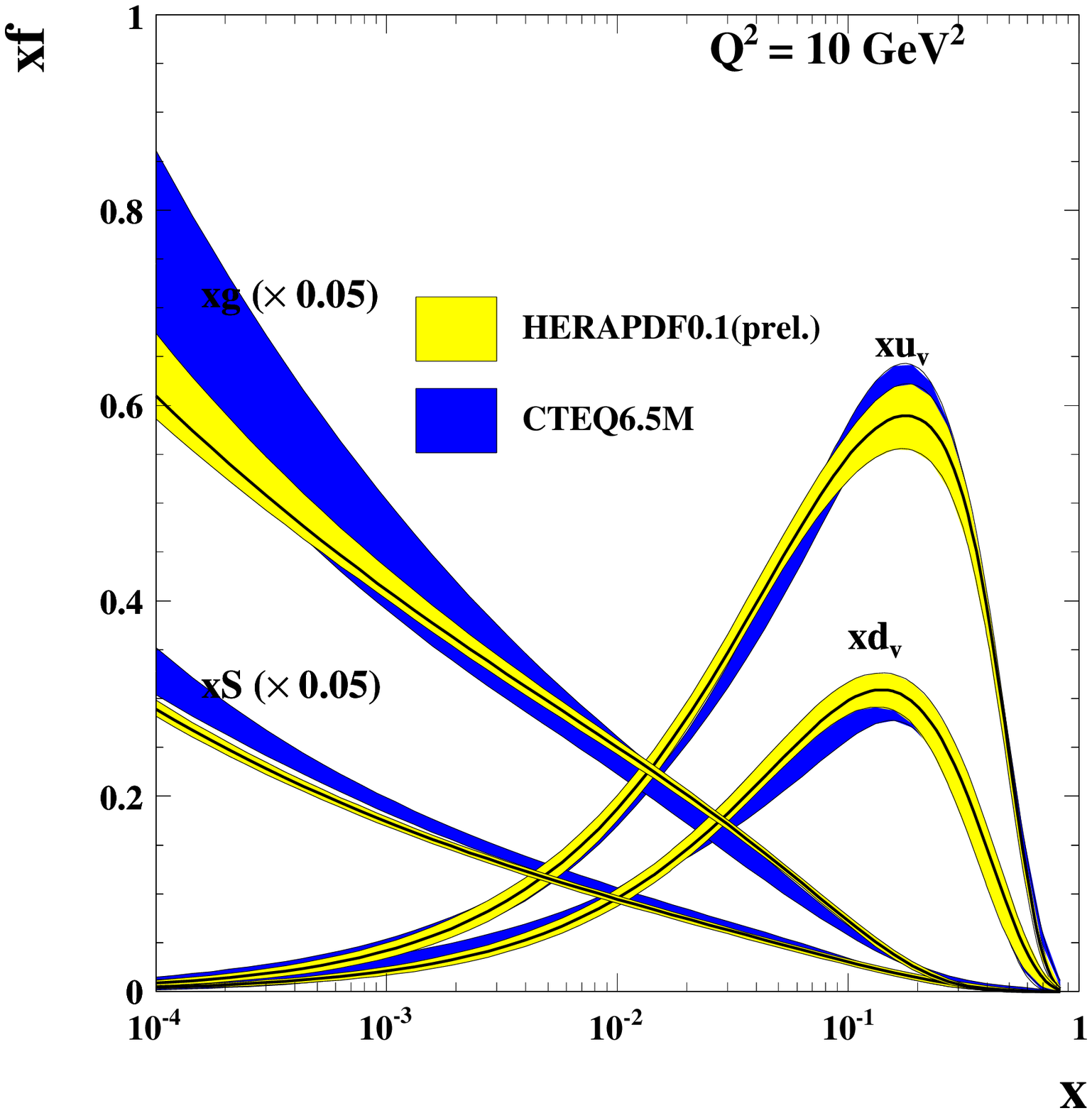}
      \hspace{-0.7cm}
      \includegraphics[width=.53\textwidth]{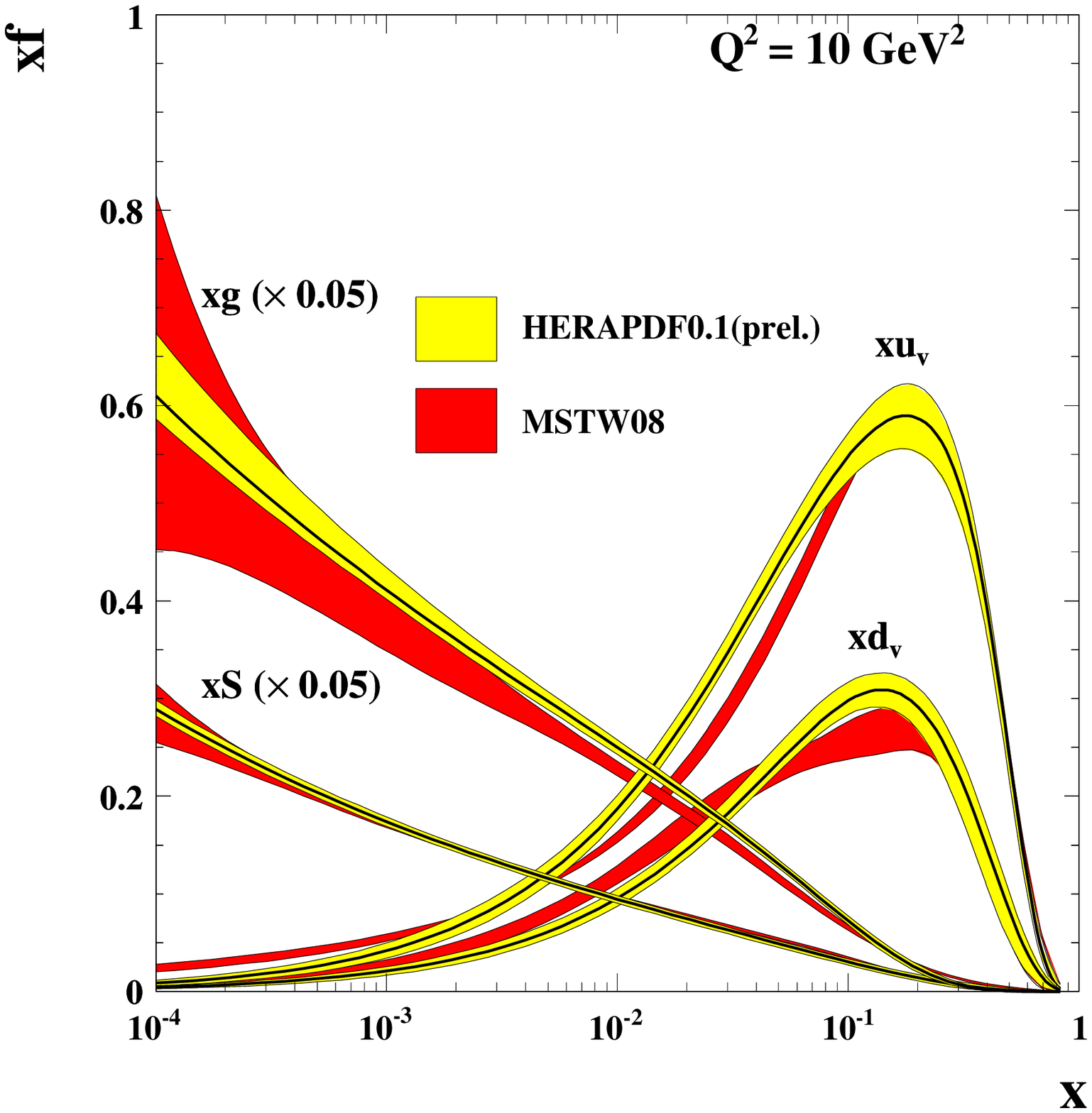}
      \hspace{-0.7cm}
      \vspace{-0.7cm}
    \end{center}
    \caption{\label{XQ10} ($x$, $Q^2$) PDFs at $\rm Q^2=10~GeV^2$ obtained from CTEQ6.5, MSTW08 and HERAPDF01(prel.)\cite{HeraIchep08}.}
  \end{figure}

The impact at higher scale of some approximations can be tested using different
sets of PDFs, because the underlying hypotheses on the functional form are different.
As an example, the effects of the underlying hypotheses can be seen in
Fig.~\ref{XQ10}, where some differences are visible at $x\approx 10^{-2}$ for
valence quark distributions. The hypotheses on the shape of PDF at $Q_0^2$ have
to be tested when measurements become more and more precise. The LHC will play a
major role in testing the PDF functional form.

Third, the QCD evolution of these densities is performed using different
evolution equations: DGLAP for $Q^2$ evolution, BFKL for $\ln 1/x$ evolution. 
The evolution equation can also be computed at
different perturbative orders (LO, NLO, NNLO in $\alpha_S$ for DGLAP and 
leading log (LL), next-to-leading log (NLL) in $\log 1/x$ for BFKL), 
including the appropriate
treatment of the mass of heavy quarks and the evolution of the strong
coupling constant at low $Q^2\approx 2 m_q^2$. In addition, saturation
effects could be strong at low $Q^2$ and very low $x$. Due to the large
($x$, $Q^2$) domain accessible at LHC, many measurements are sensitive to
these effects, and thus can bring constraints on evolution equations.

In the following, we describe which measurements at LHC can bring information
on the proton structure, and thus help not only to reduce the PDF
uncertainties, but also to test the hypotheses on the initial shape
of PDFs at $Q_0^2$ and to test the evolution equations. Prospects for
improvements and their difficulties will be presented in this context.

%%%%%%%%%%%begin CHR%%%%%%%%%%%%%%%
Before describing the expected results from LHC, we will give the QCD
results obtained at Tevatron. The HERA results are described in
detail in Ref.~\cite{reviewhera}.

\section{State of the art at the Tevatron}

\subsection{Inclusive jet cross section measurements at the Tevatron}

The first measurement sensitive to PDFs which can be performed at Tevatron
is the inclusive jet cross section which relies on the precise
determination of the jet energy calibration. We will describe briefly how the
jet energy is obtained at Tevatron since similar methods can be used at
LHC. The jet energy scale
is determined mainly using $\gamma+$jet events. In the D0 collaboration as an
example, the
corrected jet energy is obtained using the following method
\begin{eqnarray}
E_{jet}^{corr} = \frac{E_{jet}^{uncorr} - Off}{Show \times Resp}
\end{eqnarray}
where $E_{jet}^{corr}$ and $E_{jet}^{uncorr}$ are respectively the corrected and uncorrected
jet energies. The offset corrections ($Off$) are related to uranium noise
and pile-up and are determined using minimum-bias and zero-bias data. 
The showering corrections
($Show$) take into account the energy emitted outside the jet cone because of
the detector and dead material and, of course, not the
physics showering outside the jet cone which corresponds to QCD radiation
outside the cone. The jet response ($Resp$) is the largest correction, and can
be subdivided in few corrections. The first step is to equalize the calorimeter
response as a function of rapidity, and the jet response is then measured for
the central part of the calorimeter only using the $p_T$ balance in $\gamma+$jet
events. Some additional small corrections related to the method biases are
introduced. One important additional correction deals with the difference in
response between quark and gluon jets. The difference was studied both in data
and in Monte Carlo (using for instance the $\gamma+$jet and the dijet samples
which are respectively quark and gluon dominated) and leads to a difference of
4 to 6\% as a function of jet $p_T$, which is not negligible if one wants a
precision on jet energy scale of the order of 1\%. This has an important
consequence. The jet energy scale is not universal but sample dependent. QCD
jets (gluon dominated) will have a different correction with respect to the $t
\bar{t}$ events for instance which are quark dominated. The CDF collaboration
follows a method which is more Monte Carlo oriented using beam tests and single
pion response to tune their Monte Carlo. At LHC, it will be possible to use
$Z+$jets which do not suffer from the ambiguity of photon identification in the
detector.

The uncertainties reached by the D0 collaboration concerning the determination
of jet energy scale are of the order of 1.2\% for jet $p_T$ between 70-400 GeV
in a wide range of rapidity around zero (the uncertainty is of the order of 2\% for a
rapidity of 2.5). This allows to make a very precise measurement of the jet
inclusive cross section as a function of their transverse momentum.

\begin{figure}[htb]
\begin{center}
\epsfig{file=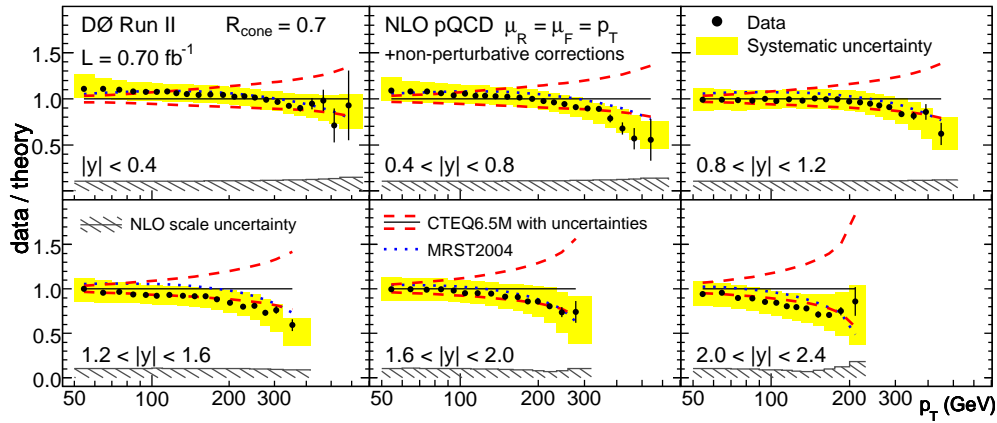,height=2.2in}
\caption{Data over theory for the inclusive $p_T$ cross section measurement from
the D0 collaboration using the 0.7 jet cone. Data are compared to NLO QCD calculations using the
CTEQ6.5M parametrisation.}
\label{inclusive1}
\end{center}
\end{figure}

\begin{figure}[htb]
\begin{center}
\epsfig{file=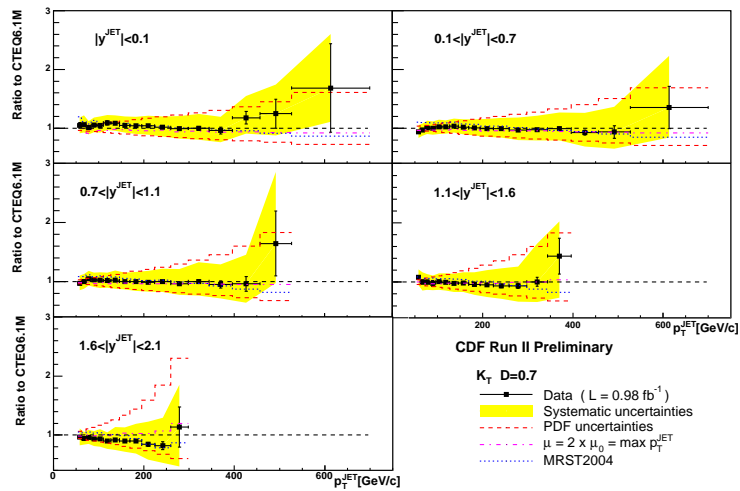,height=3.1in}
\caption{Data over theory for the inclusive $p_T$ cross section measurement from
the CDF collaboration using the $k_T$ algorithm. Data are compared to NLO QCD calculations using the
CTEQ6.1 parametrisation.}
\label{inclusive2}
\end{center}
\end{figure}

The measurement of the inclusive jet cross section~\cite{inclusive} was performed by the D0 and
CDF collaborations at Tevatron using a jet cone algorithm with a cone size
of 0.7 (D0 and CDF) and the $k_T$ algorithm (CDF). Data are corrected to hadron
level (D0) or parton level (CDF). The motivation of this measurement is double:
it is sensitive to beyond standard model effects such as quark substructure and
to PDFs, especially the gluon density at high $x$. Historically, the excess
observed by the CDF collaboration in 1995 concerning the inclusive jet $p_T$
spectrum compared to the parametrisations was suspected to be a signal of quark
substructure but it was found that increasing the gluon density at high $x$
could accomodate these data. This raises the question of PDFs versus beyond
standard model effects, and the interpretation of data in general.
Data are compared with NLO QCD calculations
using either CTEQ6.5M~\cite{cteq} for D0 or CTEQ6.1 for CDF (the uncertainties of the
CTEQ6.5M parametrisation are two times smaller). A good agreement is found over
six orders of magnitude. The ratio data over theory for the D0 and CDF
measurements are given in Figs.~\ref{inclusive1} and \ref{inclusive2}. A good
agreement is found between NLO QCD and the D0 or CDF measurements with a
tendency of the CTEQ parametrisation to be slightly lower than the data at high
jet $p_T$. The MRST2004~\cite{cteq} parametrisation follows the shape of the measurements.
Given the precision obtained on jet energy scale, the uncertainties obtained 
by the D0 collaboration are lower than the PDF ones and will allow to constrain
further the PDFs and specially the gluon density at high $x$ (the 
uncertainties of the measurement performed by the CDF collaboration are about two times
larger). An update of the CTEQ and MRST PDFs using these latest data are expected
soon. The D0 collaboration took also special care of the uncertainty
correlation studies, by giving the effects of the 24 sources of systematics in
data.
 
In addition, the CDF collaboration measured the dijet mass cross
section~\cite{dijetmass} above
180 GeV, and up to 1.2 TeV. No excess was found with respect to NLO QCD
calculations and this measurement allows to exclude excited quarks below 870
GeV, $Z'$ (resp. $W'$) below 740 (resp. 840) GeV~\footnote{Stronger limits 
on $W'$ and $Z'$ mass limits come from lepton based searches.}, 
and technirho below 1.1 TeV.

The question rises if PDFs can be further constrained at LHC using inclusive
measurements. The PDF uncertainties are typically of the order of 15\% for a jet
$p_T$ of 1 TeV, and 25\% of 2 TeV for $1<|\eta_{jet}|<2$ (without taking into
account the new Tevatron measurements which we just discussed). A typical
uncertainty of 5\% (resp. 1\%) on jet energy scale leads to a systematic
uncertainty on 30 to 50\% (resp. 6 to 10\%) on the jet cross section. A precise
determination of the jet energy scale at LHC will thus be needed to get
competitive measurements at LHC. In the following, we will discuss more
clever ways to find observables less sensitive to PDF uncertainties but still
sensitive to beyond standard model effects.

\subsection{Multijet cross section measurements at the Tevatron and at HERA}

The measurement of multijet cross sections at the Tevatron and at HERA (and
later on at LHC) is fundamental to constrain the PDFs and to tune the Monte
Carlo, since it is a direct background entering in many searches for Higgs
bosons or new particles at the LHC. We can quote for instance the search for
Higgs bosons in association with $t \bar{t}$, the measurement of the $t \bar{t}$
production cross section, the search for $R$-parity violated SUSY (which can
lead up to 8-10 jets per event...).

\subsubsection{Measurement of $\Delta \Phi$ between jets in D0}
The advantage of the measurement of the difference in azimuthal angle 
between two leading jets in an inclusive QCD sample as was performed by the DO
collaboration is
that there is no need of precise knowledge of jet
energy scale (the measurement is dominated by the knowledge of jet angles)
and this can be performed at the beginning of data taking at the LHC for
instance when the detectors are not yet fully calibrated. 
The $\Delta \Phi$ spectrum was measured in four
different regions in maximum jet transverse momentum, and a good agreement was
found with NLO calculations using either the CTEQ and MRST parametrisations except at very high $\Delta \Phi$ where soft
radiation is missing~\cite{deltaphi}. PYTHIA~\cite{pythia} shows a disagreement at small
$\Delta \Phi$, showing a lack of initial state gluon radiation, while
HERWIG~\cite{herwig} shows a good agreement with data. It will be important to redo
this kind of measurements at the beginning of LHC.

\subsubsection{Measurement of $\gamma +$jet cross sections}

The D0 collaboration measured the inclusive production of isolated $\gamma +$
jets in different detector regions requiring a central photon and a central or a
forward jet. It distinguished the cases when the photon and
the jet are on the same or opposite side. The cross section has been found in
disagreement with NLO QCD expectations both in shape and normalisation and the
reason is still unclear~\cite{gammajet}. It is worth noticing that the transverse
momentum of the photon is not very high for that measurement,
and may be the problem is related to the fact that it is not performed where
perturbative QCD can be trusted. We will come back on that kind of measurement when
we discuss the possibilities at LHC.

\subsubsection{Jet shape measurements in CDF}
The jet shape is dictated by multi-gluon emission from primary partons, and is
sensitive to quark/gluon contents, PDFs and running $\alpha_S$, as well as
underlying events. We define $\Psi$ which is sensitive to the way the energy is
spread around the jet center
\begin{eqnarray}
\Psi (r) = \frac{1}{N_{jets}} \Sigma_{jets} \frac{P_T(0,r)}{P_T^{jet}(0,R)}
\end{eqnarray}
where $R$ is the jet size.
The energy is more concentrated towards the jet center for quark than for gluon
jets since there is more QCD radiation for gluon jets (which
means that $\Psi$ is closer to one for quark jets when $r \sim 0.3 R$ for
instance). The CDF collaboration measured $\Psi(0.3/R)$ for jets with $0.1 < |y|
< 0.7$ as a function of jet $p_T$ and found higher values of $\Psi$ at high
$p_T$ as expected since jets are more quark like~\cite{shape} and this
is well described by QCD expectations. This measurement also helps
tuning the PYTHIA and HERWIG generators since it is sensitive to underlying
events in particular.

\begin{figure}[htb]
\begin{center}
\epsfig{file=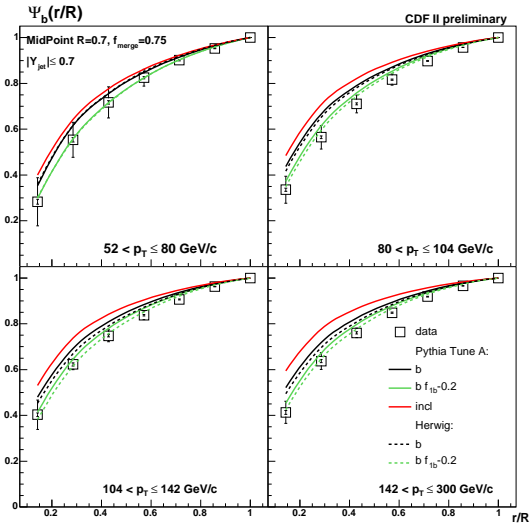,height=2.8in}
\caption{Measurement of the $b$-jet shapes and comparison with the predictions
of the PYTHIA and HERWIG Monte Carlo (see text).}
\label{bjets}
\end{center}
\end{figure}

The CDF collaboration also studied the jet shapes for $b$-jets in four different
$p_T$ bins~\cite{shapeb} since it is sensitive to the $b$-quark content
of the proton, and the result
is given in Fig.~\ref{bjets}. The default PYTHIA and HERWIG Monte Carlo in black
full and dashed lines respectively are unable to describe the measurement.
Compared to the inclusive jet shape depicted in Fig.~\ref{bjets} in full red line
for PYTHIA, the tendency of the $b$-jet shape is definitely the right one,
leading to smaller values of $\Psi$ as expected, but the measurement leads to a
larger difference. The effect of reducing the single $b$-quark fraction by 20\%
leads to a better description of data as it shown in green in Fig.~\ref{bjets}.
The fraction of $b$-jets that originate from flavour creation (where a single
$b$-quark is expected in the same jet cone) over those that originate from
gluon splitting (where two $b$-quarks are expected in the same jet cone)  is
different in Monte Carlo and data. This will be an important measurement to perform
again at LHC since it is a direct background to searches for the Higgs boson
and for new phenomena.

The CDF collaboration also measured the $b \bar{b}$ dijet cross section as a
function of the leading jet $p_T$ and the difference in azimuthal angle between
the two jets and it leads to the same conclusion, namely that PYTHIA and HERWIG
underestimates the gluon splitting mechanism~\cite{dijetmass}.

\subsection{A parenthesis: underlying events at Tevatron and LHC}
This section is not directly related to the proton structure but understanding
underlying events is a necessary step prior constraining further the proton
structure. We will not mention further this aspect in the following but wanted to
stress it while discussing the main results from Tevatron.
The CDF collaboration measured underlying events at Tevatron and used these
measurements to tune in particular the PYTHIA generator. $pp$ or $p \bar{p}$
interactions are namely not as simple as interactions in $ep$ colliders. In
addition to the hard scattering producing dijets, high $p_T$ leptons...,
spectator partons produce additional soft interactions called underlying events.
The main consequence is that it introduces additional energy in the detector not
related to the main interaction which needs to be corrected. 

\begin{figure}[htb]
\begin{center}
\epsfig{file=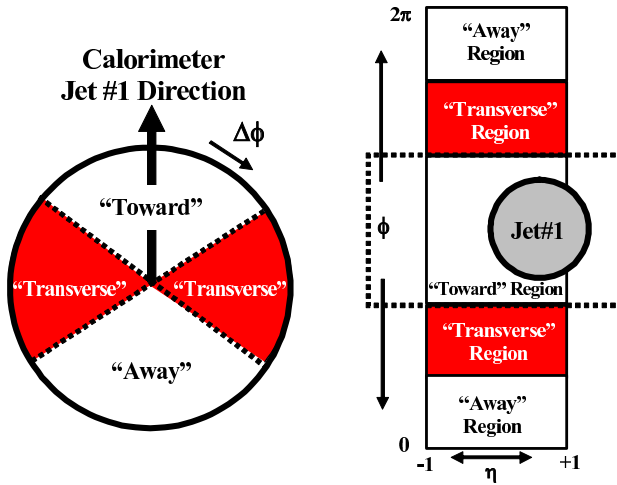,height=2.5in}
\caption{Definition of the
``toward", ``away" and ``transverse" regions in the case of dijet events
as an example.}
\label{underlyingb}
\end{center}
\end{figure}

To study this kind of events, the idea is quite simple. It is for instance
possible to use dijet events and we can distinguish in azimuthal angle three
different regions: the ``toward" region around the leading jet direction defined
by a cone of 60 degrees around the jet axis, the ``away" region in the opposite
direction to the jet, and the ``transverse" region the remaining regions far
away from the jet and the ``away" region, as shown in
Fig.~\ref{underlyingb}. In dijet events, the ``transverse"
region will be dominated by underlying events. The CDF collaboration measured
the charged multiplicity and the charged transverse evergy as a function of jet
transverse energy and used these quantities to tune the PYTHIA Monte Carlo
leading to the so called Tune A and Tune AW~\cite{dijetmass}. 

\begin{figure}[htb]
\begin{center}
\epsfig{file=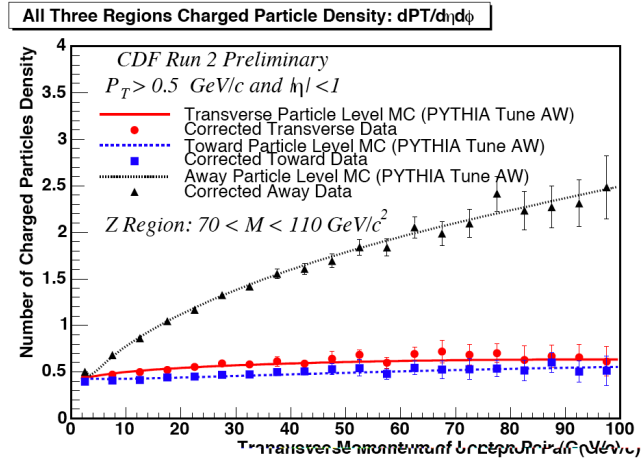,height=2.5in}
\caption{Measurement of the charged particle density for Drell Yan events in the
``toward", ``away" and ``transverse" regions compared to PYTHIA Tune AW.}
\label{underlying}
\end{center}
\end{figure}

Clean Drell Yan events can also be used to tune underlying
events~\cite{dijetmass}. The lepton
pair defines the ``toward" region while the ``away" and ``transverse" regions
are defined in the same way as for dijets. As an example, we give in
Fig.~\ref{underlying} the charged particle density as a function of the
transverse momentum of the lepton pair in the three regions compared with the
Tune AW of PYTHIA.

At LHC, one of the first measurements to be performed will be related to the
tuning of underlying events in the generators. Present tunings between the 
different Monte Carlo (PYTHIA, PHOJET, HERWIG) show differences
up to a factor six concerning the average multiplicity of charged particles as a
function of the $p_T$ of the leading jet as an example, and it is crucial to
tune the Monte Carlo to accomplish fully the LHC program.

\subsection{Measurements of the $W+$jet and $Z+$jet cross sections at the Tevatron}

The measurements of the $W+$jet and $Z+$jet cross sections are specially
important since they are a background for many searches and especially the
search for the Higgs boson. We will also study further the $W$ and $Z$ production
cross sections to show their sensitivities on PDFs and as a method to constrain 
them further in the following.

\subsubsection{Measurements of the $W+X$ cross sections}
The D0 collaboration measured the ratio of the $W+c$ to the inclusive cross
section 0.074 $\pm$ 0.019 (stat.) $\pm^{0.012}_{0.014}$ (syst.) in agreement
with NLO calculation~\cite{wcharm}. It will be important to redo this measurement 
with higher statistics since it is directly sensitive to the $s$-quark PDF.

\subsubsection{Measurement of the $Z+b$ and $W+b$ cross sections}
The motivation to measure the $Z+b$-jet cross section is quite clear: this is a
direct background for Higgs boson searches and it is also sensitive to the $b$
quark content of the proton. The measurements of the $Z+b$-jet and $W+b$-jet 
cross sections were performed by the CDF collaboration at the Tevatron 
$\sigma(Z+b~jets)=$0.86 $\pm$ 0.14 $\pm$ 0.12 pb and 
$\sigma(W+b-jets) \times BR(W \rightarrow l \nu) = 2.74 \pm 0.27 (stat.) \pm 0.42 (sys.)$ pb 
in agreement with NLO calculations and PYTHIA predictions~\cite{zbwb}. 
The CDF collaboration also compared the differential distributions in jet 
$p_T$ and rapidity as an example and the distributions are found in good 
agreeement with PYTHIA.

%%%%%%%%%%%%end CHR %%%%%%%%%%%%%%%
After reviewing briefly the present status on the proton knowledge from HERA and
Tevatron, we will now discuss what can be expected at LHC, concerning the
gluon and quark densities in the proton.

\section{The gluon density in the proton}
We review below the main physics items relying on the gluon density by studying two
aspects: how does the physics at LHC (mainly searches) depend on the gluon
uncertainty and how can we constrain it further using LHC data?

\subsection{Impact on searches for new physics}

The discovery of the Higgs boson, main objective and motivation for
the construction of the LHC, is dominantly produced $via$ gluon fusion
for the mass range 100~GeV~$< m_H <$~1~TeV. The hard cross section and
the decay modes can be accurately computed; for example, the NNLO 
QCD computation of FEHIP~\cite{fehip} claims a residual uncertainty of 1\%
on the production cross section, assuming standard model
couplings. The decay modes as computed with HDECAY~\cite{hdecay} carry an
even smaller uncertainty.

\begin{figure}[htbp]
    \begin{center}
      \includegraphics[width=\textwidth]{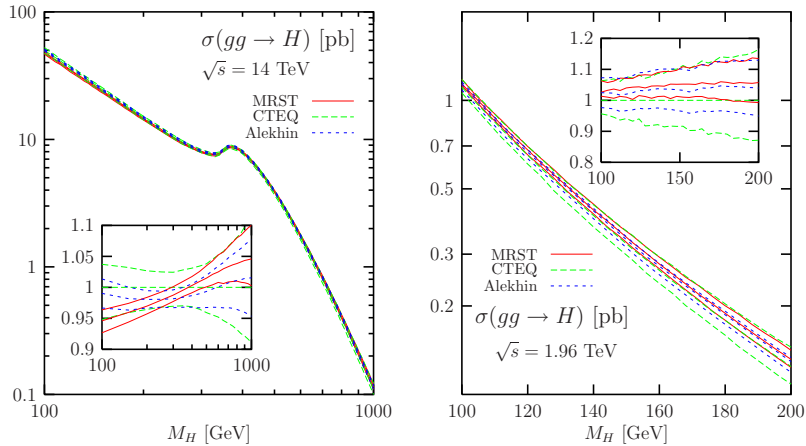}
    \end{center}
    \caption{\label{higgsprediction} Higgs boson production
    cross section and its uncertainty, as a function of $m_H$.}
  \end{figure}

The dominant residual uncertainty on $\sigma_H$ at the LHC comes from
the uncertainty on the gluon density, as illustrated in
Fig.~\ref{higgsprediction}~\cite{ferrag1}. Depending on the value of $m_H$, the
uncertainty varies between 5\% and 10\%. It is worthwhile to note, as
can be seen on the figure, that the different PDF sets used (MRST,
CTEQ, Alekhin) are sometimes marginally compatible; this examplifies the
need to consider, in addition to the fit uncertainty claimed by each
set, the framework (theory, underlying hypotheses) in which each fit
has been performed.

This level of uncertainty does not affect the discovery potential
(most often, signal and backgrounds are affected by the same
uncertainty, which cancels in the ratio; besides, other systematic
uncertainties dominate, depending on $m_H$ and the final state
considered). However, once the particle has been established and
high statistics measurements of its couplings are underway, the gluon
density will be the most significant source of theoretical uncertainty
to the measurements.

High-$E_T$ jets will be copiously produced at LHC : the expected
cross section for $E_T > 1$~TeV is still about 20 pb. While this
process has a high cross section and is dominantly coupled to initial
gluons, and as such is a natural probe of the proton PDFs, it is also
sensitive to new particles. Scalar or vector $s$-channel resonances
can appear in technicolor theories~\cite{technicolor}; graviton production is
another possibility~\cite{graviton}; as these processes display a peak in the 
invariant mass, these searches are relatively safe against PDF
uncertainties. 

\begin{figure}[htbp]
    \begin{center}
      \includegraphics[width=\textwidth]{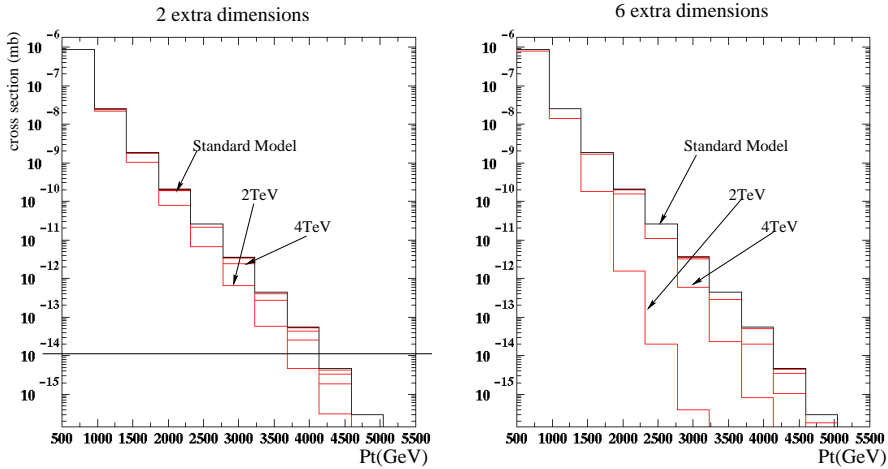}
    \end{center}
    \caption{\label{dijetprediction} Standard Model dijet
    cross section as function of $E_T$, and expected modifications in 
    the presence of extra dimensions. The successive curves represent,
    from top to bottom, the SM prediction and expected modifications of the
    spectrum in the presence of extra dimensions of size 8, 4 and 2~TeV.}
  \end{figure}

On the other hand, certain theories with large compactification radius
extra dimensions, or large number of extra dimensions, produce
a continuum of Kaluza-Klein excitations, which appears as a
modification of the slope of the differential cross section,
$d\sigma/dM$. Fig.~\ref{dijetprediction} shows what can be expected
in this case; it is easy to find unexcluded model parameters that
predict deviations of 50\% w.r.t the standard model prediction, for
$E_T > 2$~TeV.

\begin{figure}[htbp]
    \begin{center}
      \includegraphics[width=.6\textwidth]{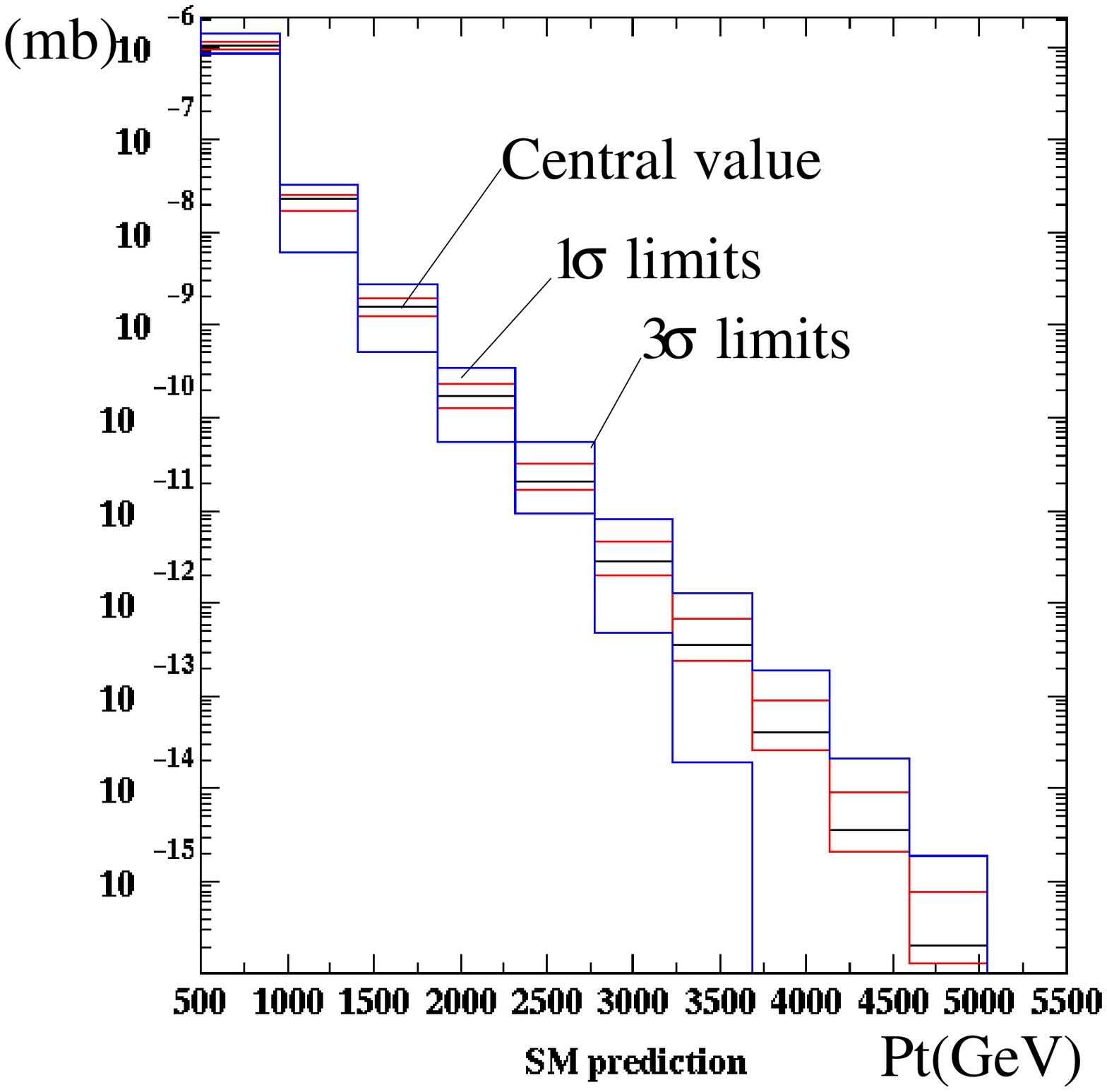}
    \end{center}
    \caption{\label{dijetsm} Standard Model dijet cross section and
    its uncertainty, as a function of $m_{JJ}$~\cite{ferrag2}. The central
    curve represents the SM prediction, and the envelopes represent
    the 1$\sigma$ and 3$\sigma$ confidence interval.}
  \end{figure}

The uncertainty induced by the proton PDFs is displayed in
Fig.~\ref{dijetsm}. According to CTEQ6.1 (used in~\cite{ferrag2}), the
cross section uncertainty grows rapidly with $m_{JJ}$, being 30\% at 2
TeV and up to factors of 10 at higher $E_T$. As we see, a good part of
this process na{\"i}ve sensitivity to new physics vanishes once PDF
uncertainties are accounted for.
  
Another approach of the same problem was studied by
ATLAS~\cite{alphasrunning}. This note presented the LHC potential to measure the
running of $\alpha_S$ through an analysis of the dijet mass spectrum
or jet $E_T$ spectrum. The note concluded that although an absolute
measurement of $\alpha_S$ is out of reach, its running could be
observed, and the standard evolution verified. At that time, PDF sets
with uncertainties were not available and the related systematic
uncertainties could ne be studied; however, from the above discussion
we can again anticipate that unless significant improvements, PDF
uncertainties compromise this prospect.

\subsection{Discussion and expected improvements at the LHC}

\begin{figure}[htbp]
    \begin{center}
      \includegraphics[width=.385\textwidth]{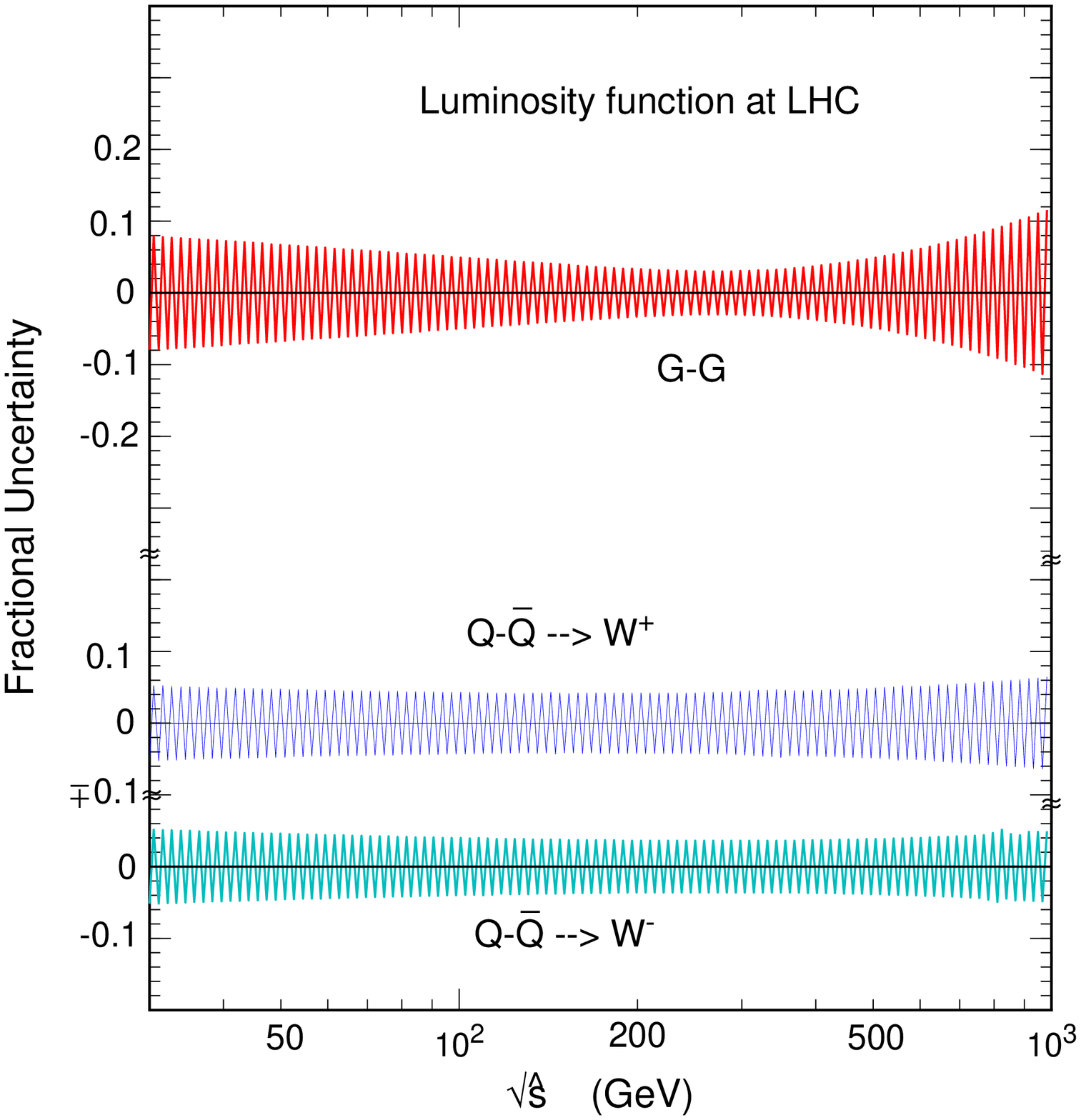}
      \includegraphics[width=.52\textwidth]{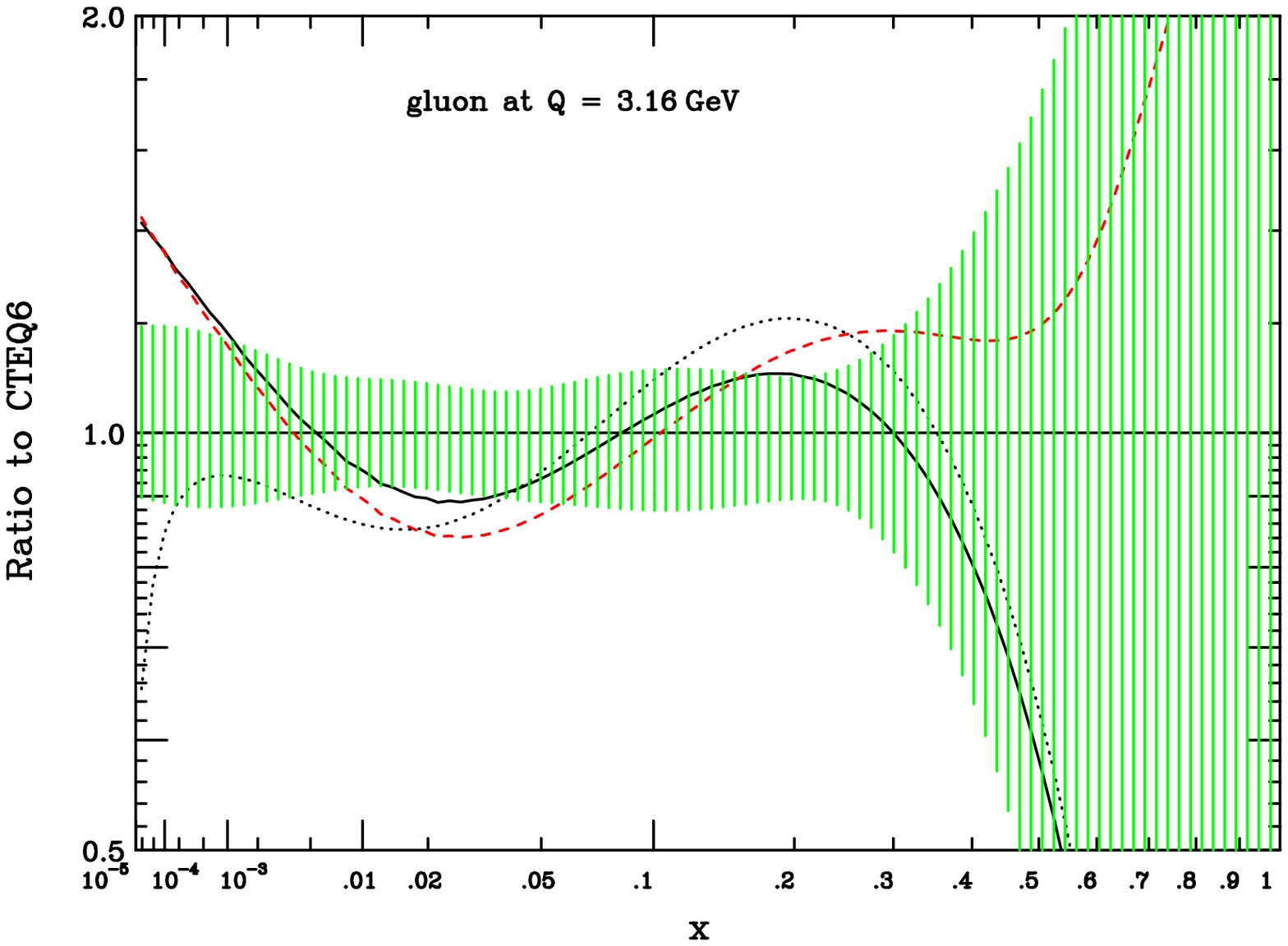}
    \end{center}
    \caption{\label{glulumiprediction} Gluon-gluon luminosity and
    its uncertainty, as a function of $\sqrt{\hat{s}}$~\cite{cteq}.}
  \end{figure}

As we already mentioned, the above examples are direct results of the
gluon density uncertainty. A convenient way to show this is displayed
in Fig.~\ref{glulumiprediction}. Writing

\begin{equation}
\sigma(M,y) \,\,\, = \,\,\, \int g(x,M) g(s/(Mx),M) \hat\sigma(M) 
\,\,\, \equiv \,\,\, {\cal L}(M,y) \hat\sigma(M),
\end{equation}

\noindent one can compute the ``luminosity'' ${\cal L}$ directly from
the PDFs, and estimate the cross section from the product of ${\cal L}$
with the hard process cross section.

Fig.~\ref{glulumiprediction}, left, shows the uncertainty on $\cal
L$ as a function of $M$. The uncertainties quoted in the
previous section are observed here, with $\delta {\cal L/L} \sim
5-10\%$ up to 1 TeV. On the right, the gluon uncertainty itself is
displayed at the initial scale ($Q \sim 3$~GeV), and one observes a
very rapid growth for $x > 0.2-0.3$. At LHC, we have essentially no
prediction for gluon initiated processes above $M \sim 3$~TeV.

The first way that comes to mind to improve the situation is the
analysis of the jet cros-section itself. There are however several
complications. First of all, as we said, the high mass or high $E_T$
spectrum should in principle be avoided because the possible
appearance of new physics effects in this region.

\begin{figure}[htbp]
    \begin{center}
      \includegraphics[width=.8\textwidth]{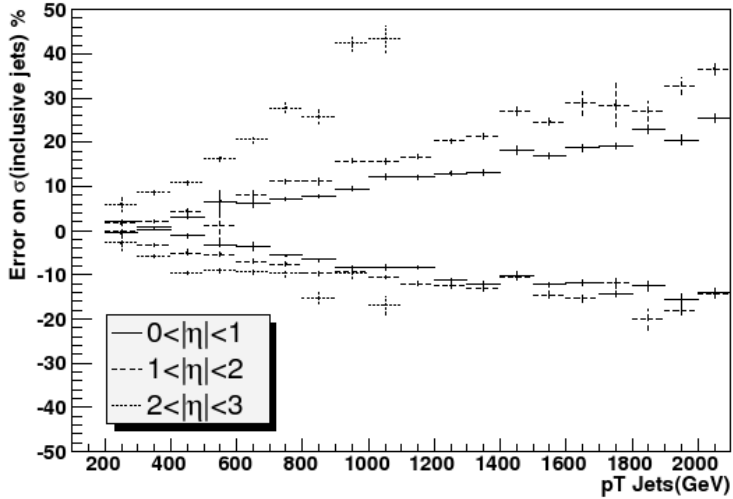}
    \end{center}
    \caption{\label{dijetprospect} Expected measurement precision of
    the dijet mass spectrum as a function of $m_{JJ}$, for three
    rapidity regions~\cite{clements1}.}
  \end{figure}

In addition, jet reconstruction involves many difficult experimental
and theoretical issues. Jet reconstruction algorithms, and the
experimental control of the jet energy scale and resolution both
affect the shape of the measured $M$ and $E_T$ spectra. The study
presented in~\cite{clements1} assesses the improvement on the gluon density
from the analysis of dijet events. Fig.~\ref{dijetprospect} displays
a projected dijet mass spectrum measurement, including the jet energy
scale uncertainty. 

\begin{figure}[htbp]
    \begin{center}
      \includegraphics[width=.6\textwidth]{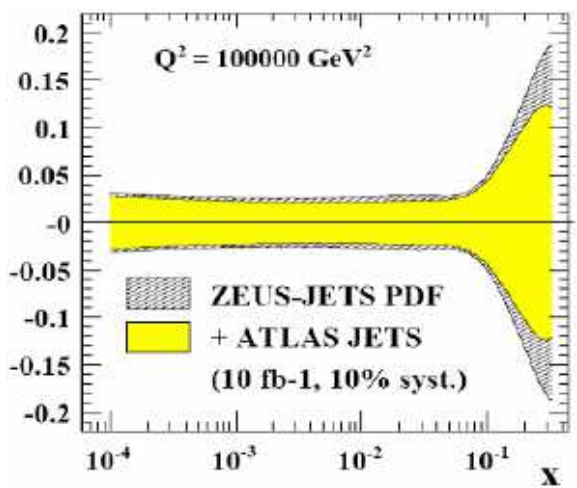}
    \end{center}
    \caption{\label{glulumiprospect} Expected improvement on the gluon
    density from the analysis of dijet events at the LHC.}
  \end{figure}

Although the statistical sensitivity is almost asymptotical, a jet
energy scale of a few percent already limits the exploitation of
the measurement result in terms of physics. Consequently, the
improvement on the gluon density from this process is rather modest as
shown in Fig.~\ref{glulumiprospect},
unless the JES uncertainty can be constrained to about 1\% above $E_T
\sim 1$~TeV - a challenging problem.
  
A complementary approach is to exploit direct photon spectra. While the
leading order cross sections $q\bar{q} \rightarrow
g\gamma/\gamma\gamma$ are coupled to quarks, the box processes $gg
\rightarrow g\gamma/\gamma\gamma$ are not strongly suppressed, and
enhanced by the very large gluon density. As a result, the gluon
initiated process dominates the overall rate at the
LHC. Moreover, since the quark densities are much better known
than the gluon density, the box process carries essentially all the
uncertainty. As a result, direct photons provide a sensitive probe to
the gluon PDF.

Another advantage of this process is the reduced sensitivity to the
jet observables. Photon reconstruction does not involve theoretically
sensitive algorithms, and the precision on the photon energy scale
will reach 1\% or better~\cite{atlasdetector}. While completely jet-free, direct
photon pairs have a low cross section, and potentially the same
physics bias as dijets, since non resonance extra-dimension effects
can appear through direct photon pairs as well.

\begin{figure}[htbp]
    \begin{center}
      \includegraphics[width=.56\textwidth]{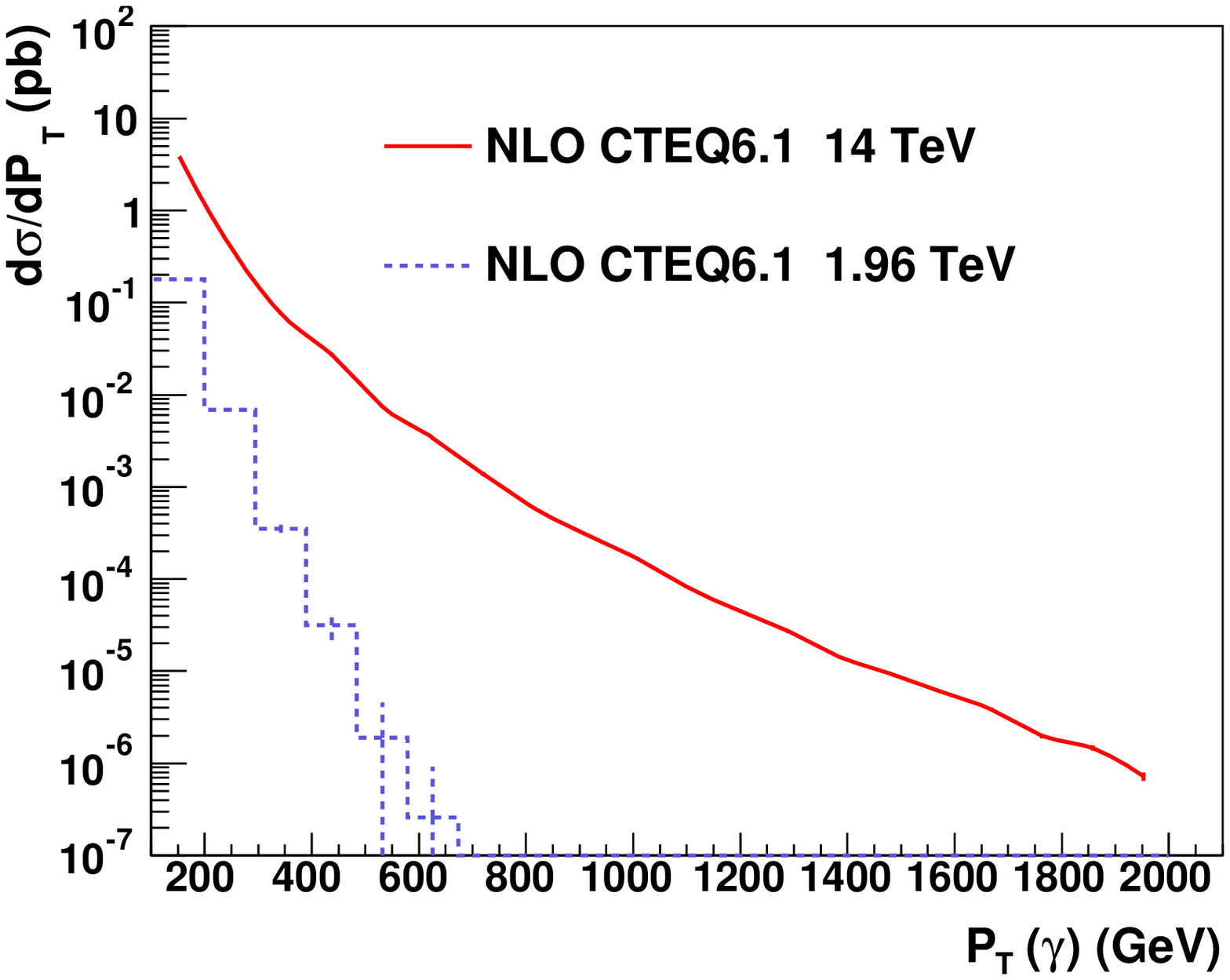}
      \includegraphics[width=.42\textwidth]{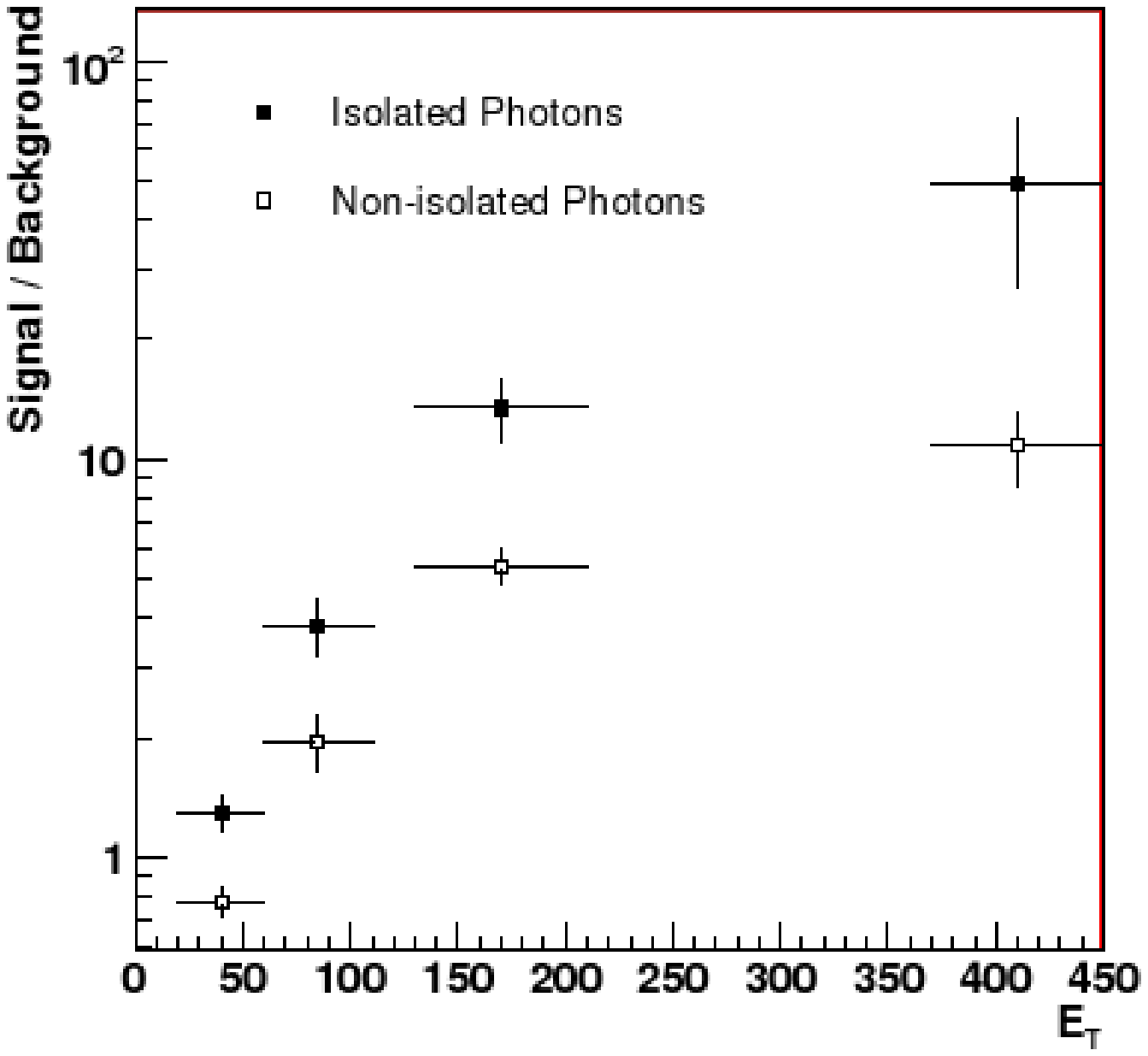}
    \end{center}
    \caption{\label{photonprospect} Left : expected direct photon production 
    cross section, at  
      Tevatron and LHC. Right : signal-to-background ratio, as a function of $E_T$.}
  \end{figure}

Photon-jet processes are a convenient midway. The cross section is
sufficient and extra-dimensional new physics are unlikely to provoke
an effect in this final state. Jet reconstruction difficulties
can be circumvented if one restricts the analysis to the photon $E_T$
spectrum.  Fig.~\ref{photonprospect} illustrates the expected
cross section of the direct photon $E_T$ spectrum at LHC, in
comparison with Tevatron, and the experimental separation between
photon and jets as a function of their transverse energy. As can be
seen, the signal purity is adequate in the high-$E_T$ region where
this measurement is relevant.

Note that while photons of moderate transverse energy, $E_T \sim
50$~GeV, such as those expected from Higgs boson decays, have
important backgrounds from jets (the expected signal-to-background
ratio is $\sim 1$), the background rate decreases rapidly
with $E_T$. Above $E_T > 500$~GeV, the photon sample is sufficiently
pure~\cite{ivanhollins}, with a signal-to-background ratio in excess of $10^2$,
not to affect the interpretation significantly. 

\begin{figure}[htp]
    \begin{center}
      \includegraphics[width=.75\textwidth]{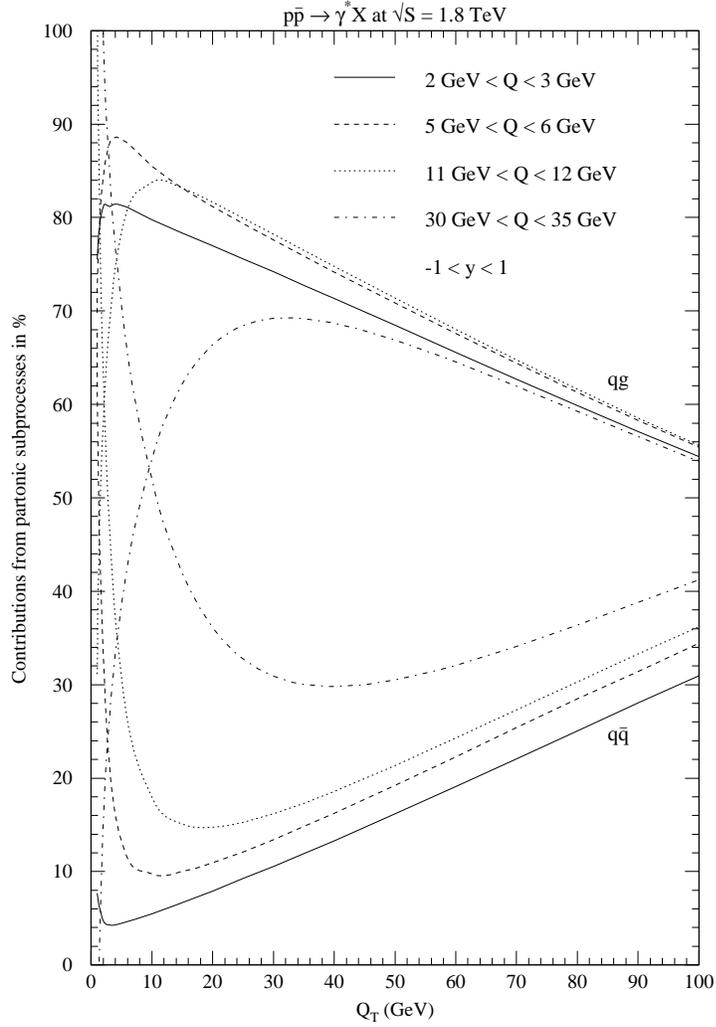}
    \end{center}
    \caption{\label{wzjet} $q\bar{q}$ and $qg$ contributions to
    the $\gamma^* \rightarrow l^+ l^-$ production cross section at the
    Tevatron, as a function of $p_{T}$ and for different $Q$~\cite{bergerklasen}.}
  \end{figure}

Finally, weak boson production provides another robust probe of the
gluon density. While na{\"i}vely a quark-induced process, the proces
receives significant gluon-induced contributions at finite
$p_T$. The hard processes $q\bar{q} \rightarrow Zg$ and $qg
\rightarrow Zq$ are of similar magnitude $per$ $se$, but the larger
gluon density favours the second. Again, because of the better
knowledge of the quark density, the gluon initiated process carries
essentially all the uncertainty and hence the potential to constrain
the gluon density. This is illustrated in Fig.~\ref{wzjet} on the
similar example of Drell-Yan production at the Tevatron, where for
virtualities of $Q \sim 5-30$~GeV, and lepton pair $p_T$ above 10 GeV
the gluon initiated process dominates~\cite{bergerklasen}.

As above, jet reconstruction issues can be avoided by measuring only
the leptonic system. In $Z$ events, the electron or muon pair momentum
allows to select the high-$p_T$ region; in $W$ events, one has to rely
on missing transverse energy as well, making this process
experimentally more complicated.

As we will see in the next section, the vector boson $p_T$
distribution is uncertain not only due to PDF uncertainties
$stricto$~$sensu$, but also due to the mechanisms of repeated soft
gluon emission. However, this particular uncertainty mostly concerns
the region of moderate transverse momentum, $p_T < 50$~GeV. Above this
threshold, the spectrum is not affected any more by resummation
effects and can be used to constrain the gluon density.

\subsection{Observables less sensitive to the uncertanties on the gluon density}
\begin{figure}[htp]
    \begin{center}
      \includegraphics[width=.78\textwidth]{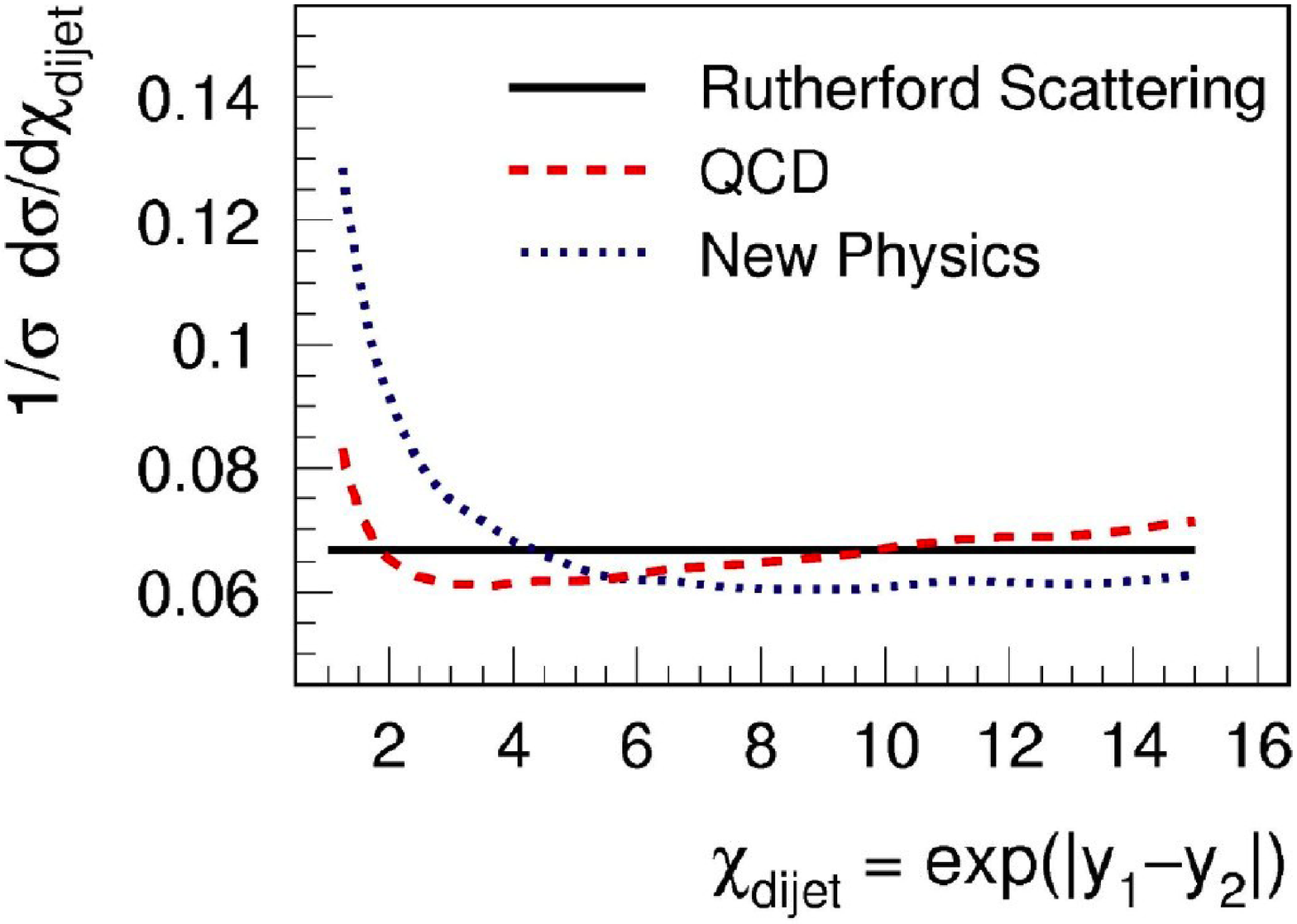}
    \end{center}
    \caption{\label{chi_d0}$\chi_{dijet}$ distribution for Rutherford
    scattering, QCD calculations and new physics (compositeness,
    extra-dimensions...) }
  \end{figure}

Another idea complementary to the discussion we just had about constraining further
the gluon density in the proton is to find other observables which are less
sensitive to PDF uncertainties but still to beyond standard model effects.
We will just quote one example of such observables $\chi_{dijet}$, related to the 
jet angular distribution~\cite{chid0} in dijet events:
\begin{eqnarray}
\chi_{dijet} = \exp(|y_1-y_2|)=\frac{1+ cos \theta^*}{1-cos \theta^*}
\end{eqnarray}
where $y_1$, $y_2$ are the rapidities of the two jets and $\theta^*$ is the
center-of-mass scattering angle. The expected distributions are given in
Fig.~\ref{chi_d0} for Rutherford scattering, QCD and new physics (compositeness,
extra-dimensions...). The distribution is flat for Rutherford scattering,
slightly shaped for QCD, and strongly enhanced at low $\chi_{dijet}$ in the case of
quark compositeness or extra-dimensions. The idea is thus to measure normalised
distributions as a function of $\chi_{dijet}$ as the one shown in Fig.~\ref{chi_d0}
since experimental and theoretical (mainly PDF related) uncertainties cancel. This
observable is a direct way to assess beyond standard model effects such as
compositeness or extra-dimensions without suffering from the uncertainties of the
gluon density at high $x$. On the contrary, the dijet mass cross section is of course sensitive to
such beyond standard model effects as an example, but is directly sensitive to PDF
uncertainties as well.

\subsection{Summary on the gluon density}

The discovery potential of new particles in gluon-initiated processes
is not strongly affected for moderate masses, up to $m \simeq
1$~TeV. We gave the example of the Higgs boson search, but the same
conclusions hold for supersymmetric particle searches. 

For higher masses, the sensitivity is strongly limited by the gluon
PDF uncertainty. At or above $m \simeq 2$~TeV, PDF uncertainties are
in excess of 50\% and prevent the interpretation of the observed
spectra in terms of new physics. A possible way out is to use dijet angular
distributions.

We have argued that jet measurements themselves are difficult to
exploit in constraining the PDFs due to jet reconstruction
difficulties, and the danger to absorb non resonant new physics into
the PDFs. However, events with photons are better determined
experimentally (in particular the energy scale). Photon-jet events
avoid large classes of non-resonant new physics, and the photon $E_T$
spectrum in these events constitute a robust probe of the gluon
PDFs. To avoid backgrounds, the high transverse energy range ($E_T >
300-500$~GeV) should be favoured.

The range $50 < p_T < 300-500$~GeV can be covered by events with weak
bosons. Avoiding the low $p_T$ region affected by resummation
uncertainties, these samples have low backgrounds and benefit from
precise reconstruction, making them good probes of the gluon density.

\section{The quark densities in the proton}

In this section, we will discuss the uncertainties related to the quark densities
in the proton. As in the previous section, we will follow two different approaches:
how are the searches at LHC dependent on the quark density uncertainties and
are there clever observables reducing their impact, as well as how can the
knowledge of quark PDFs be improved at LHC?

\subsection{W and $Z$ production}

The $W$ and $Z$ production cross sections (and their ratios) are often
regarded as precise tests of QCD. Indeed, the available NNLO
calculations~\cite{fewz} claim residual uncertainties below 1\%. 

\begin{figure}[htbp]
    \begin{center}
      \includegraphics[width=.50\textwidth]{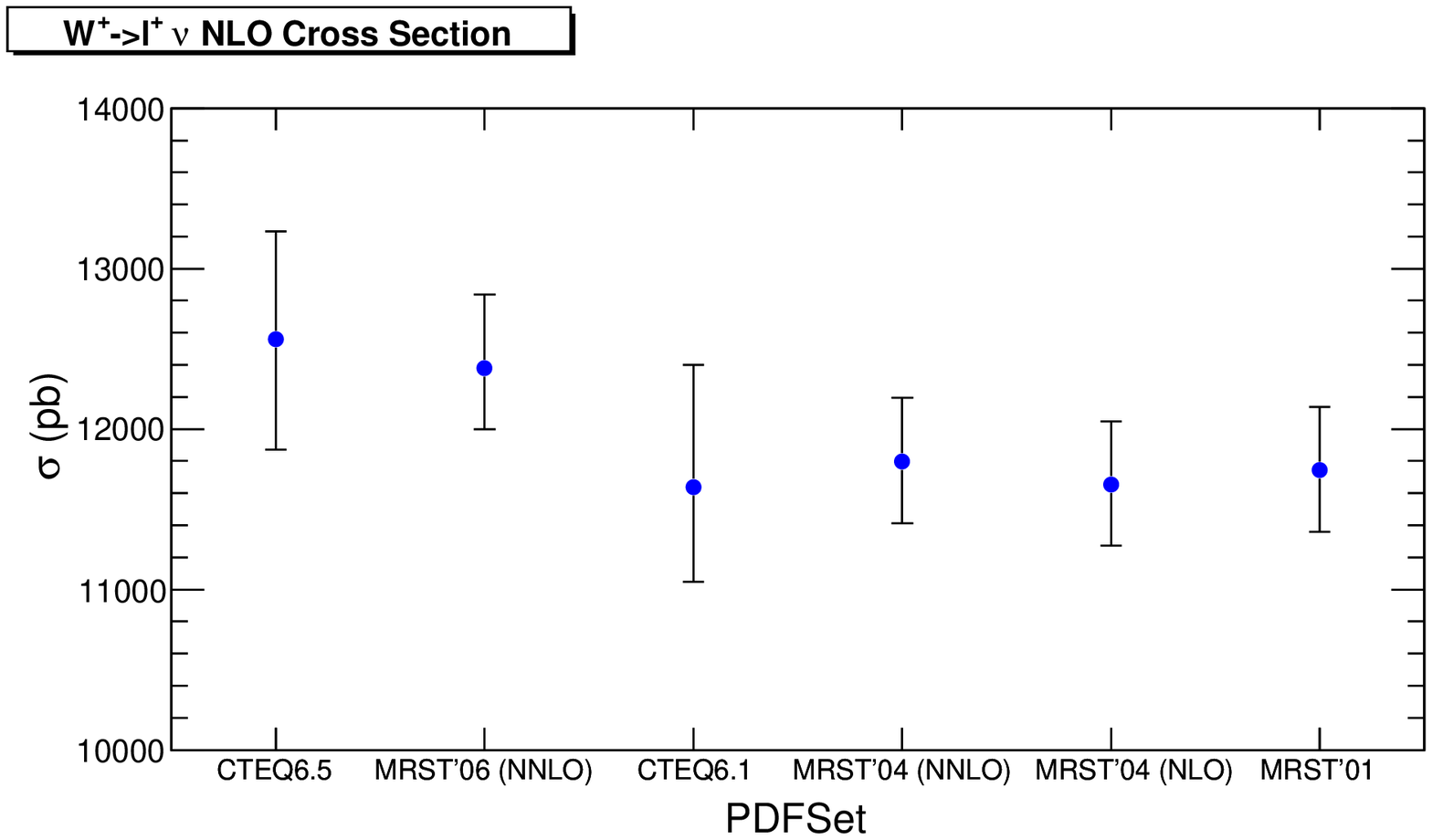}      
      \includegraphics[width=.49\textwidth]{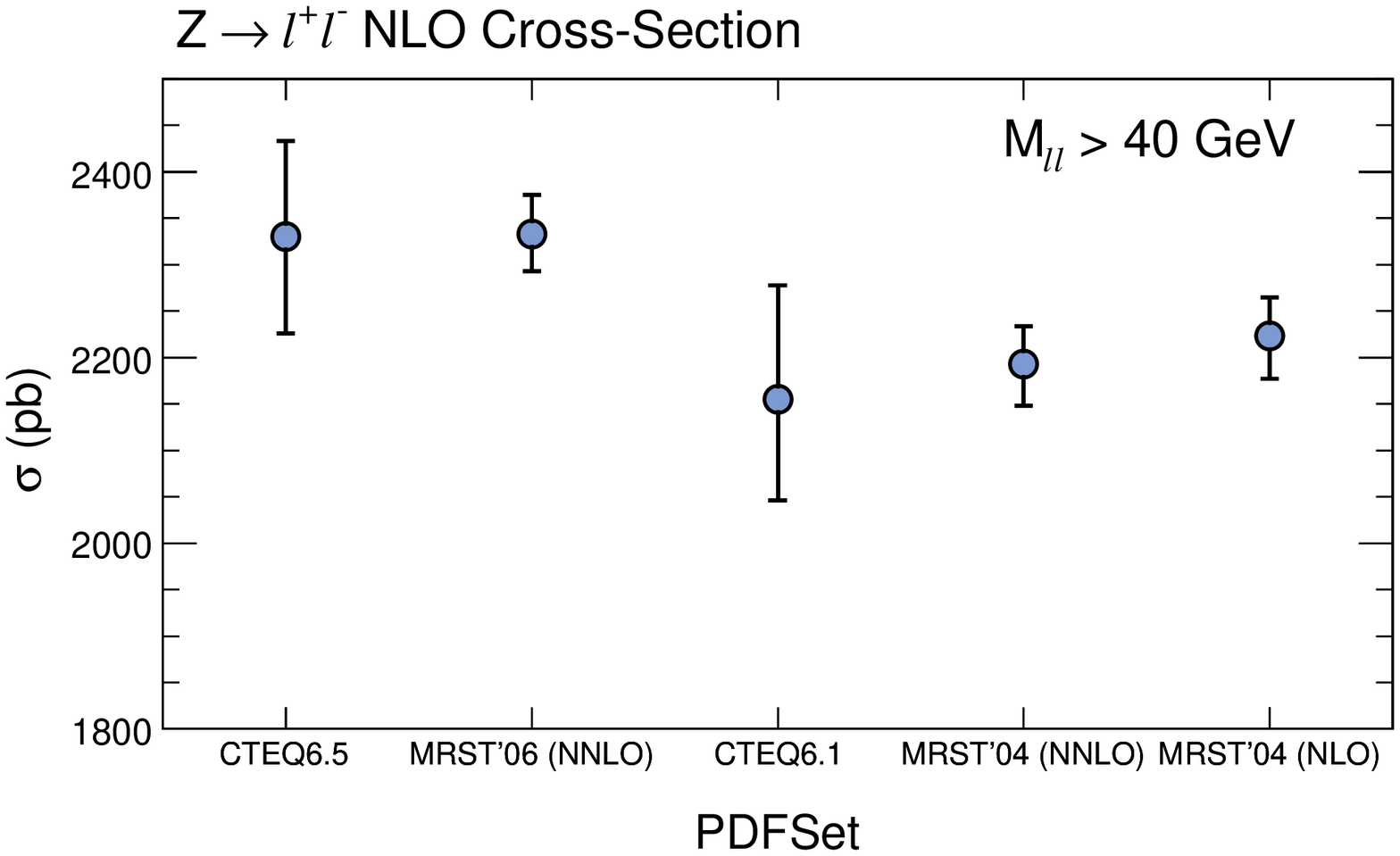}
    \end{center}
    \caption{\label{wzxsec} $W$ and $Z$ total cross sections and their
    uncertainties, as predicted 
    by different PDF sets~\cite{adamhalyo1,adamhalyo2}.}
  \end{figure}

However, it is worthwhile to mention the evolution of the $W$ and Z
cross section predictions with recently published PDF sets. The CTEQ
collaboration has produced several sets with different underlying
assumptions. A summary is proposed in Fig.~\ref{wzxsec}~\cite{adamhalyo1,adamhalyo2}. 

Compared to CTEQ6.1, CTEQ6.5~\cite{tungHQ} introduced a formalism accounting
for the masses of heavy flavour initial quarks, resulting in an 8\%
increase of the $W$ and $Z$ cross sections. While it is argued that this
number should not be taken as an uncertainty, the theoretical
framework is not unique.

In the most recent set, CTEQ6.6~\cite{cteq66}, the assumption that the strange
quark density is given by the u and d sea quark densities, $s = \bar{s} =
\kappa(\bar{u} + \bar{d})$, was released. This resulted in another
increase of the cross sections by about 2-3\%. Again, the net increase
should not be regarded as an estimate of the uncertainty, but it is
well known that the strange quark density is poorly constrained, and the
question arises whether the analysis of Ref.~\cite{cteq66}, and in particular the
choice of the strange quark initial parametrisation and the assumption
$s=\bar{s}$ allows to fully reflect the uncertainty related to this
flavour.

\begin{figure}[htbp]
    \begin{center}
      \includegraphics[width=.45\textwidth]{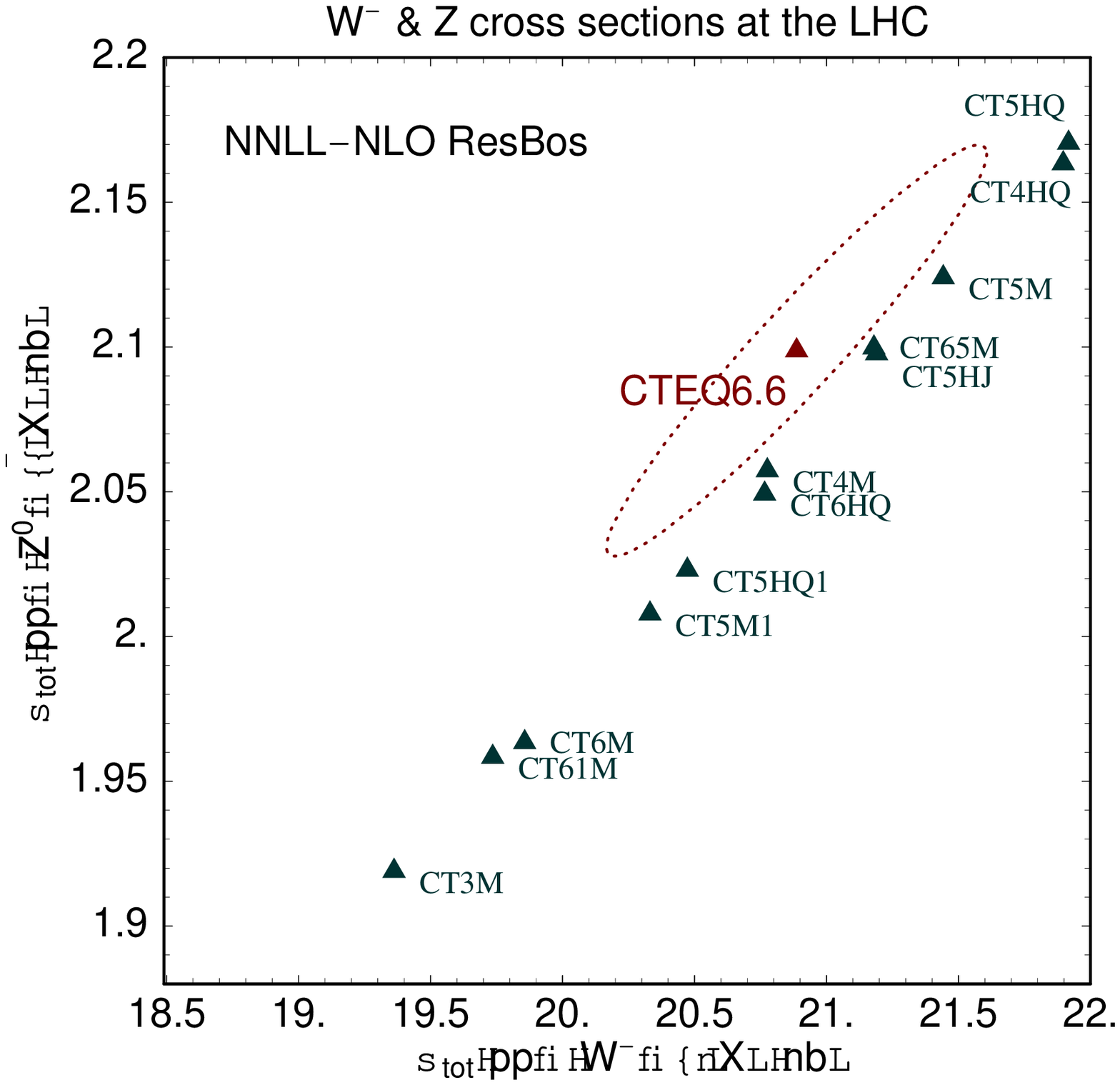}
      \includegraphics[width=.45\textwidth]{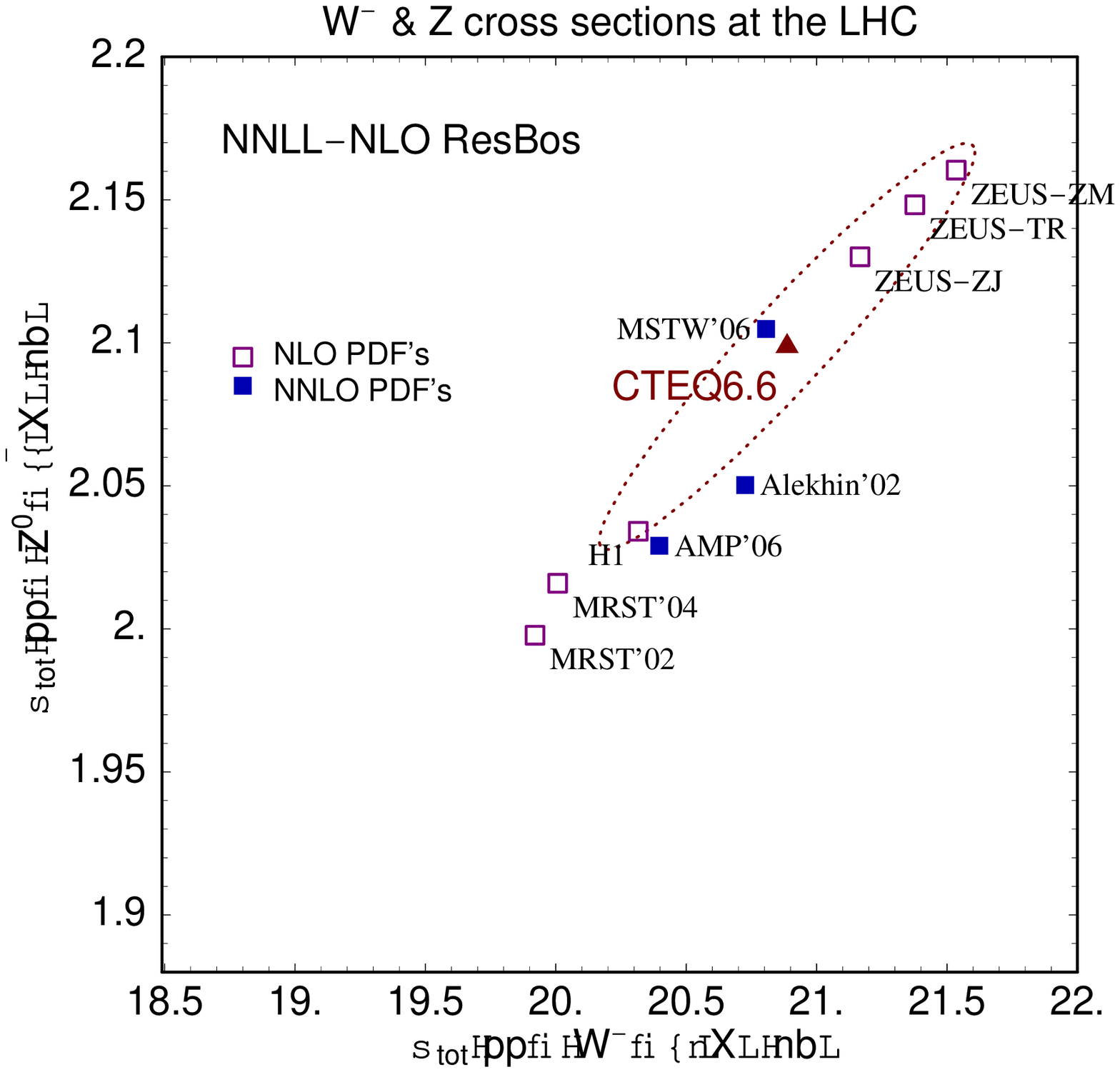}
    \end{center}
    \caption{\label{wzratio} Correlation between the $W$ and $Z$ total
    cross sections, as predicted by several PDF sets~\cite{cteq66,cteq,mrst,alekhin}.}
  \end{figure}

The release of the strange quark density plays a particular role in the
W$/$Z cross section ratio. Indeed, as shown in Fig.~\ref{wzratio},
this ratio was particularly stable in all previous PDF sets. As can be
seen, all shown predictions of this ratio are compatible with
each other (CTEQ or MRST; LO, NLO or NNLO evolution), except the
prediction of CTEQ6.6. The free strange quark density decorrelates W
production from $Z$ production. Note also that the CTEQ6.6 prediction
agrees better than earlier CTEQ versions with predictions by other
groups. This agreement might be coincidental, since the other groups all
assume fixed strangeness. This issue remains to be clarified.

From the above examples, it appears that $W$ and $Z$ production, and even
their ratio, are very sensitive to the details of the proton PDFs. An 
interpretation of these measurements in terms of genuine QCD
(i.e. $\alpha_S$ corrections to the hard process) is thus difficult,
pending significant improvements. As shown by the ratio example,
the problems do not come from the limited perturbative expansion of
the PDF evolution, but rather from the starting assumptions, in
particular the strange quark density in the low-$Q^2$ proton.

\begin{figure}[htbp]
    \begin{center}
      \includegraphics[width=.6\textwidth]{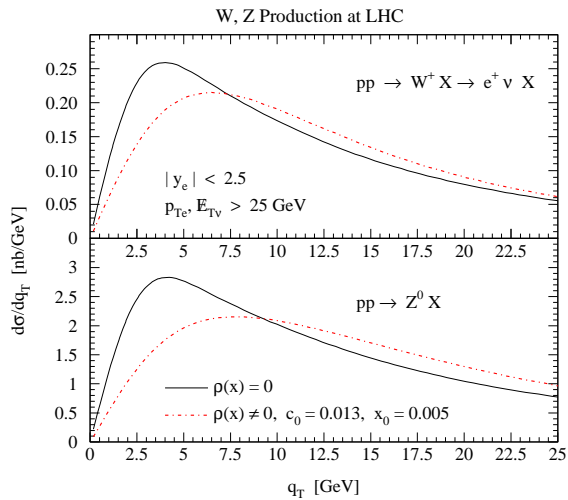} 
    \end{center}
    \caption{\label{qtb} $W$ $p_T$ distribution in the absence or
    presence of non-perturbative contributions to the soft gluon
    resummation~\cite{bergenadolsky}.} 
  \end{figure}

A more intricate example is given by $W$ and $Z$ $p_T$ distributions. While not
purely a PDF problem, the gluon emissions are determined by the
Sudakov form factors, which in turn are PDF integrals. The fact that $W$ and Z
couple to different initial partons creates again subtle differences
between the two processes. This distribution can also exhibit
uncertainties related to the evolution itself; at low $x$, BFKL-like
evolution can generate additional ``broadening'' of the $p_T$
distribution, as illustrated in Fig.~\ref{qtb}~\cite{bergenadolsky}.

\subsection{Precision electroweak measurements}

The program of testing the SM at the quantum level, through precise
measurements of the electroweak parameters, will be pursued at
LHC. Improvements in the measurements of notably $m_W$ and $m_t$ will
result in improved predictions for the Higgs boson mass;
predictions which hopefully be confronted to the observed value of
$m_H$. 

The $W$ mass measurement is affected by PDFs through acceptance
effects. Unlike the cross section example, the PDF normalization is
irrelevant; however, their $x$-dependence determines the $W$ rapidity
distribution. The kinematical distributions of the $W$ sample that
passes acceptance cuts are thus affected. As a result, the
distributions that enter the measurement of $m_W$, i.e the transverse
momentum of the charged decay lepton, $p_T(l)$, and the transverse
mass of the lepton-neutrino pair, can be mis-modeled, and this
mis-modeling can be wrongly absorbed in the mass measurement.

\begin{figure}[htbp]
    \begin{center}
      \includegraphics[width=.6\textwidth]{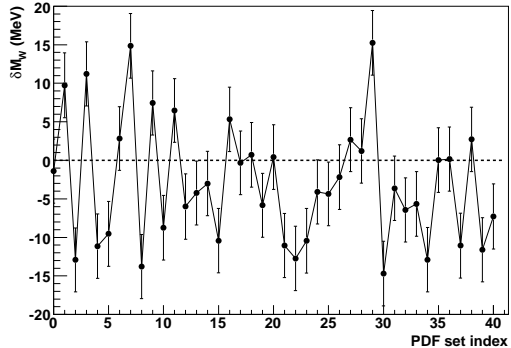}
    \end{center}
    \caption{\label{mwsyst} Expected shift on the fitted value of
    $m_W$, for PDF sets differing from the best fit by one standard
    deviation in each of the free parameters. Biases are shown as
    predicted by CTEQ6.5. The total systematic uncertainty is
    essentially obtained from the quadratic sum of all biases, and
    amounts to 25 MeV~\cite{besson}.}
  \end{figure}

The LHC prospects for the measurement of $M_W$ have been discussed
in~\cite{krasny,besson}. With current data, the PDF uncertainties include a
systematic uncertainty on $M_W$ of about 25~MeV,
cf. Fig.~\ref{mwsyst}.

The desired precision on $m_W$ is much better. The statistical
sensitivity is below 2~MeV, and given the current precision on the top
quark mass, an uncertainty $\delta m_W \sim 10$~MeV is desirable.
Therefore, such a PDF uncertainty is prohibitive.

\begin{figure}[htbp]
    \begin{center}
      \includegraphics[width=.45\textwidth]{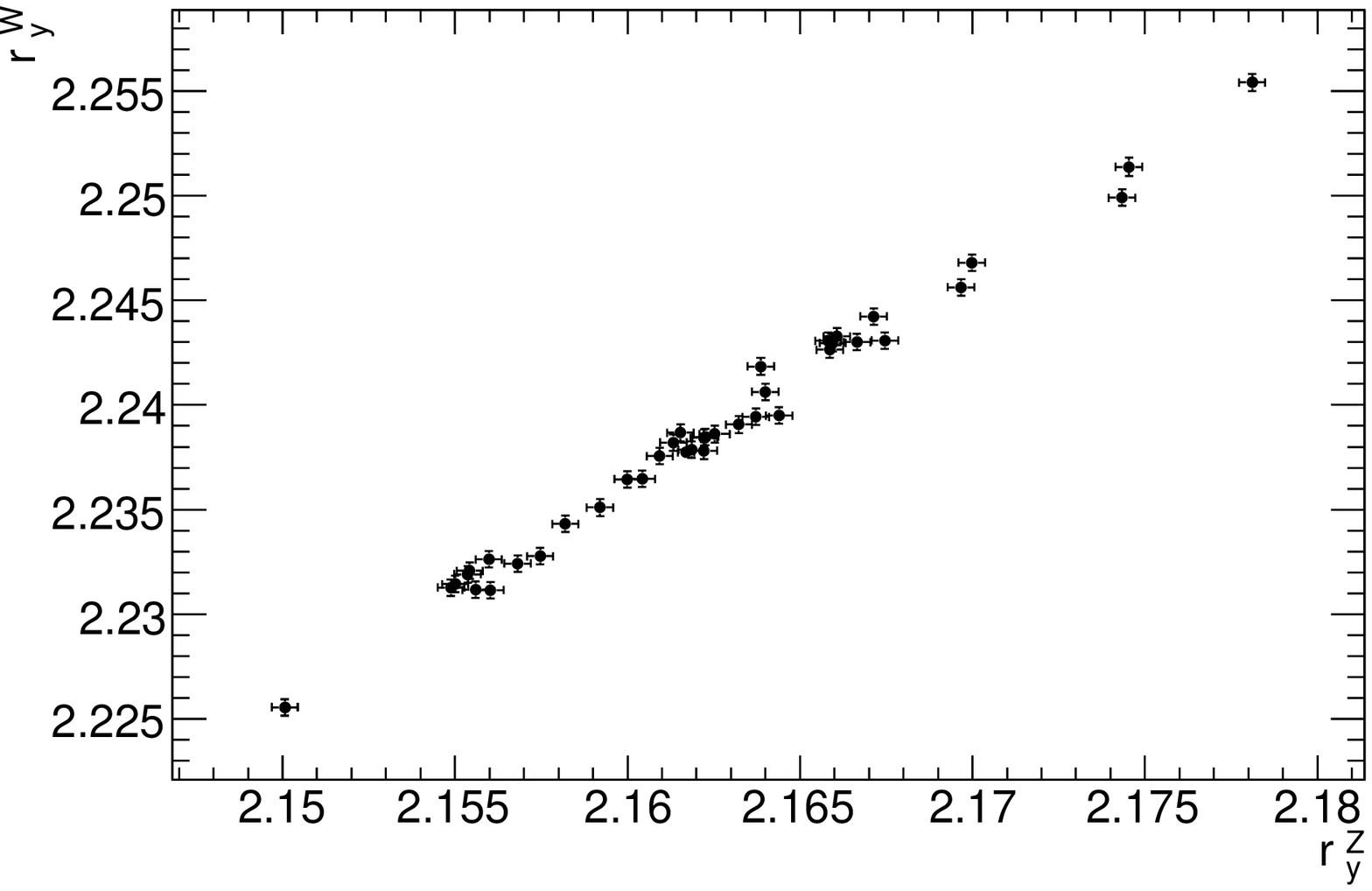}
      \includegraphics[width=.45\textwidth]{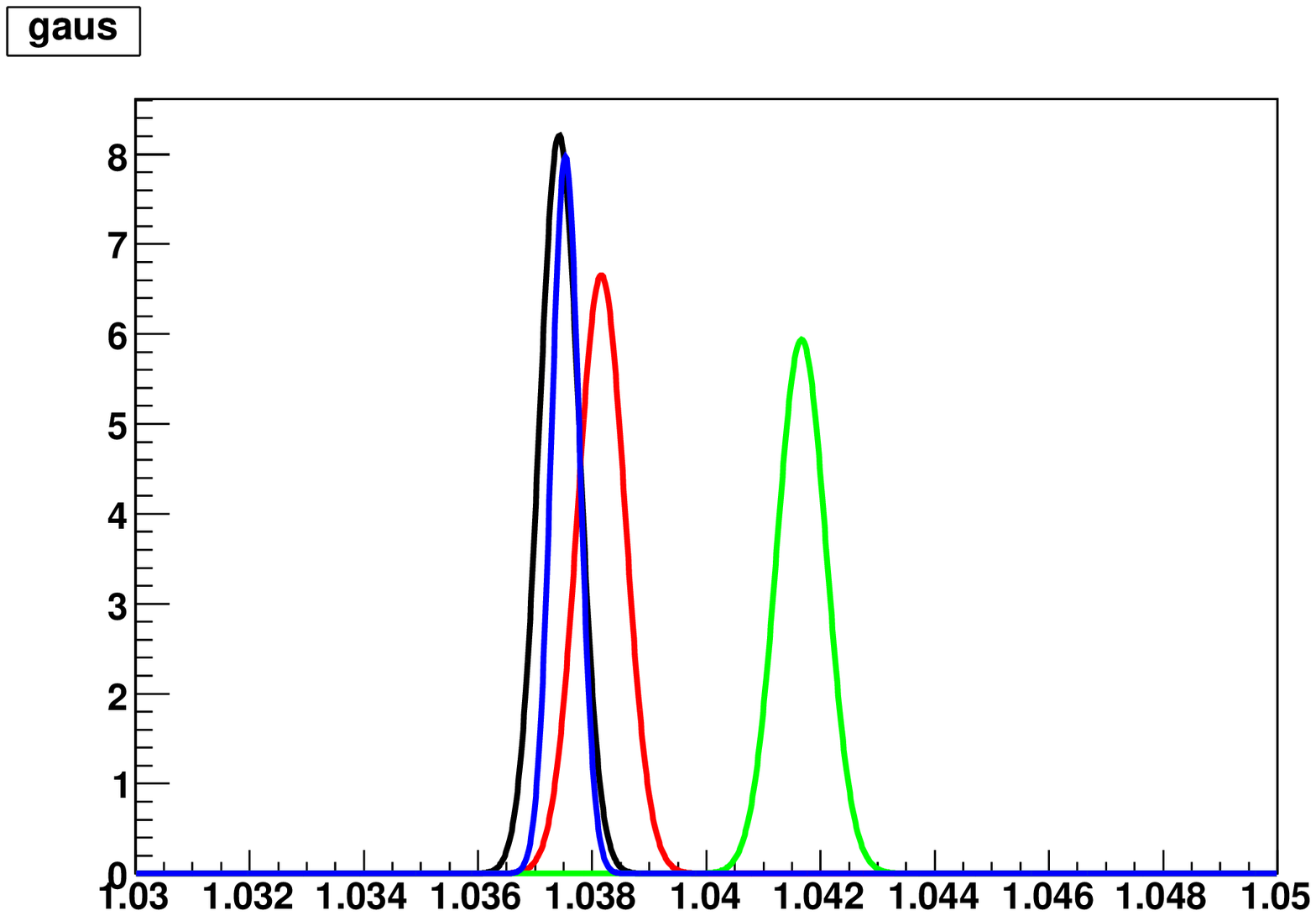}
    \end{center}
    \caption{\label{wzratio2} Left : spread (RMS) of the rapidity distributions of $W$ vs. Z
    events, plotted for the 41 CTEQ6.1 PDF sets~\cite{besson}. Right :
    ratio of these spreads for $W$ and Z, as predicted by, from left to
    right, CTEQ61~\cite{cteq}, CTEQ65~\cite{tungHQ},
    MSTW2006~\cite{mrst}, and CTEQ66~\cite{cteq66}.}  
  \end{figure}

As shown in ~\cite{besson} (cf. also Fig.~\ref{wzratio2}, left), the situation
can be greatly improved by measuring the $Z$ boson rapidity distribution
at LHC, and exploiting the expected correlation of these
distributions between $W$ and $Z$ events. In first approximation, the
measurement of the $Z$ rapidity distribution constrains the $W$ one to the
point of reducing the PDF systematic ucertainty to below 1 MeV.

However, as for the cross sections, this picture is questioned by the
CTEQ6.6 PDF sets. As shown in Fig.~\ref{wzratio2}, right, the free strange
quark density produces a decorrelation between $W$ and $Z$ distributions which
partly obscures the interpretation of the $Z$ rapidity distribution. It
is remarkable that three fits with very different theoretical
assumptions (CTEQ61 - NLO; CTEQ65 - with
improved heavy quark treatment, and MSTW2006 - NNLO), but identical
hypotheses on the initial proton parametrization, reach the same
result, whereas the CTEQ66 prediction, with its strange quark degrees of
freedom, is significantly different.

\subsection{Heavy gauge bosons}

If new, heavy gauge bosons exist and are related to the electroweak
symmetry breaking, their mass should be in the TeV range and
accessible at LHC~\cite{atlastdr,cmstdr}. In addition to the determination of
their mass, which can be performed through a straightforward
determination of the peak position in the invariant mass or transverse
mass spectra, it is also important to determine their couplings as
well as possible. This can in principle be done  through a more
difficult analysis of the lineshape, determining the interference
pattern with the gauge bosons of the Standard Model. Such analyses
would be reminiscent of the LEP2 $e^+ e^- \rightarrow f\bar{f}$
cross section measurements~\cite{lep2ffbar} which, providing data with a precision of
about 1\%, put stringent constraint on such new physics. For these to
succeed at LHC, the standard model reference cross section should
thus be known with a precision not worse than 1\%.

\begin{figure}[htbp]
    \begin{center}
      \includegraphics[width=.45\textwidth]{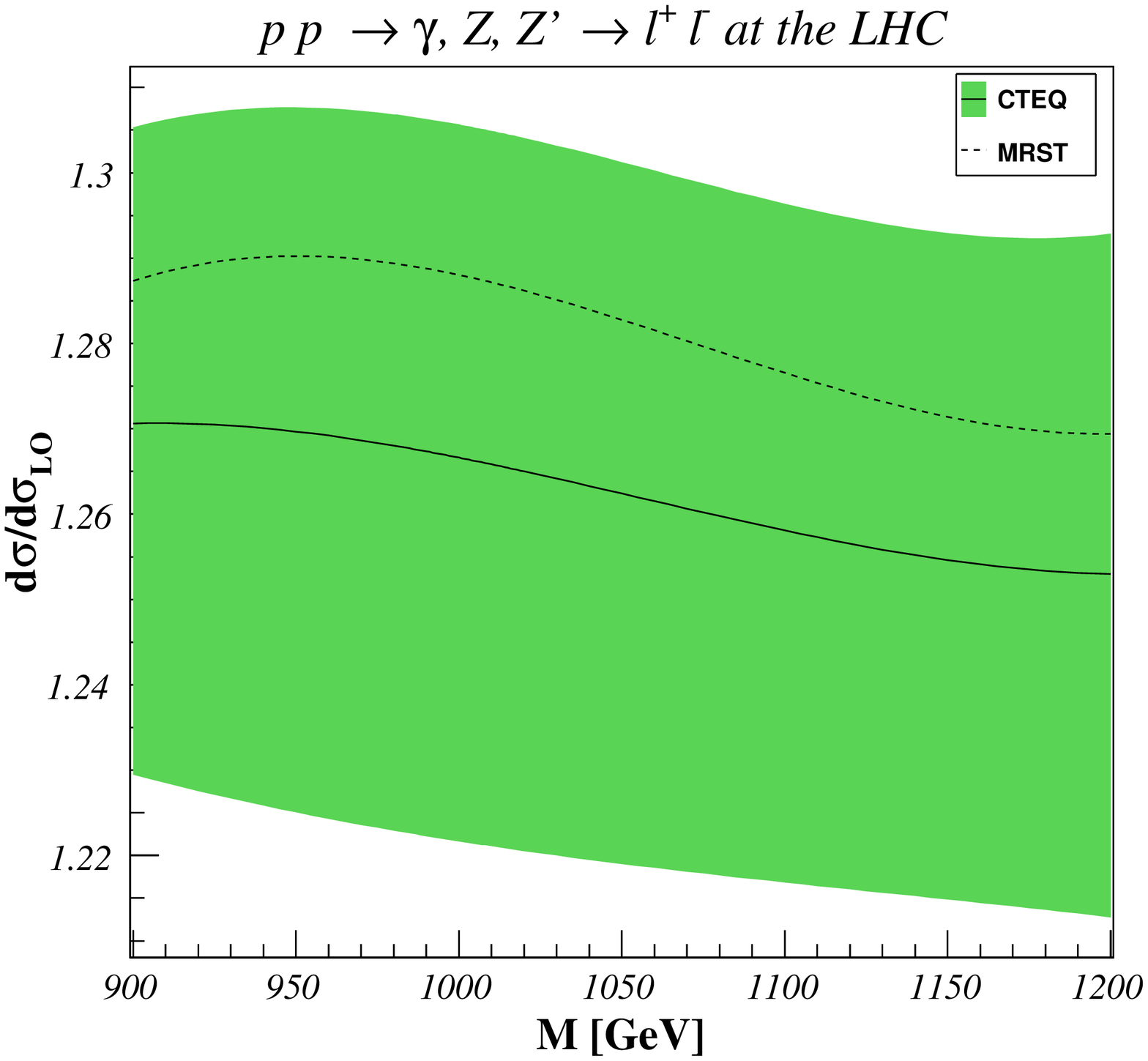}
      \includegraphics[width=.45\textwidth]{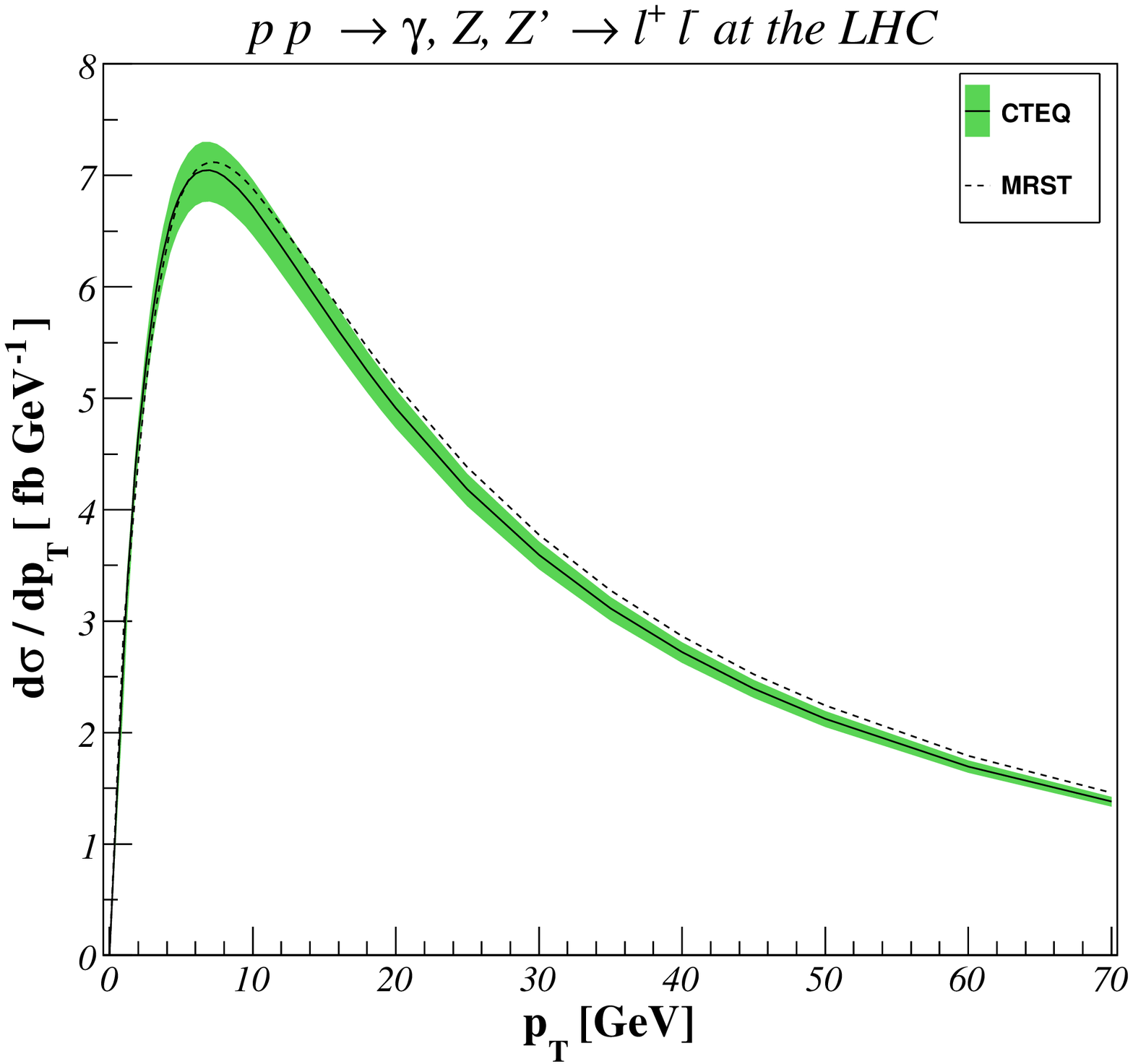}
    \end{center}
    \caption{\label{highmassDY} Left : High-mass Drell-Yan
    cross section and its uncertainty, in the region $m_{l^+ l^-} \sim
    1$~TeV. Right : expected $p_T$ distribution in this region, and
    its uncertainty.}
  \end{figure}

As can be seen in Fig.~\ref{highmassDY}~\cite{ledroitmorel}, the PDF uncertainty
in the mass range $M \sim 1$~TeV is about 3\%. While not affecting the
discovery potential, such uncertainties are too strong for the
precision measurements mentioned above to be performed. It is thus
desirable to reduce these uncertainties by a factor 3 or more.

This can be obtained by constraining the relevant $x$ range, through a
combination of measurements of high-rapidity $Z$ boson events, and of
the low-mass Drell-Yan spectrum. This will be discussed in
Section~\ref{gstarz}.

\begin{figure}[htbp]
    \begin{center}
      \includegraphics[width=.5\textwidth,angle=90]{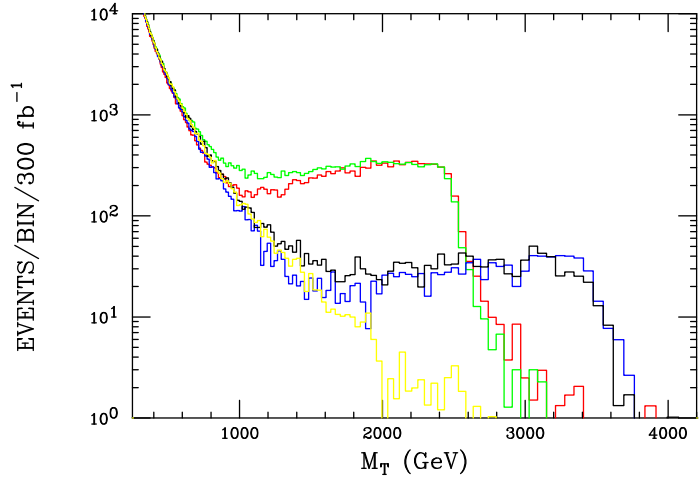}
    \end{center}
    \caption{\label{wprimehelicity} Transverse mass distributions of
    heavy W' bosons, for different mass and helicity hypotheses. The
    lower curve represents the Standard Model background, and the two
    pairs of curves correspond to $m_{W'}$=2.5 and 3.5~TeV, plotted
    for $\lambda = \pm 1$.}
  \end{figure}

A similar measurement was suggested in~\cite{rizzohelicity}, which proposes a
determination of the W' helicity, $via$ a measurement of the
transverse mass distribution well below the W' peak, as illustrated in
Fig.~\ref{wprimehelicity}. While the new jacobian peaks are clearly
visible, additional model discrimination can be obtained through an
analysis of the $M_T$ distribution well below the peak. Here again, it
is necessary to dispose of precise Standard Model benchmarks, and
similar arguments as above hold.

\subsection{Summary: opportunities for improvements at the LHC}
In the previous section we have illustrated the impact of the PDF
uncertainties on prominent example of the LHC physics programme. Many
of the questions raised could hopefully be improved at LHC  using
dedicated analyses. The cleanest way to improve on the gluon
distribution, in the high-$x$ region, is provided by the analysis of
high-$E_T$ direct photon production. In the low-$x$ region, the study
of high-$p_T$ vector boson production is easier and avoids the jet 
background when lower $E_T$ photons are used. Jet production
will provide useful information, but is subject to difficulties
related to the jet reconstruction. The sea quark distributions will be
constrained $via$ measurements of $W$ and $Z$ boson production at low
transverse momentum. The valence distributions are discussed in Section~\ref{ratios}.

Improving the knowledge on the $s$, $c$ and $b$ quark content of the
proton will need more specific measurements. While these are not new
ideas, there exists almost no literature on these subjects and it is
worthwhile to draw the attention to their importance.

The strange quark contents of the proton can hopefully be constrained
exploiting the associated production of $W$ boson with charm. This
process is dominantly produced through $sg \rightarrow cW$, hence
sensitive to the strange quark density. The process is experimentally
difficult to select, and in practive the $c$-jet can be identified through
the presence of specific $D^*$ decays. This mode provides sufficient
purity, but at the cost of a low efficiency. Further study is needed
to establish whether this process can be exploited.

The cleanest way to access the $c$ and $b$ densities is through $cg,
bg \rightarrow cZ, bZ$. The charm jet can be identified as
above. B jets are selected with good efficiency and purity using
b-tagging, but the measurement can be obscured by background processes
like $q\bar{q} \rightarrow Zg$, with subsequent ``gluon splitting'' $g
\rightarrow b\bar{b}$ (this background is also present in the charm
final states).

Such analyses have been pursued at Tevatron~\cite{cdfwcharm}. The higher
statistics at LHC should provide good prospects, but the
experimental and theoretical difficulties remain to be quantified.

\section{Another way to be less sensitive to PDFs: cross section ratios}
\label{ratios}

In this section, we will discuss another method to reduce the influence of PDF
uncertainties (both quark and gluons) by considering ratio of cross sections 
in a well chosen given kinematical domain. It follows somehow the idea of the 
$\chi_{dijet}$ observable but the discussion will be more complicated since 
the aim will be here to define and choose ratios showing a dependence on PDFs 
as low as possible. 

Measuring ratios of quantities are interesting from an experimental
point of view because correlated systematic uncertainties may cancel.
It can also provide more accurate information on PDFs since some processes
use the parton densities in the same $x$ or ($x$, $Q^2$) region. 
In a ratio of quantities related to these processes, the influence of PDFs 
could cancel. Another motivation for using ratios is to separate the effects due to
new physics from the ones due to insufficient knowledge of the PDFs. 
In this section, we develop some measurements based on ratios, for which 
the sensitivity to PDFs is reduced. Such measurements at LHC directly 
probes and then constrains the remaining PDF contribution~: $W$ charge 
asymmetry, $Z$ forward-backward asymmetry, $W$ over $Z$ cross section ratio, 
boson pair over single boson cross sections ratio and Drell-Yan cross sections ratio.

\subsection{$W$ charge asymmetry}
The $W$ charge asymmetry quantifies the imbalance between positive or 
negative $W$ bosons, produced at a given rapidity $y_W$~:
$$
{\cal A}_W(y_W)=\frac{d\sigma(W^+)/dy_W-d\sigma(W^-)/dy_W}{d\sigma(W^+)/dy_W+d\sigma(W^-)/dy_W}
$$

At LHC, the $W$-charge asymmetry can be linked to structure functions in a 
simple manner. 
In first approximation, $W$ bosons are mostly produced by up and down quarks 
and $\sigma(W^+)$ can be written as : 
$\sigma(W^+)=(u_{val}+u_{sea})\bar{d}_{sea}+q_{sea}\bar{q'}_{sea}$. 
Replacing this information in the definition above, and under the approximation 
that the {sea} quark content of the proton is the same for u, d, s, c quarks, 
one gets the following result~:
$$
{\cal A}_W(y_W)=\frac{u_{val}-d_{val}}{u_{val}+d_{val}+2q_{{sea}}}(x)
$$
At leading order, the $W$ charge asymmetry directly probes the 
difference between valence quarks. From the experimental side, it is 
preferred to measure the lepton-charge asymmetry because the $W$ boson 
cannot be fully reconstructed from its decay leptons. In this experimental 
ratio, the relation between PDFs and $W$ charge asymmetry is more complex 
but it still provides complementary measurements of valence PDFs.

\begin{figure}[htbp]
  \begin{center}
    \includegraphics[width=.8\textwidth]{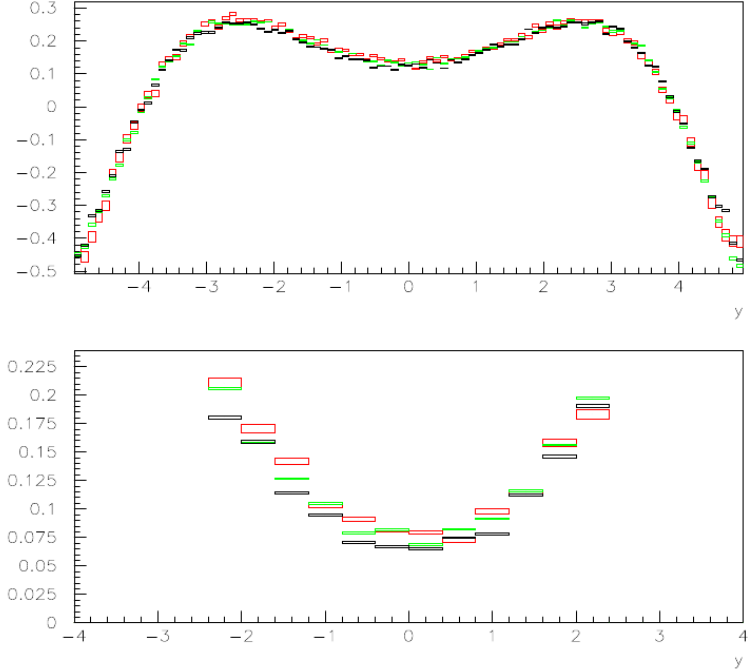}
  \end{center}
  \caption{\label{Wasym} Lepton charge asymmetry in $W$ decay as a
  function of lepton rapidity, generated using HERWIG and CTEQ6.1
  (dashed red) ZEUS-S (green) or MRST2001 (full black) PDF sets with full
  uncertainties, at the generator level (up) and after a fast
  simulation of ATLAS detector and reconstruction effects (bottom).} 
\end{figure}

The predictions of the lepton-charge asymmetry have large uncertainties 
due to the lack of data on valence quantities for both high- and small-x 
values. At high lepton rapidity, the different predictions using CTEQ6.1, 
MRST2001 or ZEUS-S PDFs are compatible but the uncertainty on the ratio 
is large ($\sim 10\%$) as shown in fig.~\ref{Wasym}. On the countrary, 
for centrally produced $W$ bosons, a 4-$\sigma$ discrepancy is present 
between the different $W$ charge asymmetry predictions using CTEQ6.1, 
MRST2001 or ZEUS-S PDFs~\cite{paperWasym}. The input parametrization of $u_v/d_v$ at $Q_0^2$ 
in global QCD fits could be improved with such measurement at LHC.

\subsection{$Z$ forward-backward asymmetry}
The forward-backward asymmetry in $Z$ events is an important measurement 
since it is sensitive to new physics effects and provides a determination 
of the mixing angle $\sin 2\theta_W$. The presence of both vector and 
axial-vector couplings of the quarks and leptons to the $Z^*/\gamma^*$ 
boson gives rise to an asymmetry in the polar emission angle $\theta^*$ 
between the quark and the lepton in the $Z^*/\gamma^*$ rest frame. 
The differential cross section for the parton level process can be written~:
$$
\frac{d\sigma}{d\cos\theta^*}
(q\bar{q}\rightarrow\gamma^*/Z\rightarrow l^+ l^-)=
{\cal A}(1 + \cos^2\theta^*) + {\cal B} \cos\theta^*
$$
The weak interaction introduces the asymmetry (${\cal B}\ne$ 0), and ${\cal A}$ and {\cal B} 
are functions of the weak isospin and charge of the incoming quarks 
and of the $Z^*/\gamma^*$ invariant mass. The asymmetry is given by 
the direction of the $Z$ with respect to the direction of the incoming 
quark, according to the definition below~:
\begin{eqnarray*}
{\cal A}_{FB} & = & \frac{\sigma(Z)_{\cos\theta^*>0}-\sigma(Z)_{\cos\theta^*<0}}{\sigma(Z)_{\cos\theta^*>0}+\sigma(Z)_{\cos\theta^*<0}}\\
\\
              & \propto & 1-{\cal K}\sin^2 \theta_W(M_Z^2))
\end{eqnarray*}

\begin{figure}[htbp]
  \begin{center}
    \includegraphics[width=.75\textwidth]{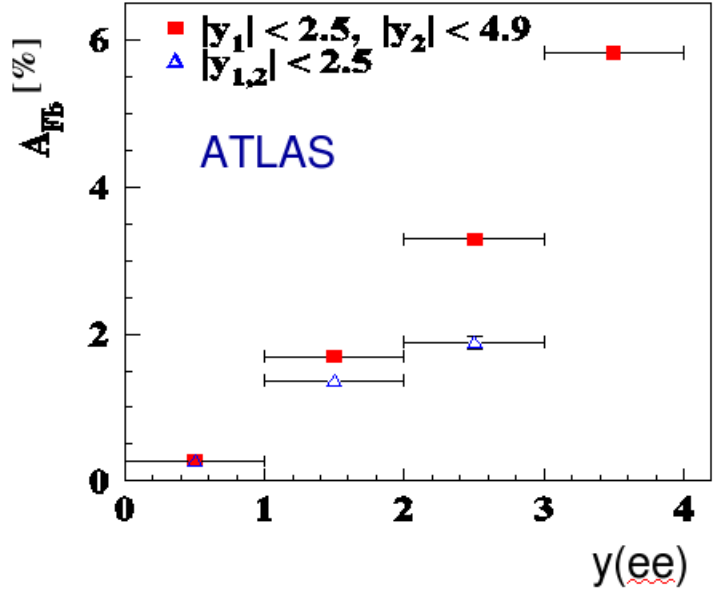}
  \end{center}
  \caption{\label{Zasym} Forward-backward asymmetry in $Z\to ee$
  events versus the reconstructed $Z$ rapidity, in the case where both electrons
  are reconstructed in the central region ($|\eta^e|<2.5$), or where 
  one electron is reconstructed in the forward calorimeter
  ($2.5<|\eta^e|<4.9$) and the other is central~\cite{atlascsc}. 
  The analysis with two forward electrons does not lead to significant 
  results because the rejection factor against fake electrons in 
  the forward calorimeter may not be high enough
  and suffers from large systematic uncertainties.} 
\end{figure}

To measure the $Z$ forward-backward asymmetry, it is necessary to tag 
the directions of the incoming quark and antiquark, which is a difficult 
task at LHC. Indeed, $Z$ bosons are mostly produced by sea partons. 
The quark and antiquark PDFs are almost equal and the direction of the 
incoming quark is no longer related to the direction of the $Z$. 
This is not the case at high $x$, for which the incoming 
quark is rather a valence quark than a sea quark. This asymmetry 
between sea and valence quarks is useful to distinguish from which 
of the two protons the quark was coming. In order to produce a $Z$ 
boson with a high-$x$ quark, the antiquark must come from a low-$x$ 
region. This produces a highly boosted $Z$ boson with a high-rapidity, 
and the direction of the incoming quark is now correlated with the 
direction of the $Z$. These bosons can be used for forward-backward 
asymmetry measurements. However, since leptons from high-rapidity 
boosted $Z$ bosons decay are at high pseudo-rapidity, they are likely 
to be reconstructed in the forward regions of the detectors. The 
analysis aiming at measuring this asymmetry needs to include the 
forward detectors, as shown in Fig.~\ref{Zasym}. Precise identification 
of forward leptons is then a difficult task, because of the high 
hadronic activity and because of the absence of tracking device. 
This requires a good understanding of forward physics, and the 
detector and reconstruction effects.\\

The main systematics come from the lepton acceptance and
reconstruction efficiency and from the PDF uncertainties. These ones
are quite large at high-$x$, and a measurement of $\rm {\cal A}_{FB}$
could directly constrain both $\rm \sin^2 \theta_W$ and the high-$x$
valence PDFs. 

\subsection{$W/Z$ production cross sections ratio}
$W$ to $Z$ bosons production cross sections ratios are motivated by both experimental and
theoretical aspects. At LHC, this ratio should have low statistical
and systematic errors. Firstly, the selection of such events relies on
isolated leptons in the same transverse momentum range, they can be
selected by the same trigger condition, and the hadronic environment is
expected to be similar. Many experimental uncertainties can cancel in
such ratios of cross sections. Secondly, both leading order processes
are similar (quark initial state, singlet final state). Higher order
corrections like initial state radiations, affect both processes in an
equal way, and many theoretical uncertainties cancel in this ratio.\\ 

\begin{figure}[htbp]
  \begin{center}
    \includegraphics[width=.8\textwidth]{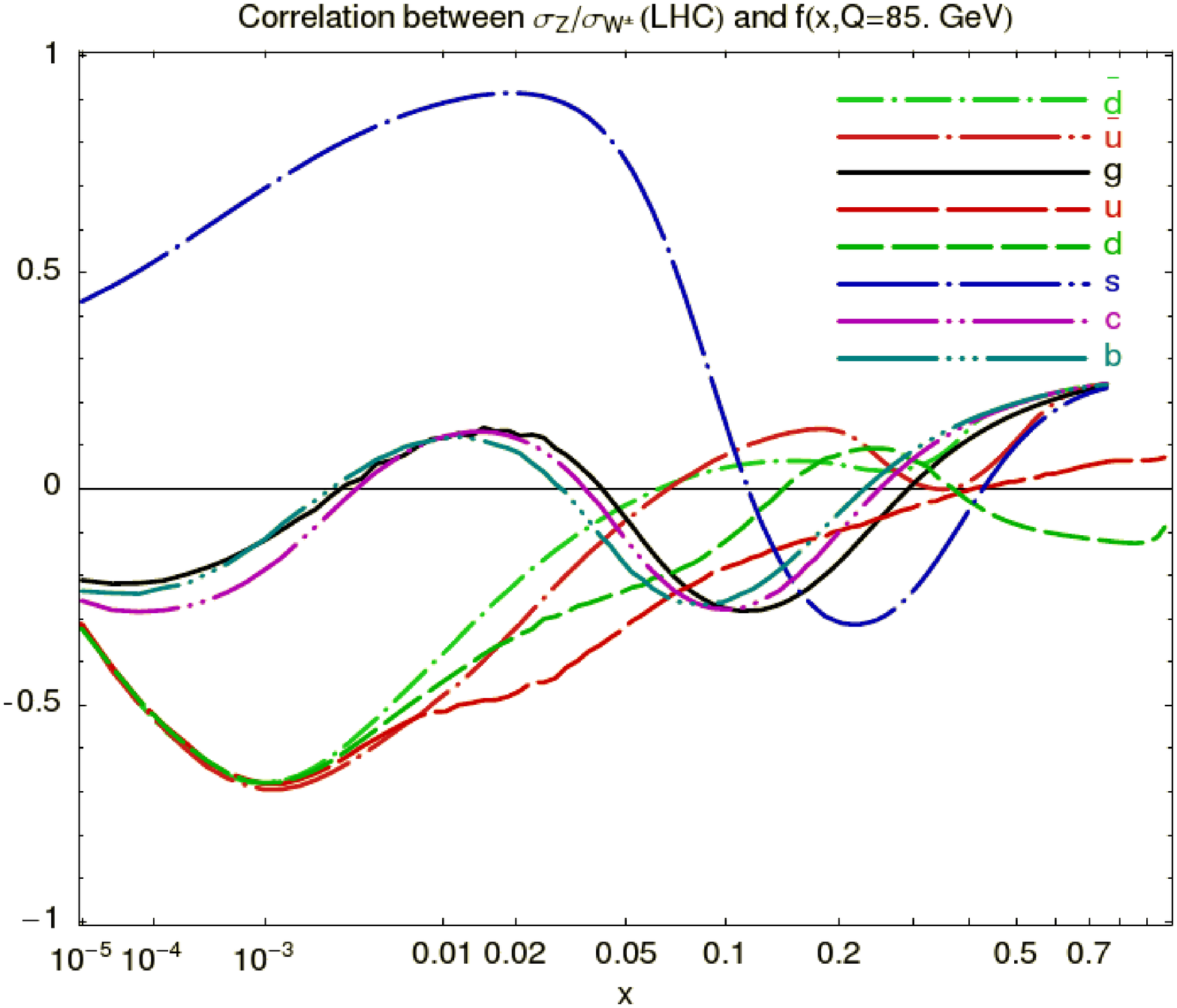}
  \end{center}
  \caption{\label{WZratio} Correlation between $\sigma(Z)/\sigma(W^\pm)$ 
and the PDF versus $x$ for different partons at $Q=$85~GeV (see text).}
\end{figure}

From the PDF side, $W/Z$ production cross sections ratio behave similarly under PDF
variations because the ($x,~Q^2$) range is the same. The remaining PDF
uncertainties are mostly due to the strange quark contribution. Indeed,
Fig.~\ref{WZratio} shows the correlation between $\sigma(Z)/\sigma(W^\pm)$ 
and the PDF versus $x$ for different partons at $Q=$85~GeV. It indicates that a change 
of the gluon, $c$ and $b$ quark PDF have little impact ($<20\%$) on the ratio 
$\sigma(Z)/\sigma(W^\pm)$ for the whole $x$ range. The variations of light quark 
($u$ and $d$) PDFs have a higher effect on the $W/Z$ production cross sections ratio, 
especially at low $x$ ($x<0.1$) where they are anti-correlated. The higher correlation 
is obtained for the strange quark distribution in the region $x~\sim 10^{-2}$, 
where it reaches 90\%. At LHC, $W$ and $Z$ bosons are mainly produced with partons 
in this region. This means that the ratio of the total $W$ and $Z$ production 
cross sections (integrated over $x$) is highly correlated with the proportion 
of strange quarks in the proton. In addition, since additional degrees of
freedom are used to describe the strange quark PDF (in CTEQ6.6 for instance), the
PDF uncertainty on the ratio is now greater than the expected
experimental errors. As a consequence, a measurement of
$\sigma(pp\rightarrow W^\pm)/\sigma(pp\rightarrow Z)$ sets strong constraints 
of the strange quark parametrization. 

\subsection{Ratio of boson pair over single boson production cross sections}
At LHC, pair production of weak bosons are one of the processes that
might reveal some signs of new physics. Their production cross section
is quite low (3.5~pb for $W^+W^-$ events in $e$ or $\mu$ final states)
compared to single $W$ production, but this signal will be precisely
studied. With a cross section measured at the percent level,
systematic uncertainties are expected to dominate the total
error~\cite{atlascsc}.\\ 

In order to reduce the effect of PDF uncertainties, the diboson
cross section can be normalized to the single weak boson
cross section~\cite{paperdittmarpauss}. For instance, $W^+W^-$ 
cross section is compared to the
$Z$ one. Indeed, these two processes are quite similar since $W^+W^-$
bosons can be produced via an off-shell $Z$ bosons~:
$q\bar{q}\rightarrow Z^*\rightarrow W^+W^-$. The initial state is
the same as for Drell-Yan process~: $q\bar{q}\rightarrow
Z^*/\gamma^*\rightarrow e^+e^-$, leading to correlated higher order
QCD corrections and PDF uncertainties. However, the scales, at which
these processes are produced, are different and this is the reason why
the cancellation of PDF uncertainty is not exact. This argument is
also valid for $ZW^\pm$ or $ZZ$ production, compared to $W^\pm$, but
the statistical error will be larger.

\subsection{Drell-Yan production cross sections ratio \label{gstarz}}
With millions of $Z/\gamma^*$ produced with 1~fb$^{-1}$ of LHC data, 
the statistical error on the Drell-Yan production cross section is 
expected to be smaller than the percent. The limitation comes from 
systematics, among which a large error is due to PDF, around 6-8\%, 
even at high mass $\rm M>200~GeV/c^2$. The idea is to exploit the 
$Z/\gamma^*$ mass and rapidity spectrum and to make ratios of 
cross sections when the initial quarks have the same kinematics.\\

Let us consider a quark and anti-quark that produce a $Z$ boson with 
a rapidity $y$. The momentum fractions of these partons are
$x_1$$=M_Z/\sqrt{s}\cdot e^{-|y|}$ and $x_2=M_Z/\sqrt{s}\cdot e^{+|y|}$. 
But these momentum fractions can also be encountered in other 
$\gamma^*/Z$ processes. Symetric $q \bar{q}$ collisions with two partons 
carrying the momentum fraction $x_1$ can produce $\gamma^*/Z$ 
particles with the invariant mass $m=\sqrt{x_1 x_1 s}=M_Z\cdot e^{-|y|}$ 
and rapidity of 0. In the same way, symetric $q \bar{q}$ collisions with 
two partons carrying the momentum fraction $x_2$ can produce 
$\gamma^*/Z$ particles with the invariant mass $M=\sqrt{x_2 x_2 s}=M_Z\cdot e^{+|y|}$ 
and rapidity of 0. In other words, the same quark momenta have 
been found in three cross sections~: $\sigma_{Z/\gamma^*}(M_Z, y)$, 
$\sigma_{Z/\gamma^*}(m, y=0)$ and $\sigma_{Z/\gamma^*}(M, y=0)$ 
where $y=\ln M/M_Z$ and $m=M_Z^2/M$. Uncertainties on quark 
kinematics could be reduced in the following ratio, involving these 
three cross sections~:
$$
R(M)=\frac{\sigma(m, y=0)\cdot \sigma(M, y=0)}{\sigma^2(M_Z, y)}
$$
where $y=\ln M/M_Z$ and $m=M_Z^2/M$.\\

\begin{figure}[htbp]
  \begin{center}
    \includegraphics[width=.8\textwidth]{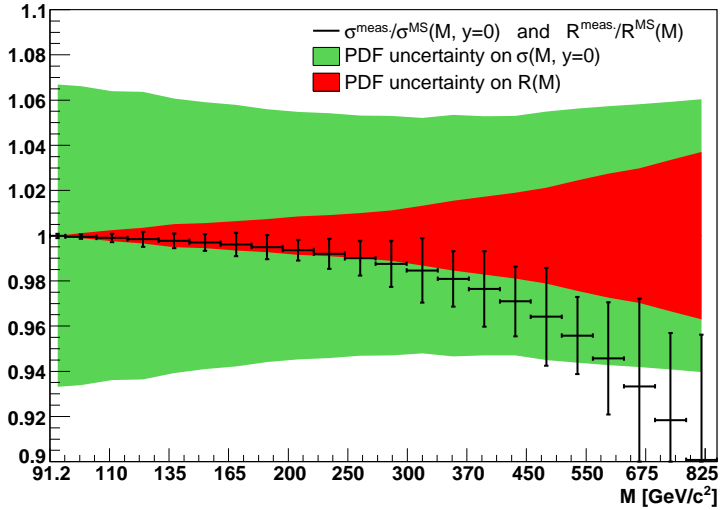}
  \end{center}
  \caption{\label{RatioSigma} Pseudo-measurement to Standard Model 
prediction ratios of $\sigma_{Z/\gamma^*}(M, y=0)$ or $R(M)$ with 
statistical error bars. The pseudo-measurements use $\rm 30~fb^{-1}$ 
of LHC data, and a 2~TeV SSM $Z'$ have been added in the simulation. 
The central values of  these two measurement are the same and the 
statistical uncertainties are very close, so only one set of error 
bars is shown. The uncertainty bands due to PDF on 
$\sigma_{Z/\gamma^*}(M, y=0)$ and $R(M)$ are also represented.}
\end{figure}

With only one quark flavour and with scale invariance, 
the PDF completely cancel and so their uncertainties. 
This is no longer valid in the real case but the prediction 
of $R(M)$ is still more precise than the high mass Drell-Yan 
cross section $\sigma(M, y=0)$. Fig.~\ref{RatioSigma} shows 
how these errors vary for different $Z/\gamma^*$ invariant masses. The PDF 
uncertainties can be reduced by more than a factor two, 
leading to a higher sensitivity to non-Standard Model processes.

An example of new physics sensitivity is shown on 
Fig.~\ref{RatioSigma}. Pseudo-measurement of $\sigma_{Z/\gamma^*}(M, y=0)$ 
and $R(M)$, including a 2~TeV SSM $Z'$ are compared to 
the Standard Model predictions. The statistical and PDF-induced 
uncertainties are also displayed. A  measurement of $R(M)$ 
shows a sensitivity of a 2~TeV SSM $Z'$ since $M>200~GeV/c^2$, 
while in a $\sigma_{Z/\gamma^*}(M, y=0)$ cross section analysis, 
no significant deviation is seen, except for $M>600~GeV/c^2$. 
Thus, it seems possible to explore a larger range of $Z'$ 
models, that may not be discovered by direct peak searches like 
non-resonant or wide $Z'$.

This method has other advantages. If a $Z'$ peak is observed, 
this ratio of cross sections can be used to measure the 
$\gamma^*/Z/Z'$ interference term at lower masses, in order to 
give additional constrains to the underlying $Z'$ model. 
Finally, this method can be applied to any $s$-channel processes 
like $W^\pm$ production. $W'$ searches or $s$-channel single-top 
cross sections can be normalized to $W^\pm\rightarrow l\nu$ 
to obtain more precise measurements.

\subsection{Summary}
At LHC, most measurements will be limited by systematic uncertainties. 
Experimental systematics can be reduced in ratios of quantities. 
In this section, appropriate ratios have been presented because their 
sensitivity on theoretical uncertainties has been reduced compared to 
individual cross sections. The total error is reduced at a few percent level, 
showing that precision measurements at hadron colliders are possible.

%%%%%%%%%%%%%%% CHR begins %%%%%%%%%

\section{Beyond the DGLAP evolution equation: looking for BFKL and saturation effects}

In this section, we will face another aspect: we discuss potential issues with 
respect to the standard DGLAP~\cite{dglap} 
QCD evolution equation which is used in standard PDF analysis, and dedicated
observables which can show effects beyond the standard DGLAP equation. 
There will be two different parts: we will first discuss potential observables
sensitive to the missing $\log 1/x$ terms in the DGLAP evolution equation and
included in the BFKL equation~\cite{bfkl}, and second, the possible influence of the saturation phenomenon
at the LHC.

\subsection{Looking for BFKL effects at the LHC: Mueller-Navelet jets}

\begin{figure}
\centerline{\includegraphics[width=0.85\columnwidth]{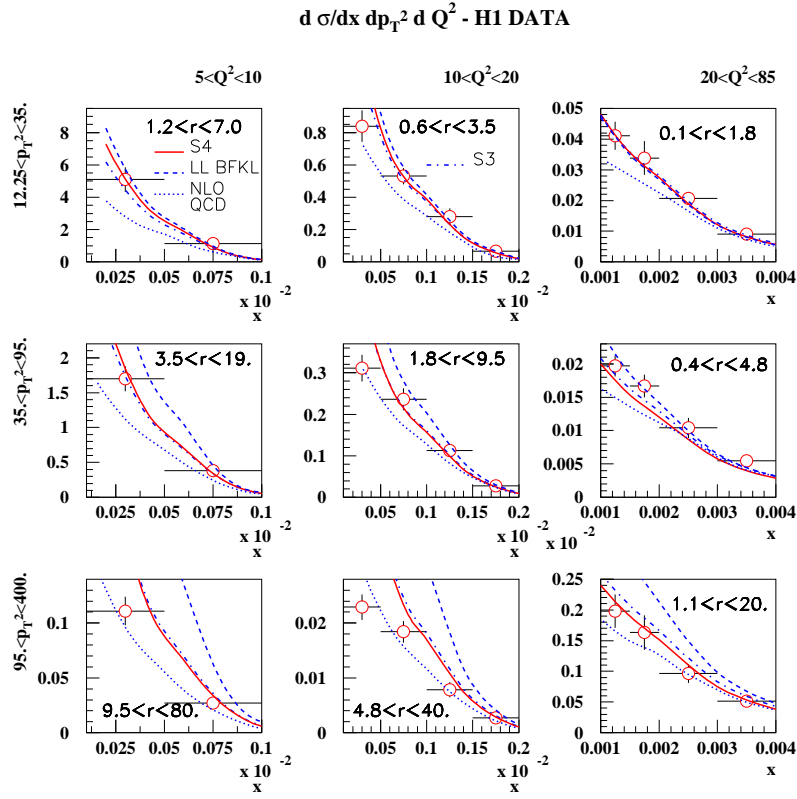}}
%\centerline{\includegraphics[width=0.45\columnwidth]{Fig4b.eps}}
\caption{Comparison between the H1 measurement of the triple differential cross
section with predictions for BFKL-LL, BFKL-NLL and DGLAP NLO calculations
(see text).}
\label{triple}
\end{figure}

\begin{figure}
\centerline{\includegraphics[width=0.45\columnwidth]{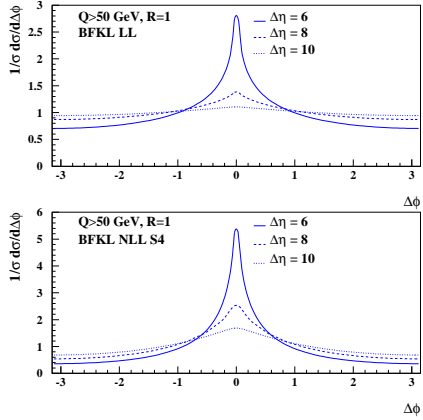}}
%\centerline{\includegraphics[width=0.45\columnwidth]{Fig4b.eps}}
\caption{The Mueller-Navelet jet $\Delta\Phi$ distribution for LHC kinematics in the BFKL framework at 
LL (upper plots) and NLL-S4 (lower plots) accuracy for $\Delta\eta=6,\ 8,\ 10.$}
\label{Figlhc}
\end{figure}

\begin{figure}
\centerline{\includegraphics[width=0.45\columnwidth]{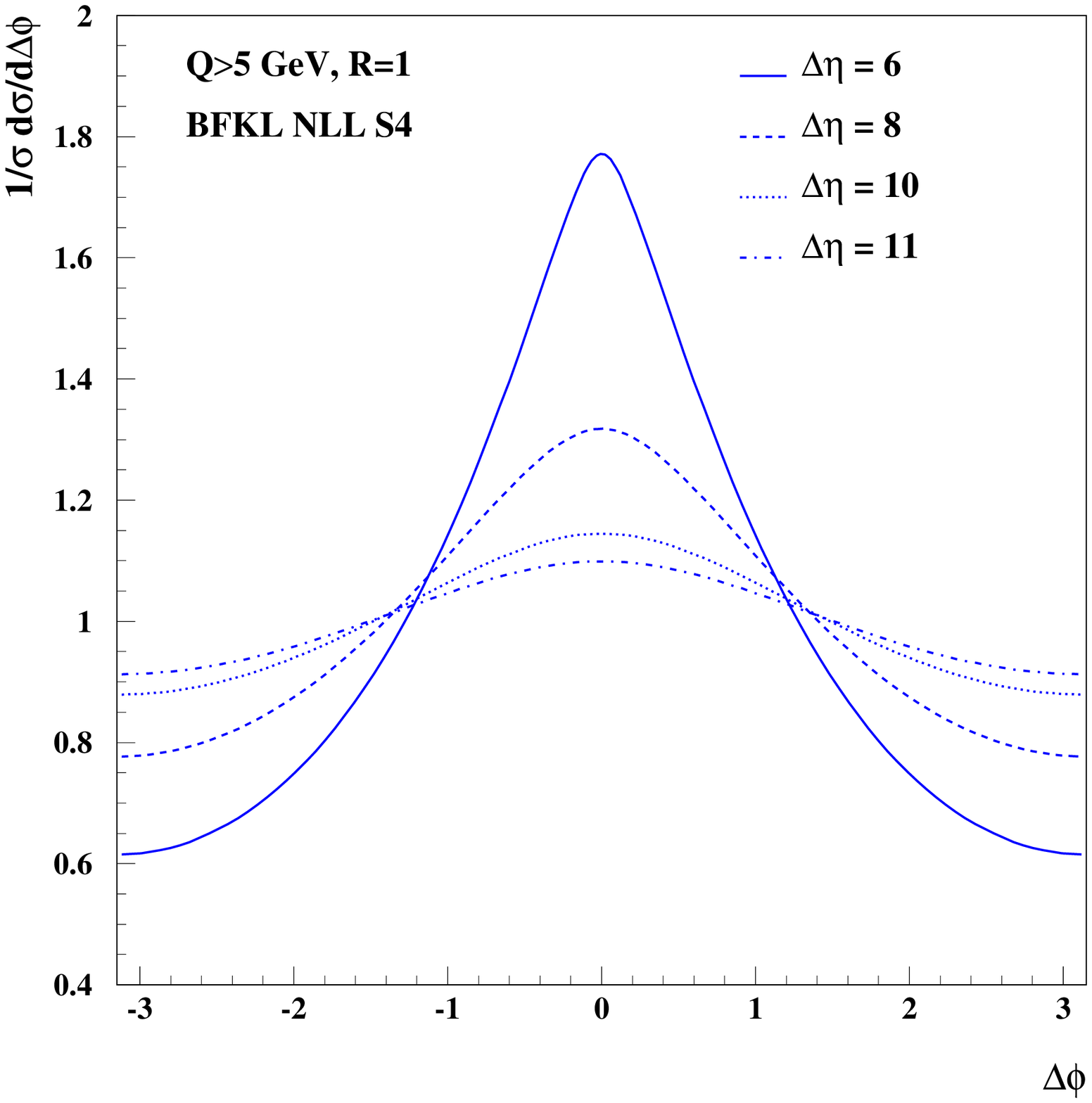}}
\caption{Azimuthal correlations between jets with $\Delta \eta=$6, 8, 10
and 11 and $p_T>5$ GeV in the CDF acceptance. This measurement will represent a
clear test of the BFKL regime.}
\label{Fig3}
\end{figure}

While it is well
known that the BFKL~\cite{bfkl} low-$x$ resummation is not required in inclusive cross section
measurements such as the proton structure functions at HERA or jet cross
sections at Tevatron and LHC (but may be needed to determine the parton
distributions at the initial scale $Q_0^2$), the situation is quite different 
if one looks in a
given phase space where the $\log 1/x$ terms become important. This is the case
for instance when one considers forward jet production at HERA. The idea is
quite simple: one considers jet production as far away as possible in rapidity
from the scattered electron, and the transverse energy of the jet is requested
to be close to the $Q^2$ of the virtual photon. Because of the $k_T$-ordering of
the gluons along the ladder for the DGLAP evolution equation, the cross section
predicted by the DGLAP evolution is small. On the contrary, the BFKL evolution
equation predicts much higher cross sections compatible with the experimental
observations since there is enough phase space to produce many
gluons~\cite{mr,fwdjet}. We will just quote one of the significant forward jet
measurements performed at HERA, the measurement of the triple differential cross
section in the H1 collaboration. The triple differential cross section 
$d\sigma/dxdk_T^2dQ^2$ shown in Fig.~\ref{triple} is an interesting 
observable as it has been measured in 9 different $p_T^2$ and $Q^2$ regions
where $p_T$ is the transverse momentum of the forward jet.
The H1 data are compared to NLO DGLAP calculations which fail to describe the
data in the low $p_T$ region when $r=p_T^2/Q^2$ is close to 1, precisely in the
region where
BFKL resummation effects are expected to be large. Fig.~\ref{triple} also shows the BFKL-LL
predictions which fail to describe the data when $r$ goes away from 1. On the
contrary, the BFKL NLL calculation~\footnote{The BFKL NLL calculation were
performed using two different schemes called S3 and S4 used to remove spurious
singularities of the BFKL NLL calculation~\cite{salam}.}
including the $Q^2$ evolution via the renormalisation group equation leads to a
good description of the H1 data on the full range. We note that the higher order
corrections are small when $r \sim 1$, when the BFKL effects are
supposed to dominate. By contrast, they are significant as expected when $r$ is different from
one, ie when DGLAP evolution becomes relevant. 

At hadronic colliders, similar processes can occur for the so-called
Mueller-Navelet jets which are ideal processes to study BFKL resummation 
effects~\cite{mnjet}.
Two jets with a large interval in rapidity and with similar
tranverse momenta are considered. For this kind of events, the cross section
predicted by the DGLAP evolution is small because of the $k_T$-ordering of the 
gluons along the ladder. The BFKL cross section can be large because there is
enough phase space to emit the gluons. 
A typical observable to look for BFKL effects
is the measurement of the azimuthal correlations between both jets. The DGLAP
prediction is that this distribution should peak towards $\pi$ - ie jets
are back-to-back- whereas
multi-gluon emission via the BFKL mechanism leads to a smoother distribution.
The relevant variables to look for azimuthal correlations are the following:
\begin{eqnarray}
\Delta \eta &=& y_1 - y_2  \nonumber \\
y &=& (y_1 + y_2)/2 \nonumber \\
Q &=& \sqrt{k_1 k_2} \nonumber \\
R &=& k_2/k_1  \nonumber 
\end{eqnarray}
The azimuthal correlation for BFKL reads:
\begin{eqnarray}
2\pi\left.\frac{d\sigma}{d\Delta\eta dR d\Delta\Phi}
\right/\frac{d\sigma}{d\Delta\eta dR}=
1+ \nonumber 
\frac{2}{\sigma_0(\Delta\eta,R)}\sum_{p=1}^\infty \sigma_p(\Delta\eta,R) \cos(p\Delta\Phi)
\nonumber 
\end{eqnarray}
where in the NLL BFKL framework,
\begin{eqnarray}
\sigma_p&=& \int_{E_T}^\infty \frac{dQ}{Q^3}
\alpha_s(Q^2/R)\alpha_s(Q^2R) \nonumber 
\left( \int dy x_1 f_{eff}(x_1,Q^2/R)x_2f_{nll}(x_2,Q^2R) 
\right) \nonumber
\\
&~& \int_{1/2-\infty}^{1/2+\infty}\frac{d\gamma}{2i\pi}R^{-2\gamma}
\ e^{\bar\alpha(Q^2)\chi_{nll}(p, \gamma, \bar{\alpha})\Delta\eta} . \nonumber
\end{eqnarray}
$\chi_{nll}$ is the effective resummed NLL BFKL kernel, and $f_{eff}$ are the
effective parton densities in the proton.
As expected, the
$\Delta \Phi$ dependence is less flat than for BFKL LL and is closer to the
DGLAP behaviour~\cite{mnjet}. In Fig.~\ref{Figlhc}, we display the observable 
$1/\sigma d \sigma/d\Delta \Phi$ as a function
of $\Delta\Phi$ for LHC kinematics. The results are displayed for different values of 
$\Delta\eta$ and at both LL and NLL accuracy. In general, the 
$\Delta\Phi$ spectra are peaked around $\Delta\Phi\!=\!0,$ which is indicative of jet emissions occuring back-to-back. 
In addition the $\Delta\Phi$ distribution flattens with increasing 
$\Delta\eta\!=\!y_1\!-\!y_2$. Note the change of scale on the vertical axis 
which indicates the magnitude of the NLL corrections with respect to the 
LL-BFKL results. 

A measurement of the cross section 
$d\sigma^{hh\!\to\!JXJ}/d\Delta\eta dR d\Delta\Phi$ at the Tevatron (Run 2) or in the future
at LHC will allow for a 
detailed study of the BFKL QCD dynamics since the DGLAP evolution leads to much less jet angular
decorrelation (jets are back-to-back when $R$ is close to 1). In particular, measurements with 
values of $\Delta\eta$ reaching 8 or 10 will be of great interest, as these could allow to distinguish between 
BFKL and DGLAP resummation effects and would provide important tests for the relevance of the BFKL formalism. 

To illustrate this result, we give in Fig.~\ref{Fig3} the azimuthal
correlation in the CDF acceptance. The CDF collaboration installed the
mini-Plugs calorimeters aiming for rapidity gap selections in the very forward
regions and these detectors can be used to tag very forward jets. A measurement
of jet $p_T$ with these detectors would not be possible but their azimuthal
segmentation allows a $\phi$ measurement. In Fig.~\ref{Fig3}, we display the jet
azimuthal correlations for jets with a $p_T>5$ GeV and $\Delta \eta=$6, 8, 10
and 11. For $\Delta \eta=$11, we notice that the distribution is quite flat,
which would be a clear test of the BFKL prediction. Mueller-Navelet jets might
also be a possible way to look for saturation effects~\cite{mr} when jets are
widely separated in rapidity.

Another measurement sensitive to BFKL resummation effects is the cross section
of dijet events where there is a gap devoid of any energy between the two jets.
The production cross section was measured by the D0 collaboration and was found
in good agreement with BFKL LL calculations~\cite{cox}. BFKL NLL calculations at
Tevatron and LHC are in progress~\cite{cox}.

\subsection{Saturation physics at HERA and LHC}
\begin{figure}[htb]
\begin{center}
\epsfig{file=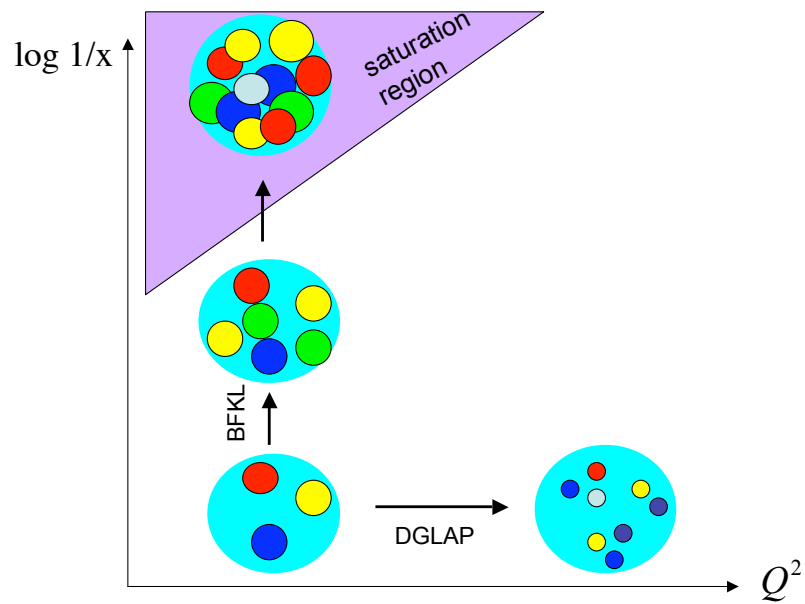,height=12cm}
\caption{QCD at hadronic colliders}
\label{scheme}
\end{center}
\end{figure}

In this section, we will discuss some preliminary approaches related to
saturation physics at HERA and briefly the implications at LHC. A sketch of
the proton structure in ($x$, $Q^2$) is given in Fig.~\ref{scheme}. 
The LHC will allow to probe scales in the proton which were never reached at
present, by accessing values of $x$ down to 5.10$^{-7}$ and $Q^2$ up to
10$^8$ GeV$^2$. For a comparison, the Tevatron only reaches $Q^2 \sim$ 2.10$^5$
GeV$^2$. When $x$ decreases, the number of
gluons increases. At some point, the number is so large that they start
overlapping each other, and one cannot longer neglect the interactions between
the different gluons. This is the domain of saturation. The domain of full saturation where
the standard equations will not hold is yet to be discovered experimentally and
this is one of the challenges for LHC. One of the already significant 
implications of the
saturation models is that the proton structure function does not depend
independently on $x$ and $Q^2$ but on scaling variables which are combinations
of $x$ and $Q^2$~\cite{usscaling,qf}. The type of the predicted scaling depends on the
considered equations: fixed coupling constant, running coupling constant, pomeron loops...
It is also worth noticing that saturation models such as the one described
in Ref.~\cite{golec} lead to a common description of diffractive and
non-diffractive data and we will come back on these models when we discuss the
diffractive results.

Geometric scaling~\cite{Stasto:2000er} is a remarkable 
property 
verified by data on high energy deep inelastic scattering (DIS). 
One can  represent with 
reasonable 
accuracy the cross section $\sigma^{\gamma^*p}$ by the formula
$\sigma^{\gamma^*p}(Y,Q)=\sigma^{\gamma^*}(\tau)\  ,$
where 
$Y$ the total rapidity in the ${\gamma^*}$-proton system and 
$\tau = \log Q^2-\log Q_s(Y) =  \log Q^2-\lambda 
Y\ $ 
is the scaling variable. 

\begin{figure}
\begin{center}
\epsfig{file=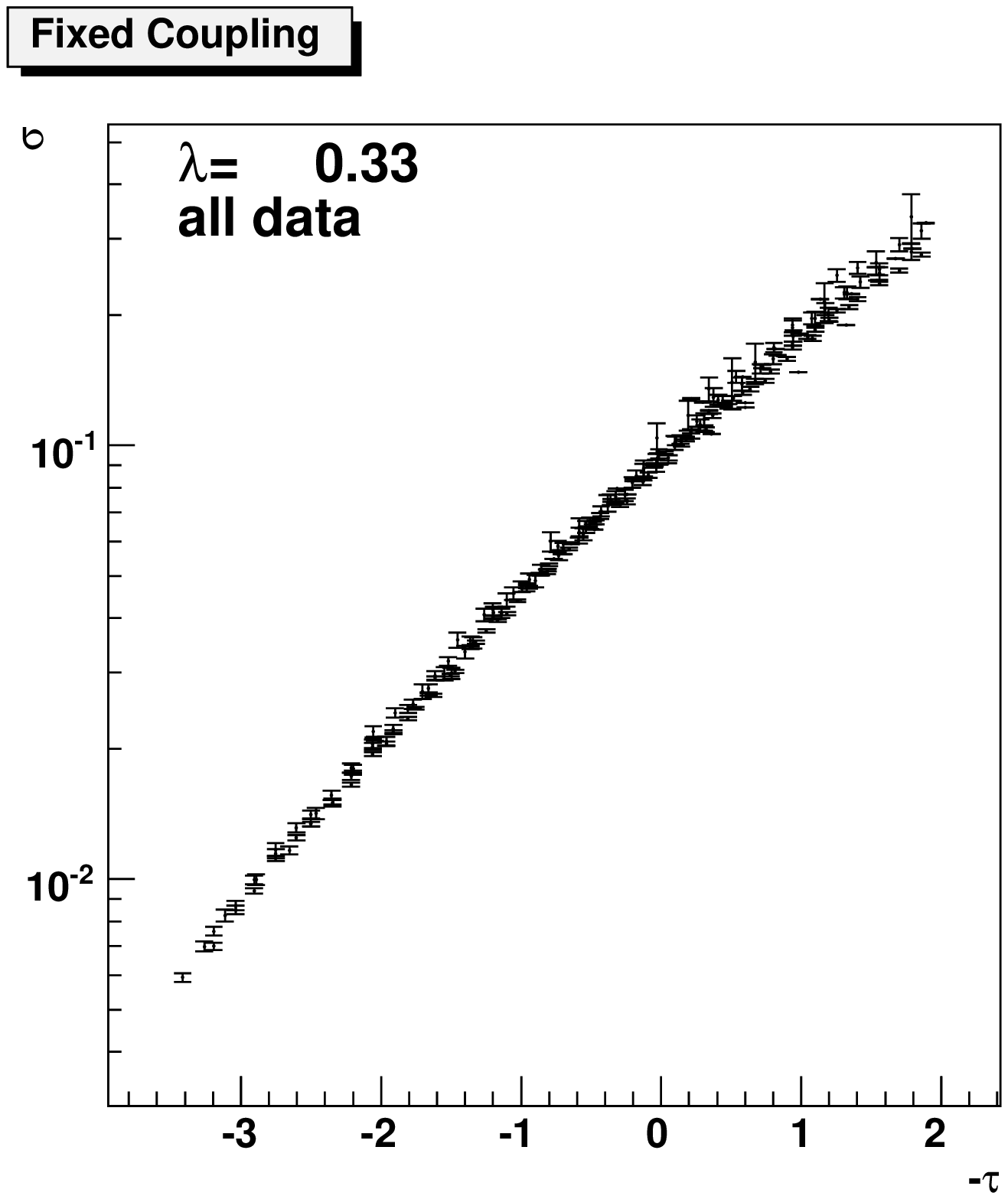,width=9.cm}
\caption{Scaling curve 
for ``Fixed Coupling" for the proton structure function $F_2$ measured 
in fixed target experiments and at HERA. A $Q^2>3$ 
and $x<10^{-2}$ cut was applied to the data.}
\label{F2_fixed_3}
\end{center}
\end{figure}

\begin{figure}
\begin{center}
\begin{tabular}{cc}
\epsfig{file=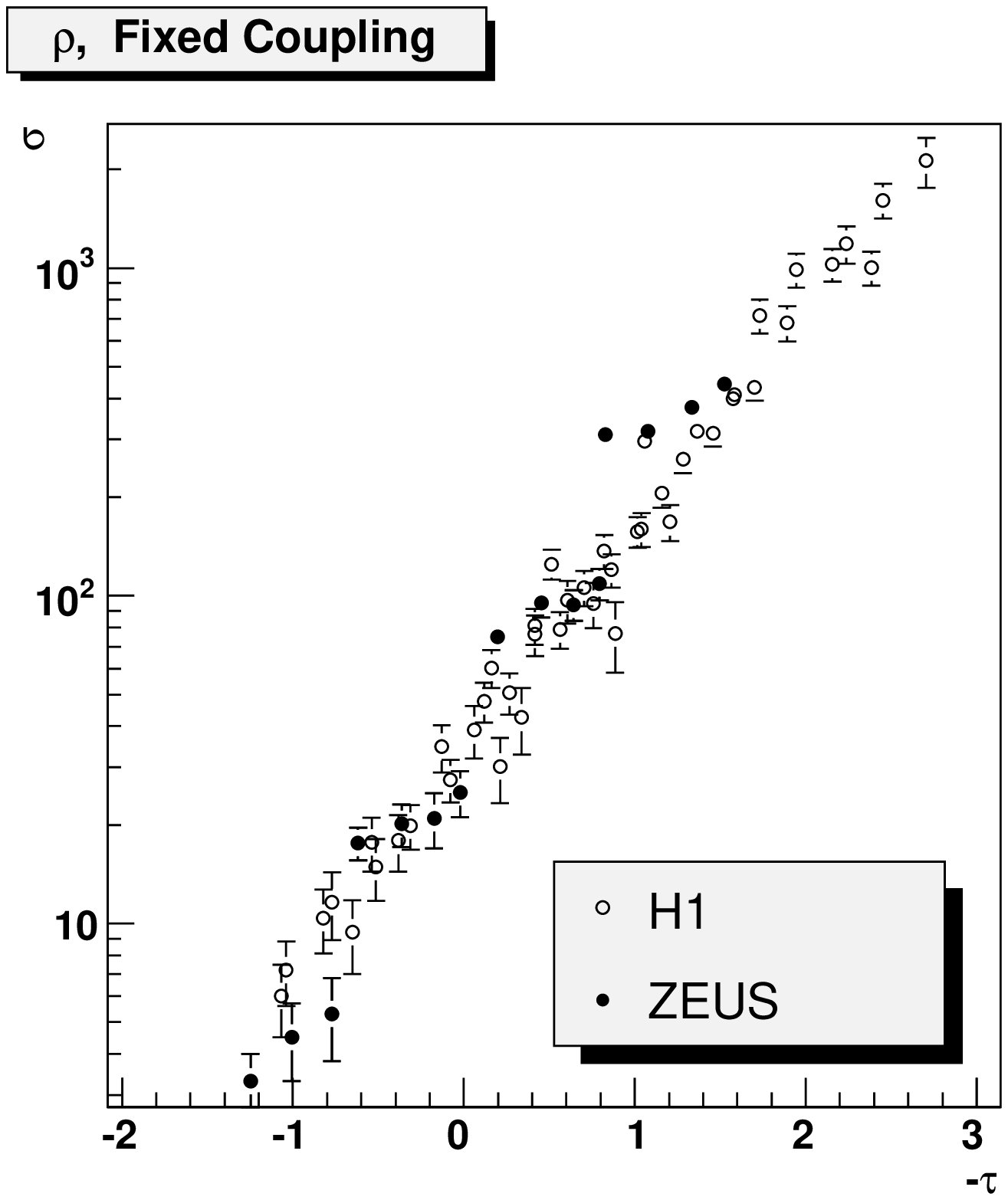,width=7.cm} &
\epsfig{file=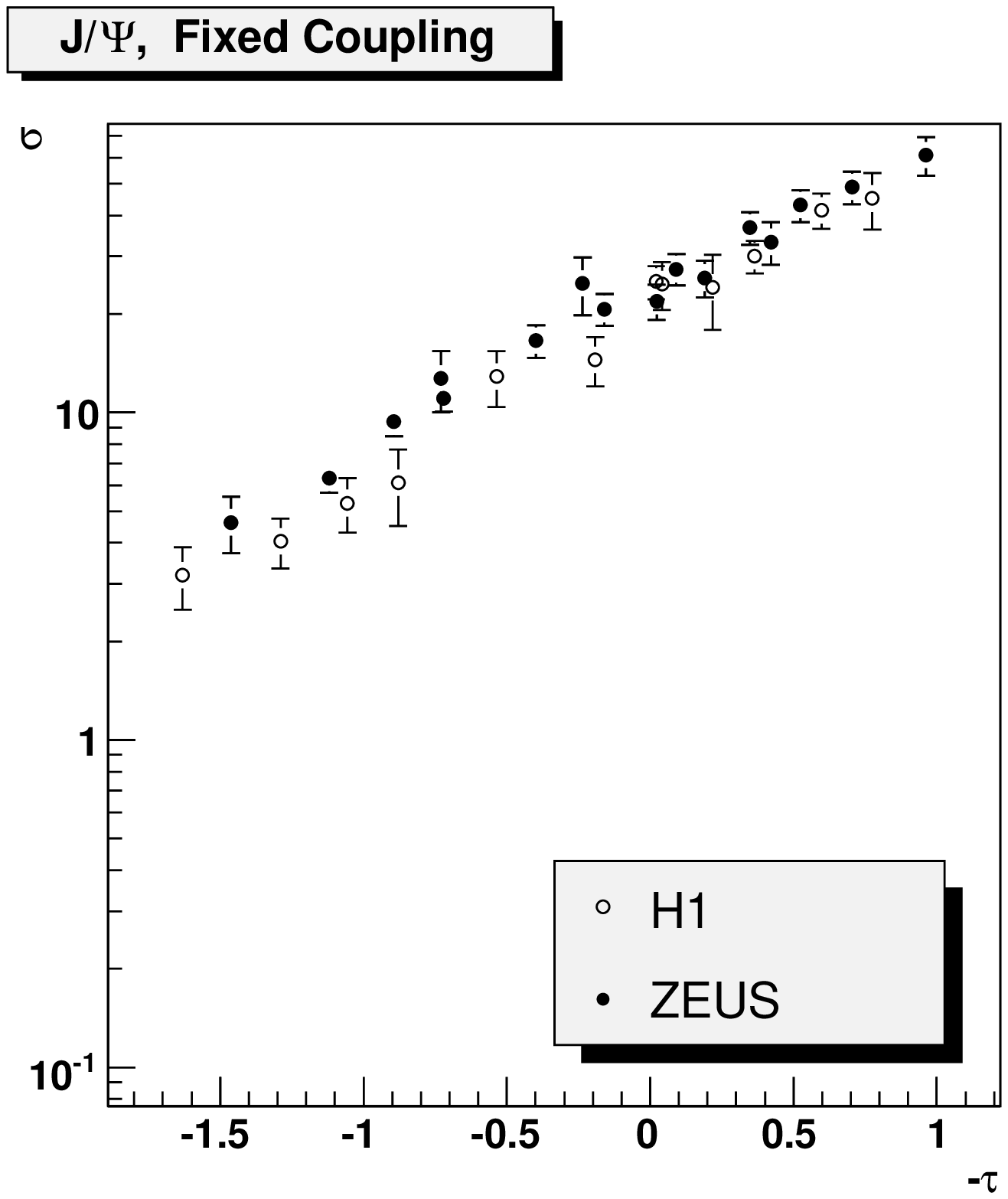,width=7.cm}\\
\epsfig{file=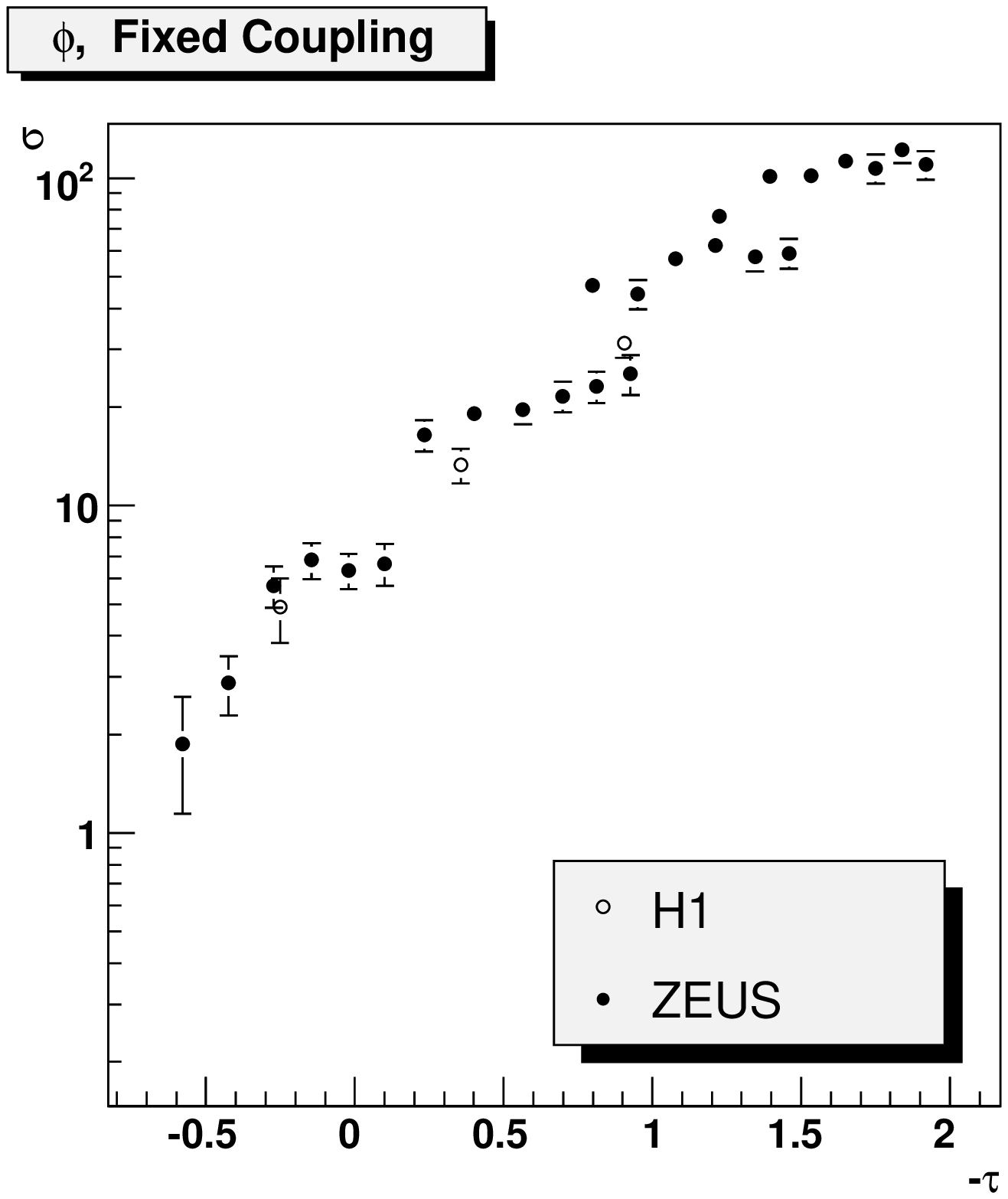,width=7.cm} &
\epsfig{file=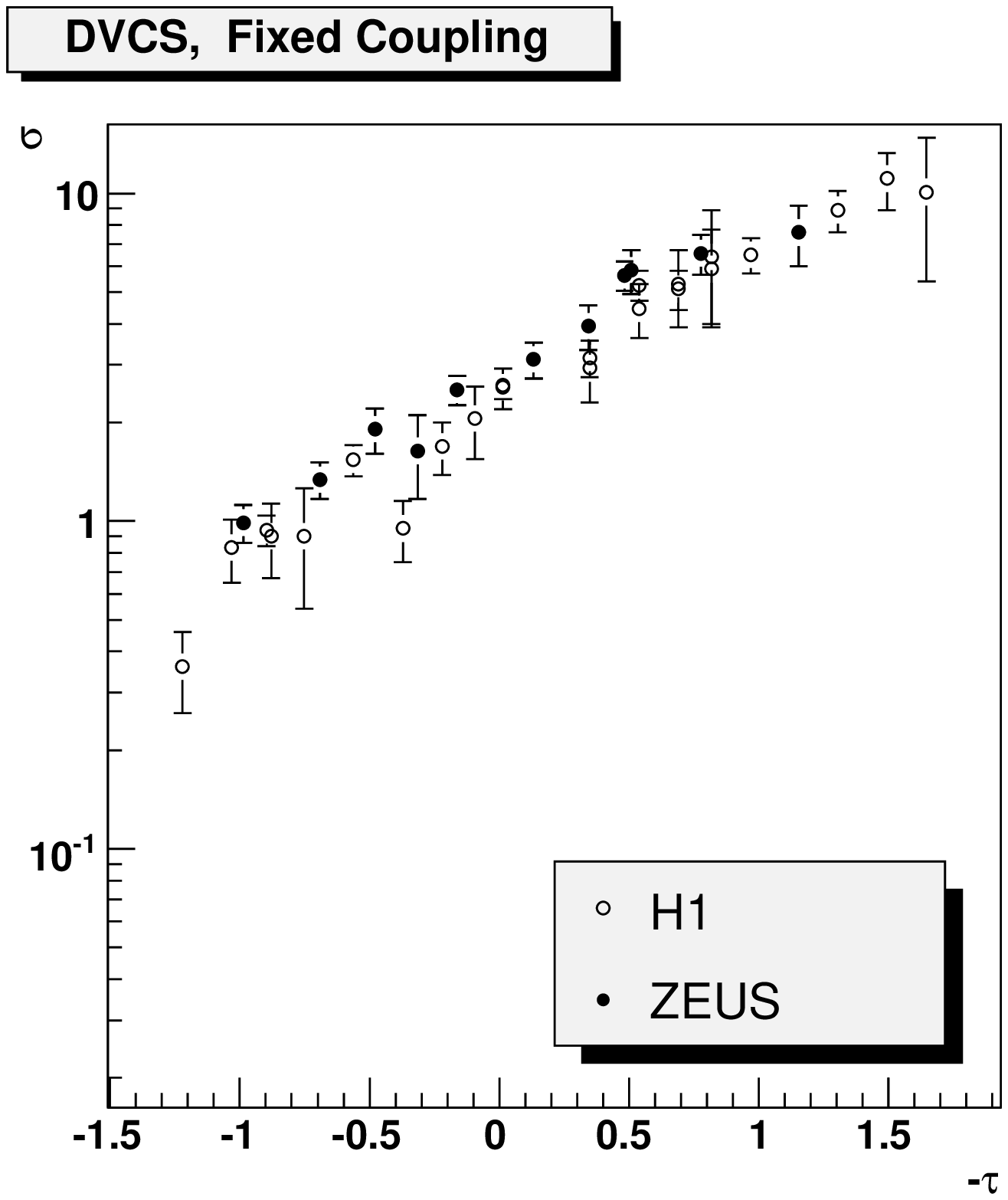,width=7.cm}\\
\end{tabular}
\caption{Scaling curves obtained for
``Fixed Coupling",
for vector meson data. }
\label{vm_fixed}
\end{center}
\end{figure}

A way to introduce theoretically saturation in the BFKL equation was developped
originally in the Balitsky-Kovchegov equation~\cite{balitsky} (BK).
When $\alpha_S$ is constant, it is possible to show that the
solution of the BK equation at high energies 
does not depend independently on
rapidity $Y=\log 1/x$ and $L=\log Q^2$ but on a combination of both, $\tau = L - \lambda Y$. This is
called ``fixed coupling" (FC).
When $\alpha_S$ is running ($\alpha_S \sim 1/\log Q^2$),
an approximate solution of the BK equation is found with a scaling in 
$(L-\lambda \sqrt{Y})$ called running coupling. Other forms of scalings
are also possible.
The experimental
aspects of scaling were studied in Ref.~\cite{usscaling}, and scaling was found
for proton structure function $F_2$, the diffractive structure function $F_2^D$,
the vector meson and DVCS production cross sections. The results were studied
quantitavely using the ``quality factor" approach~\cite{qf}. 
As an example, we give the results of the fixed coupling scaling for the proton
structure function data in Fig.~\ref{F2_fixed_3} and for vector mesons and DVCS
in Fig.~\ref{vm_fixed}. We will also discuss the description of diffractive
inclusive data using the saturation formalism in a next section.

It is worth studying the impact of saturation effects at LHC. While most of
the measurements will be done at higher $Q^2$ (Higgs boson, searches for new
phenomena) and will not be influenced by saturation effects, dedicated
measurements such as Mueller-Navelet jets might be a way to assess saturation
effects~\cite{mr}. On the other hand, it is worth noticing that the saturation
scale at HERA or at the LHC is quite low. It was estimated for instance using
inclusive $F_2$ measurements that the saturation scale is close to 1 GeV$^2$ at
HERA, and is expected to be around 1-2 GeV$^2$ at the LHC. The fact that this
scale is close to the non perturbative region makes it difficult to observe
direct consequences of saturation at LHC, and only indirect measurements
such as scaling properties can be an indication of the presence of saturation at
a lower scale. On the contrary, the proton-gold interactions at LHC 
might be a better way of observing saturation effects. The saturation scale is
expected to be higher for such events (about 4-5 GeV$^2$) and thus entering the
perturbative region. This domain is certainly worth of further studies at the
LHC.

%%%%%%%%%%%%% CHR FINISH %%%%%%%%%%%%%%%%

\section{Conclusion on the proton structure}

We have reviewed some important measurements at LHC, which are motivated
to improve the knowledge of quark and gluon distributions in the proton.
They have a reduced sensitivity to new physics effects, because the impact
of the PDF uncertainties is that a
small deviation from the Standard Model predictions could be absorbed
into the PDFs. Improvements of parton distributions are expected at three different
levels~: reduction of the uncertainties on the parameters used in global
QCD fits, reduction of the dependence on input functional form at $Q_0^2\sim m_p^2$,
and reduction of the theoretical uncertainties on evolution equations.\\

In order to reduce the PDF uncertainties, global QCD fits could benefit from
additional measurement sensitive to parton distribution at unexplored ($x$, $Q^2$)
values. Differential cross sections using jets, photons or heavy gauge bosons
versus the system invariant mass, transverse momentum or rapidity have a strong
potential to constrain PDFs. Statistics is expected to be large at LHC so these
measurements will be limited by experimental systematics that must be reduced.
Some subjects are still not covered at LHC. Heavy flavour quark distributions
still suffer from large statistical and systematic uncertainties, but they could
be constrained by $W/Z+c$ or $W/Z+b$ cross section measurements. These processes
will be one of the main backgrounds to Higgs boson or new physics searches,
their understanding is thus crucial for discoveries. Other measurements are
aimed to constrain the high-$x$ region while being safe of new physics. Low
mass dijet or Drell-Yan cross sections at high rapidity can constrain the
high-$x$ gluon or sea quark ditributions, i.e. where the uncertainties are large.

Finally, when quoting a PDF-induced uncertainty, one needs to ascertain whether
these underlying hypotheses affect the result or can be ignored. We have shown
that the input functional form at $Q_0^2\sim m_p^2$ or the choices of evolution
equations can lead to large differences between predictions on observables, and
thus must be more constrained. These assumptions can be tested with asymmetry
measurements or ratios of cross sections, especially built to cancel most of PDF
uncertainties and thus to probe a reduced set of parton distributions. As a
consequence, these observables have a better sensitivity to beyond Standard Model
phenomena.

The discovery of the electroweak symmetry breaking mechanism and the observation
of new particles are the main goal of the LHC. Direct searches and precision
measurement both require a good control of the background and the signal itself
because PDF-induced uncertainties can be larger than the size of new physics
effects. Two methods have been discussed in this chapter~: the reduction of the
uncertainties on gluon and quark distributions via interesting measurements,
or the measurement of observables less sensitive to PDF uncertainties. Both
solutions improve the discovery potential of LHC.

We will now move to another important component of the proton structure 
which we did not
mention until now and which is related to diffractive events.

%%%%%%%%%%%%%%%%%%% DIFFRACTION

%\input{heradiff}

\section{What is diffraction: the example of HERA}

\begin{figure}
\begin{center}
\vspace{10.cm}
\hspace{-7cm}
\epsfig{file=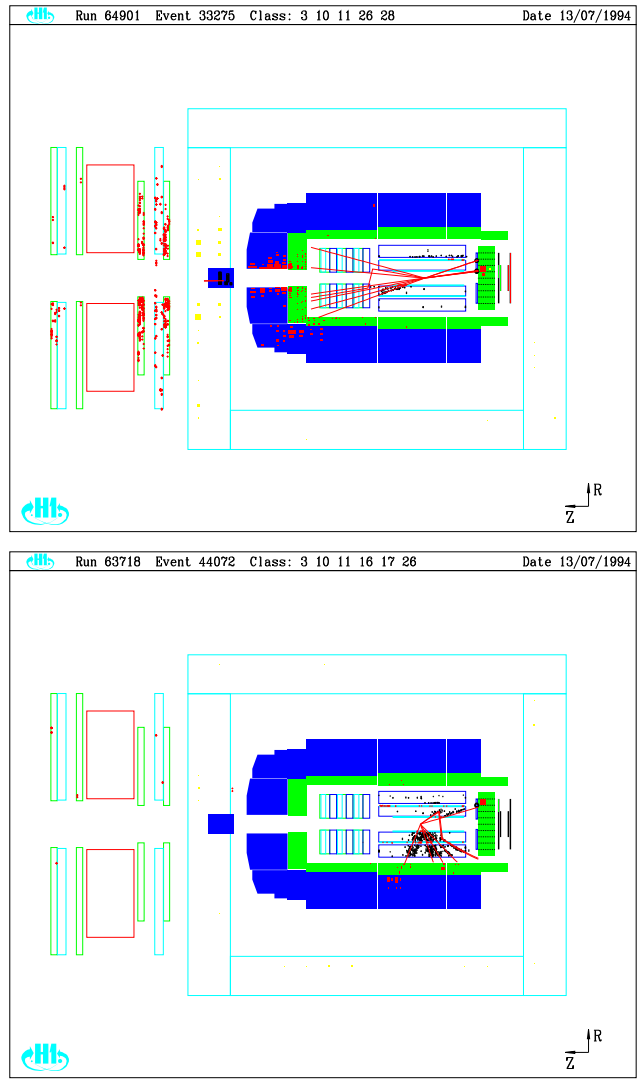,width=5.2cm}
\vspace*{-2cm}
\caption{Standard and diffractive events in the H1 experiment. For the diffractive event,
the electron is visible in the backward detector, there is
still some hadronic activity (jets) in the LAr calorimeter, but no energy above
noise level is deposited in the forward part of the LAr calorimeter or in the
forward muon detectors. In other words, there is no color exchange between the
proton and the produced jets.}
\label{fig1}
\end{center}
\end{figure}

In the following sections, we first describe diffraction at HERA before showing
the results from Tevatron and discussing the possible measurements at LHC.
As detailed in the previous sections, the advanced metrology of PDFs at LHC is a very important topic, 
not only in order to understand better the structure
of the proton, but also to determine more precisely the background to searches
for the Higgs boson or supersymmetric particles. 
However, as mentioned above in the context of saturation,
diffraction is also an important class of processes to scrutenize and understand the structure of the proton.
In fact, an important fraction of the total cross section at HERA or LHC energies is driven
by diffractive reactions, which then deserve specific studies. Let us start by giving 
a basic description of a diffractive event in HERA experiments.
A typical standard DIS event is shown in the upper plot of Fig.~\ref{fig1} is $ep \rightarrow eX$
where electron and jets are produced in the final state. We
notice that the electron is scattered in the H1 backward detector~\footnote{At
HERA, the backward (resp. forward) directions are defined as the direction
of the outgoing electron (resp. proton).} 
whereas some hadronic activity is present in the forward region of the detector
(in the LAr calorimeter and in the forward muon detectors). The proton is thus
completely destroyed and the interaction leads to jets and proton remnants directly observable
in the detector. The fact that much energy is observed in the forward region is
due to colour exchanges between the scattered jet and the proton remnants.
In contrast, for about 10\% of the events, the situation is completely
different. Such events appear like the one shown in the bottom plot of Fig.~\ref{fig1}.
The electron is still present in the backward detector, there is
still some hadronic activity (jets) in the LAr calorimeter, but no energy above
noise level is deposited in the forward part of the LAr calorimeter or in the
forward muon detectors. In other words, there is no color exchange between the
proton and the produced jets. As an example, this can be explained if the proton stays intact
after the interaction. These events amount to about 10\% of the total deep inelastic event
production at HERA in the acceptance of the measurement ---
they are called diffractive --- and about 30 \% of the total cross section 
at LHC. Thus, they can not be ignored with the assumption that the dynamics of those
reactions follow exactly the standard QCD equations that govern the PDF behaviour. The possible explanations
of the underlying dynamics of such processes is described in the following. We also show why 
their specific analysis is an essential aspect of understanding 
the proton structure at high gluon densities.

\begin{figure}
\begin{center}
\includegraphics[width=10cm,height=8.5cm]{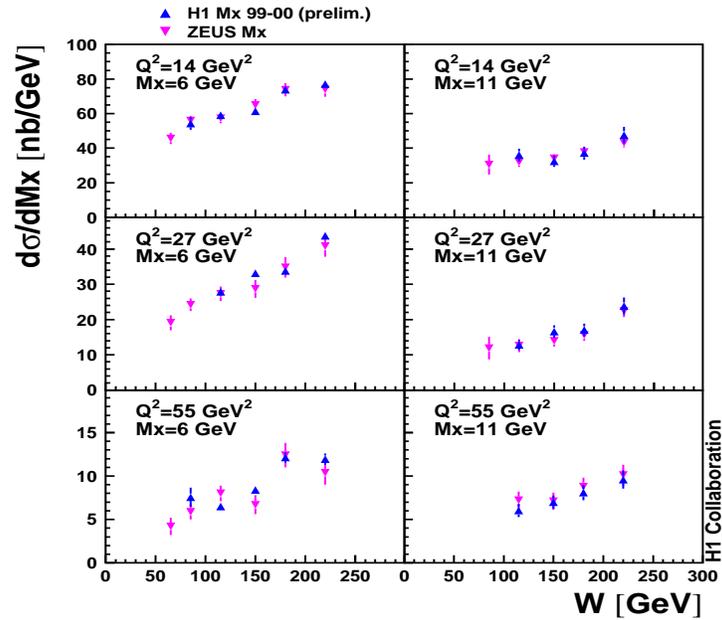}
\caption{The  cross section of the diffractive process $\gamma^* p \rightarrow p' X$
measured by the H1 and ZEUS collaborations, 
differential in the mass of the diffractively produced hadronic system $X$ ($M_X$),
is presented as a function of the centre-of-mass energy of the $\gamma^*p$ system $W$.
Measurements at different values of the virtuality
$Q^2$ of the exchanged photon are displayed.
}
\label{figdata}
\end{center}
\vspace{-0.5cm}
\end{figure}

\begin{figure}
\begin{center}
%\vspace{-1.5cm}
%\hspace{-6cm}
\epsfig{file=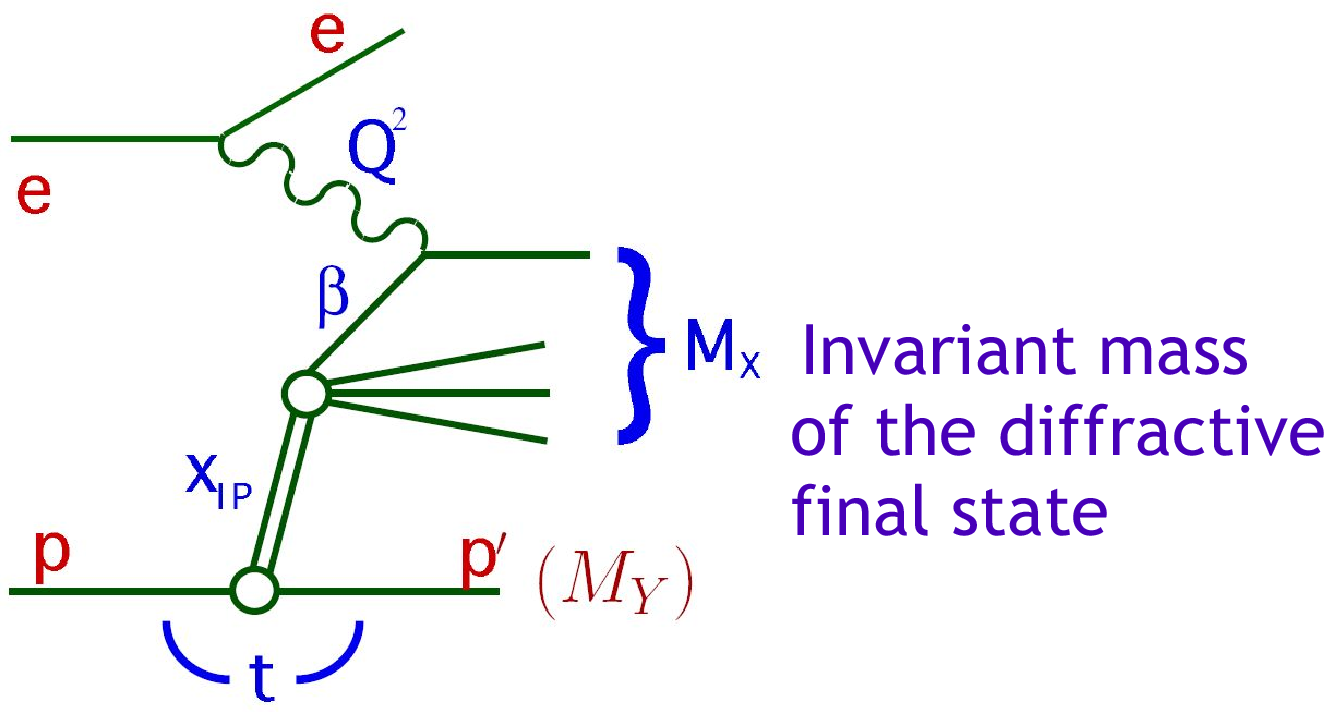,width=6cm,angle=270}
\caption{Scheme of a diffractive event at HERA.}
\label{diff-diffdisdiag}
\end{center}
\end{figure}

One of the first experimental results to be considered is the diffractive
cross section as a function of the energy dependence of the $\gamma^*p$ system, which
has been measured at HERA by H1~\cite{f2d97,f2d97b,manu} and
ZEUS~\cite{zeuslast,zeus,zeusb} experiments over a wide kinematic range (see
Fig.~\ref{figdata}). 
In order to describe the diffractive processes described in Fig.~\ref{diff-diffdisdiag}
where there is no colour exchange between the proton in the final state and the
scattered jet,
we have to introduce new variables in addition to the ones used to
describe the inclusive DIS such as $Q^2$, $W$, $x$ and $y$. Namely,
we define $\xpom$, which is the
momentum fraction of the proton carried by the colourless object (called the
pomeron), and $\beta$, the momentum fraction of the pomeron carried by the
interacting parton inside the pomeron, if we assume the pomeron to be made of
quarks and gluons:
\begin{eqnarray}
\xpom &=& \xi = \frac{Q^2+M_X^2}{Q^2+W^2} \\
\beta &=& \frac{Q^2}{Q^2+M_X^2} = \frac{x}{\xpom}.
\end{eqnarray}

\begin{figure}[t]
\begin{center}
\epsfig{file=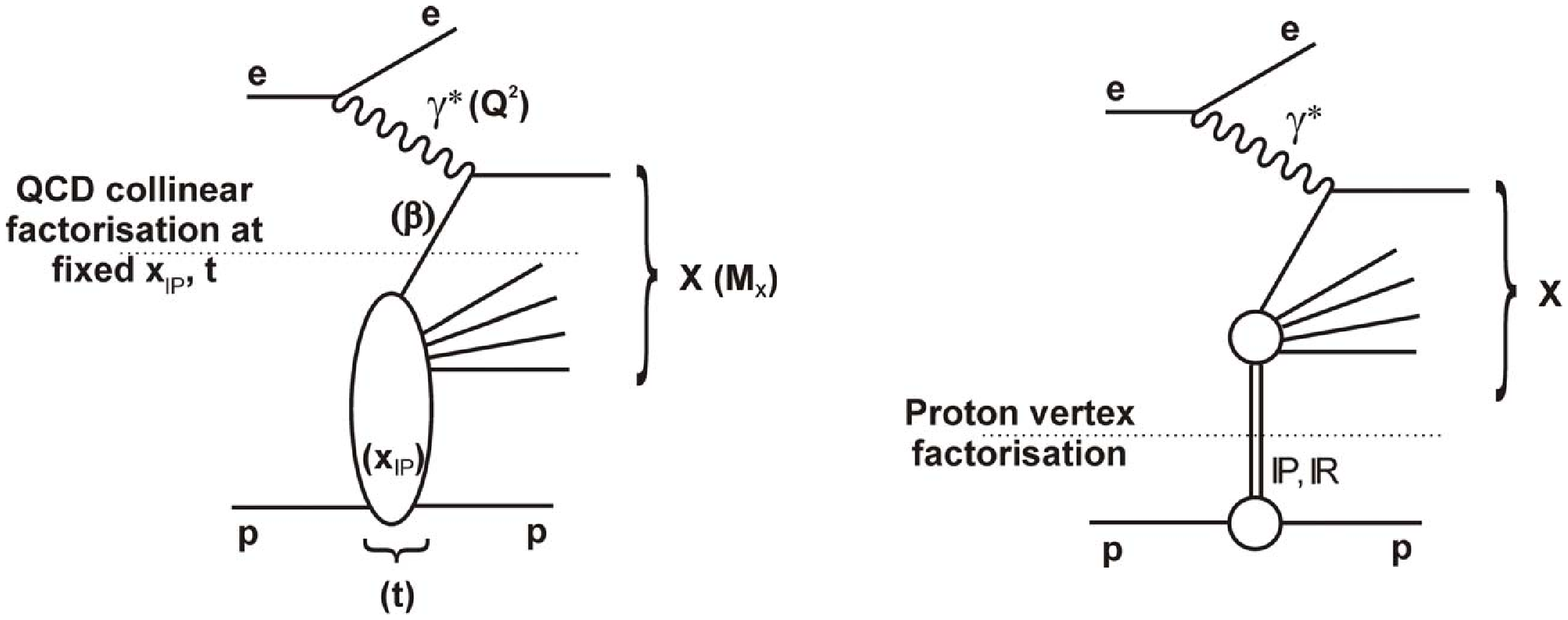,width=5cm,angle=270}
\caption{Diffractive factorisation}
\label{fact}
\end{center}
\end{figure}

In order to make quantitative predictions,
we need to distinguish two kinds of factorisation at HERA. The first factorisation is the
QCD hard scattering collinear factorisation at fixed $\xpom$ and $t$
(see left plot of Fig.~\ref{fact})~\cite{collins}, namely
\begin{eqnarray}
d \sigma (ep \rightarrow eXY) = f_D(x,Q^2,\xpom,t) \times
d \hat{\sigma} (x,Q^2)
\end{eqnarray}
where we can factorise the flux $f_D$ from the cross section $\hat{\sigma}$.
This factorisation was proven recently, and separates the $\gamma q$ coupling to
the interaction with the colourless object. 
The Regge factorisation at the proton vertex allows to factorise 
the $(\xpom,t)$ and $(\beta,Q^2)$ dependence, or in other words the hard
interaction from the pomeron coupling to the proton (see right plot of Fig.~\ref{fact}).

The measurement of the diffractive structure function is shown in Figs.~\ref{figdata}
and \ref{bekw} (next section):
\begin{eqnarray}
\frac{d^3 \sigma^D}{d \xpom dQ^2 d \beta} = \frac{2 \pi \alpha_{em}^2}{\beta Q^4}
\left( 1-y+\frac{y^2}{2} \right) \sigma_r^D(\xpom, Q^2, \beta).
\end{eqnarray}
We notice that the measurement has been performed with high precision over a
wide kinematical domain: $0.01 < \beta < 0.9$, $3.5 < Q^2 < 1600$ GeV$^2$,
$10^{-4}<\xpom<5.10^{-2}$ for H1 as an example.
We also observe that the diffractive cross section
shows a hard dependence in the centre-of-mass energy of the $\gamma^*p$ system $W$.
Namely, we get a behaviour of the form $\sim W^{ 0.6}$  for the diffractive cross section, 
compatible with the dependence expected
for a hard process. This  first observation allows further studies of the diffractive process in the context of
perturbative QCD, as diffractive PDFs or dipole models, which are described in the following.

%\clearpage

%\newpage

%\input{diffmodels}

\section{Different models of diffraction}

In this section, we 
describe the diffractive interactions and their link with the proton structure.
We should not forget that diffractive processes represent a
sizable fraction of the total cross section at HERA, Tevatron and LHC energies. 
Diffraction is also a natural process to obtain a better understanding 
of the fundamental issues of 
saturation. 
We present below four different interpretations of diffractive events.
The challenge for all models is not only to describe diffractive data from HERA but also
diffraction at Tevatron and then LHC. 
In this review, we introduce the
different models in the context of HERA and later on,
at hadronic colliders. A detailed quantitative
comparison of the different models can be found in Ref.~\cite{us}.

\subsection{Diffraction via a Pomeron made of quarks and gluons}

The requirement of a separation of the diffracted final state from the
target leads, at high energy, to the presence of a large rapidity gap
between the two systems. In the parton model, there is no mechanism for producing
large rapidity gaps other than by fluctuations in the hadronization
process which are short range in rapidity. Therefore diffractive 
dissociation as such has to be
introduced by hand.
The idea of Ingelman and Schlein was
to postulate that the pomeron has a partonic structure which may be
probed in hard interactions in much the same way as the partonic
structure of hadrons~\cite{is,us,lolopic}. They suggested that the partonic
structure of the pomeron would manifest itself in the production of
high transverse momentum jets associated with single diffractive
dissociation, for example in $pp$ scattering. The trigger for such a
reaction would consist of a quasi elastically scattered proton and the
presence of high $p_T$ jets in final state of the the dissociated
hadron. The jets would be accompanied by remnants of the pomeron
and of the diffracted hadron. 

As we mentionned already, according to Regge theory, we can factorise the 
$(\xpom,t)$ dependence from the $(\beta,Q^2)$ one for each trajectory
(Pomeron and Reggeon). The first  diffractive structure
function measurement from the H1 collaboration~\cite{firstf2d} showed that the
assumption of plain factorization between the $\xpom$ and ($\beta$, $Q^2$)
dependences was not true. The natural solution as observed in soft
physics was that two different
trajectories, namely pomeron and secondary reggeon, were needed to describe the
measurement, which lead to a good description of the data. The diffractive structure
function then reads:
\begin{eqnarray}
F_2^D \sim f_p(\xpom) (F_2^D)_{Pom}(\beta, Q^2) + 
f_r(\xpom) (F_2^D)_{Reg}(\beta, Q^2)
\end{eqnarray}
where $f_p$ and $f_r$ are the pomeron and reggeon fluxes, and $(F_2^D)_{Pom}$
and $(F_2^D)_{Reg}$ the pomeron and reggeon structure functions. The flux
parametrisation is predicted by Regge theory:
\begin{eqnarray}
f(\xpom,t) = \frac{e^{B_Pt}}{\xpom^{2 \alpha_P(t) -1}}
\end{eqnarray}
with the following pomeron trajectory
\begin{eqnarray}
\alpha_P (t)= \alpha_P(0) + \alpha'_P t.
\end{eqnarray}

The next step is to perform Dokshitzer Gribov Lipatov Altarelli Parisi (DGLAP)
\cite{dglap} fits to the pomeron structure function based on the 
Ingelman and Schlein model of the pomeron. If we assume that the
pomeron is made of quarks and gluons, it is natural to check whether the DGLAP
evolution equations are able to describe the $Q^2$ evolution of these parton
densities. As necessary for DGLAP fits, a form for the input distributions is assumed
at a given $Q_0^2$ and is evolved using the DGLAP evolution equations to a
different $Q^2$, and fitted to the diffractive structure function data at
this $Q^2$ value.

\begin{figure}
\begin{center}
\epsfig{file=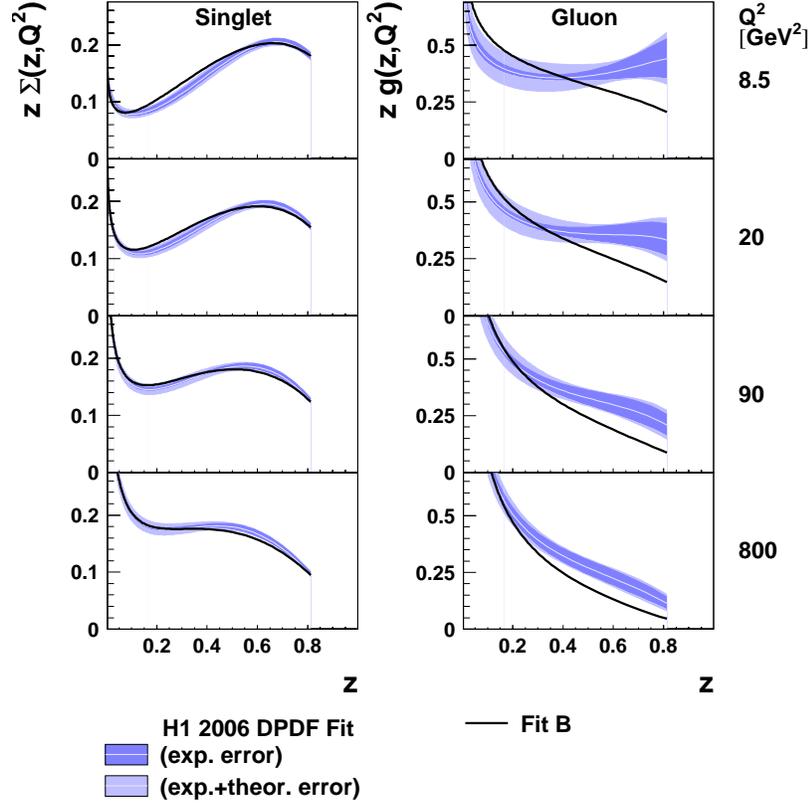,width=11cm}
\vspace{12cm}
\caption{Extraction of the parton densities in the pomeron using a DGLAP NLO fit
(H1 collaboration).}
\label{fig:pdfs}
\end{center}
\end{figure}

The DGLAP QCD fit allows to get the parton distributions in the pomeron as a
direct output of the fit, and is displayed in Fig.~\ref{fig:pdfs} as a blue shaded
area as a function of $\beta$. We first note that the gluon density is much
higher than the quark one, showing that the pomeron is gluon dominated. We also
note that the gluon density at high $\beta$ is poorly constrained which is shown
by the larger shaded area. 
Another fit was also performed by the H1 collaboration 
and is displayed as a black line in Fig.~\ref{fig:pdfs}. This shows further that the
gluon is very poorly constrained at high $\beta$ and some other data sets such
as jet cross section measurements are needed to constrain it further. The H1
collaboration showed that the jet data have the tendency to favour the lowest values of
the gluon density (black line in Fig.~\ref{fig:pdfs}).

As we show in the following, these quark and gluon densities
in the pomeron are essential ingredients to predict diffractive cross
sections at Tevatron and LHC
that we describe in the next subsections.

%%%%%%%%%%%%%%%%%%%%%%%%%%%%%%%
\subsection{Dipole models}
Another model to describe inclusive diffraction at HERA is based on
dipole models.
It is useful to look at $ep$ scattering in a frame
where the virtual photon moves very fast (for instance in the proton
rest frame, where the $\gamma^*$ has a momentum of up to about 50 TeV at
HERA).  The virtual photon can fluctuate into a quark-antiquark pair,
forming a small color dipole.
Because of its large Lorentz boost, this virtual pair has a lifetime
much longer than a typical strong interaction time. 
Since the interaction between the pair and the proton is
mediated by the strong interaction, diffractive events are possible.
An advantage of studying diffraction in $ep$ collisions is that, for
sufficiently large photon virtuality $Q^2$, the typical transverse
dimensions of the dipole are small compared to the size of a hadron.
The interaction between the quark and the antiquark, as well as the
interaction of the pair with the proton, can be treated perturbatively.
With decreasing $Q^2$ the color dipole becomes larger, and at very low
$Q^2$ these interactions become so strong that a description in terms
of quarks and gluons is no longer justified, and the
diffractive reactions become very similar to those in hadron-hadron
scattering.

\begin{figure}[htbp]
\begin{center}
\epsfig{figure=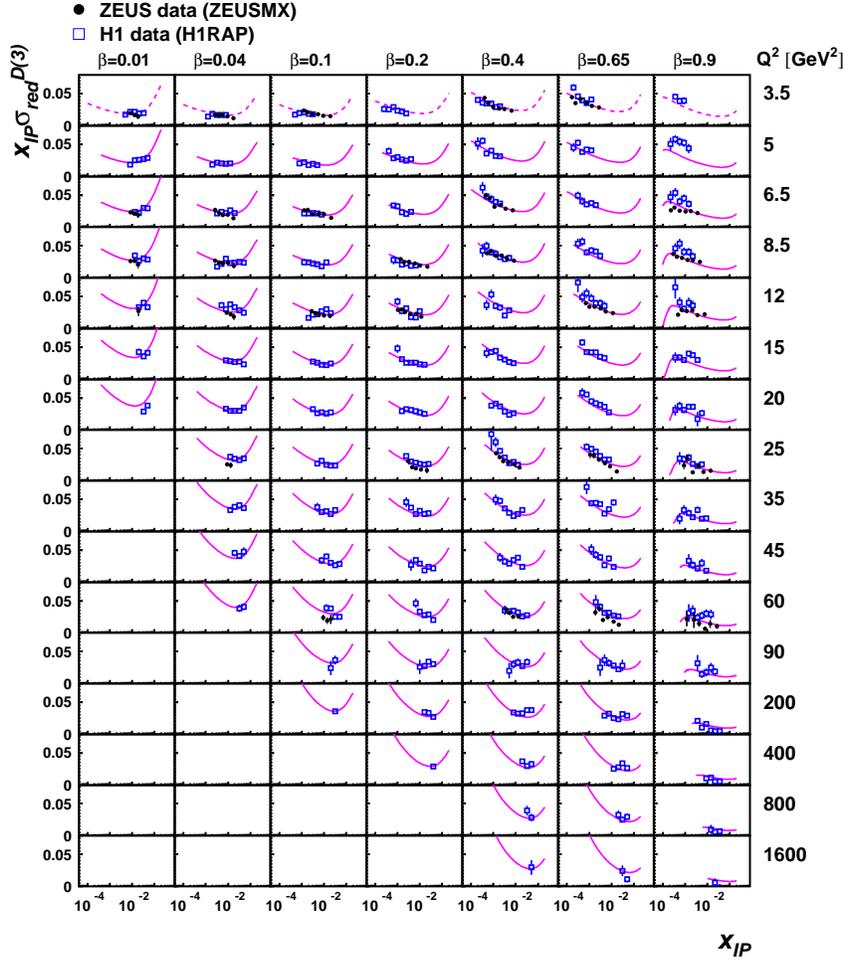,width=12.0cm,
 bbllx=10pt,bblly=70pt,bburx=580pt,bbury=715pt}
\end{center}
\vspace*{-1cm}
\caption{ Comparison of H1 and ZEUS data sets with the prediction of 
the dipole model.
Only the statistical part of the uncertainty is shown for the data points on this plot.
A dashed line is drawn for the prediction of the fit on points not included in the analysis.}
\label{bekw}
\end{figure}

The original dipole model assumes the simplest perturbative description of the Pomeron by a
 two-gluon ladder \cite{dipole1,bartels}. 
A parametrisation of the diffractive
structure function in
terms of three main contributions is proposed. The first term describes the 
diffractive production of a $q \bar{q}$ pair from
a transversely polarised photon, the second one the production of 
a diffractive $q \bar{q} g$ system, and the third one the production of a
$q \bar{q}$ component from a longitudinally polarised photon. The dipole model
leads to a good description of data. In Fig.~\ref{bekw}, we give the
comparison between the H1 and ZEUS data and the dipole model which leads to a
good description of the $F_2^D$ measurement on the full range.
Other extensions
of the dipole model containing for instance higher order contributions
such as $q \bar{q} gg$, $q \bar{q} ggg$, etc., exist and lead also
to a good description of data~\cite{dipole1}. Unfortunately, by definition,
it is difficult to transpose the dipole model to hadronic colliders.

\subsection{Description of $F_2^D$ using saturation models}
\begin{figure}[htbp]
\begin{center}
\epsfig{figure=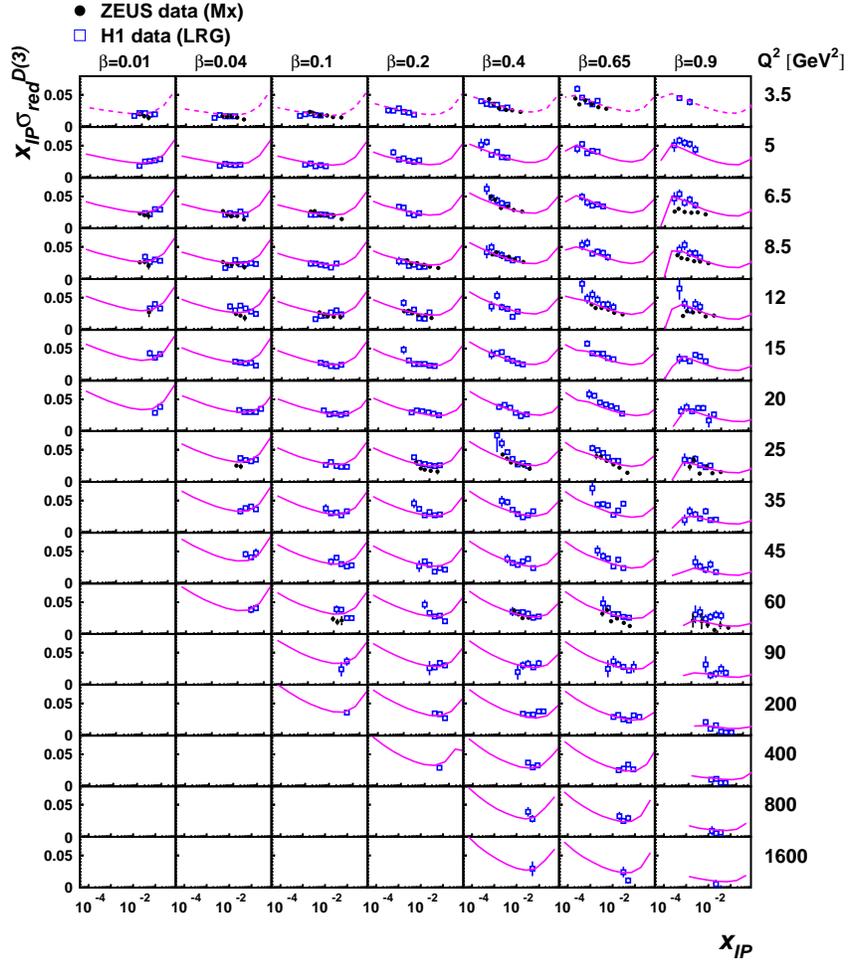,width=12.0cm,
 bbllx=10pt,bblly=70pt,bburx=580pt,bbury=715pt}
\end{center}
\vspace*{-1cm}
\caption{ Comparison of H1 and ZEUS data sets with the prediction of 
the saturation model.
Only the statistical part of the uncertainty is shown for the data points on this plot.
A dashed line is drawn for the prediction of the fit on points not included in the analysis.}
\label{fig:gbw_red_cs}
\end{figure}

We have already introduced in a previous section the concept of saturation. Below,
 we  discuss the practical implementation of this concept by Golec-Biernat and 
W\"usthoff~\cite{golec}, which has been formulated in the color dipole picture. 
In this formalism, both the inclusive and diffractive cross sections may 
be calculated.
The diffractive structure function $F^{D(3)}_2$ is the sum of three 
contributions~\cite{golec} :

\begin{equation}
\label{eq:gbw_f2dsum}
F_2^{D(3)}(Q^2,x_{I\!\!P},\beta)\,=\,F_T^{q\bar{q}}+
F_L^{q\bar{q}}+F_T^{q\bar{q} g}.
\end{equation}
The dipole cross section has the following form:
\begin{equation}
\hat\sigma (x,r)\,=\,\sigma_0\,\left\{
1\,-\,\exp\left(-r^2\, Q^2_{\rm sat}(x)/4  \right) \right\},
\qquad Q^2_{\rm sat}(x) = \left(\frac{x_0}{x}\right)^{\lambda}
\end{equation}
which introduces three parameters : the maximal possible value of the dipole 
cross section $\sigma_0$ and two parameters characterizing the saturation 
scale $Q^2_{\rm sat}(x)$  that is $\lambda$ and $x_0$. In
Fig~\ref{fig:gbw_red_cs}, we give the comparison between the H1 and ZEUS $F_2^D$
data and the saturation model which leads to a good description of data. It is
worth noticying that both the inclusive $F_2$ and the diffractive $F_2^D$
measurements can be described within the same framework of saturation models.
Since these original ideas, many theoretical and pheneomelogical developments
occured which lead to a good description of data at HERA~\cite{golec}. It
is worth noticing that this is one of the only models which aim at a global
description of HERA, RHIC and also Tevatron and LHC in a given phase space where
the gluons dominate. 
This model provides an essential tool to examine the prior effects of saturation
that takes place at the microscopic level of the dipole amplitude.
For this sake, diffractive processes show the best sensitivity and
this is why they are so precious in analyzing the structure of the proton.

\subsection{Soft colour interaction}
This alternative model assumes that diffraction
is not due to a colourless exchange at the hard vertex (called pomeron) but
rather to string rearrangement in the final state during
hadronisation~\cite{sci}. In this
kind of model, there is a probability (to be determined by the experiment) that
there is no string connection, and so no colour exchange, between the partons
in the proton and the scattered quark produced during the hard interaction.
We  discuss further this model when we discuss the measurements at the
Tevatron.

\section{Limits of diffractive hard-scattering factorization:
hadron-hadron collisions}
\label{survival}

A natural question to ask is whether one can use the diffractive PDFs
extracted at HERA to describe hard diffractive processes in hadron-hadron
collisions, and especially to predict the
production of jets, heavy quarks or weak gauge bosons at the Tevatron.  

From a theoretical point of view, diffractive hard-scattering 
factorization does not apply to
hadron-hadron collisions because of additional interactions between the
particles in initial and final states, 
but it will be interesting to study experimentally
how factorization is broken.
The breakdown of factorisation occurs 
because of interactions between spectator partons of the colliding
hadrons. The contribution of these interactions to the cross section
does not decrease with the hard scale. Since they are not associated
with the hard scattering subprocess, factorization between the 
parton-level cross section and the
parton densities of one of the colliding hadrons is no longer true. These
additional interactions are generally soft, and we have at present to rely on
phenomenological models to quantify their effects~\cite{royon}.  
The yield of diffractive events in hadron-hadron collisions is expected to be
lower
because of these soft interactions between spectator partons
(often referred to as reinteractions or multiple scatterings).  
They can produce additional final state particles which fill the would-be
rapidity gap (hence the notion of gap survival probability).  When
such additional particles are produced, a very fast proton can no longer
appear in the final state because of energy conservation.  Diffractive
factorization breaking is thus intimately related to multiple scattering
in hadron-hadron collisions. Understanding and describing this
phenomenon is a challenge in the high-energy regime that will be reached
at the LHC~\cite{royon}. It is also worth noticying that the time scale when
factorisation breaking occurs is completely different from the hard interaction
one. Factorisation breaking is due to soft exchanges occuring in the initial and
final states which appear at a much longer time scale than the hard 
interaction. In that sense, it is expected that the survival probability will
not depend strongly on the type of hard interaction and its kinematics. In other words,
the survival probability should be similar if one produces jets of different
energies, vector mesons, photons, etc, which can be cross checked experimentally
at Tevatron and LHC.

\clearpage

\begin{figure}
\begin{center}
\epsfig{file=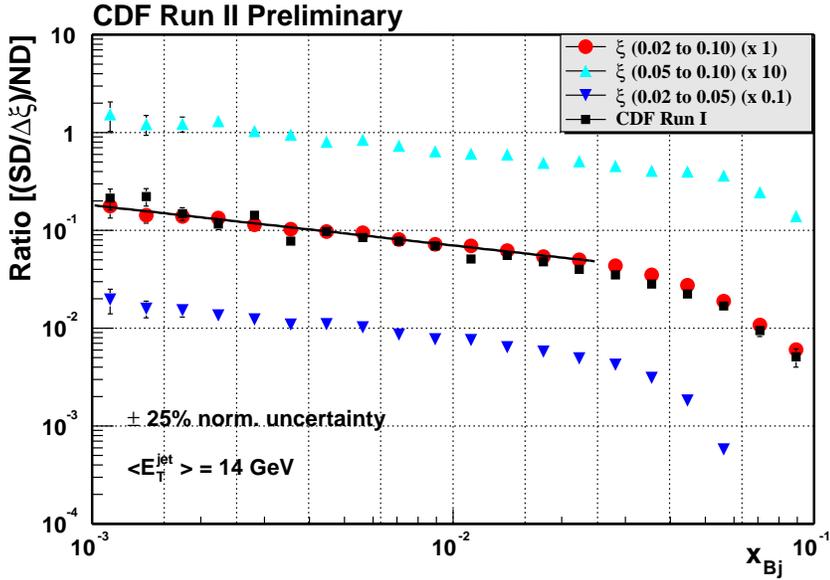,width=12cm}
\caption{Test of factorisation within CDF data alone.}
\label{fig3}
\end{center}
\end{figure}

We can also notice that
the collision partners, in $pp$ or $p\bar{p}$ reactions, are both
composite systems of large transverse size, and it is not too
surprising that multiple interactions between their constituents can
be substantial.  In contrast, the virtual photon in $\gamma^* p$
collisions shows a small transverse size, which disfavors multiple
interactions and enables diffractive factorization to hold.  According
to our discussion, we may expect that for
decreasing virtuality $Q^2$ the photon behaves more and more like a
hadron, and diffractive factorization may again be broken. 

We  now study how factorisation is broken
experimentally in two steps: is factorisation
observed within Tevatron data alone (or in other words, does the survival probability
or the soft interactions depend on the occuring hard interaction) and is
factorisation broken as expected between Tevatron and HERA?

\begin{figure}
\begin{center}
\epsfig{file=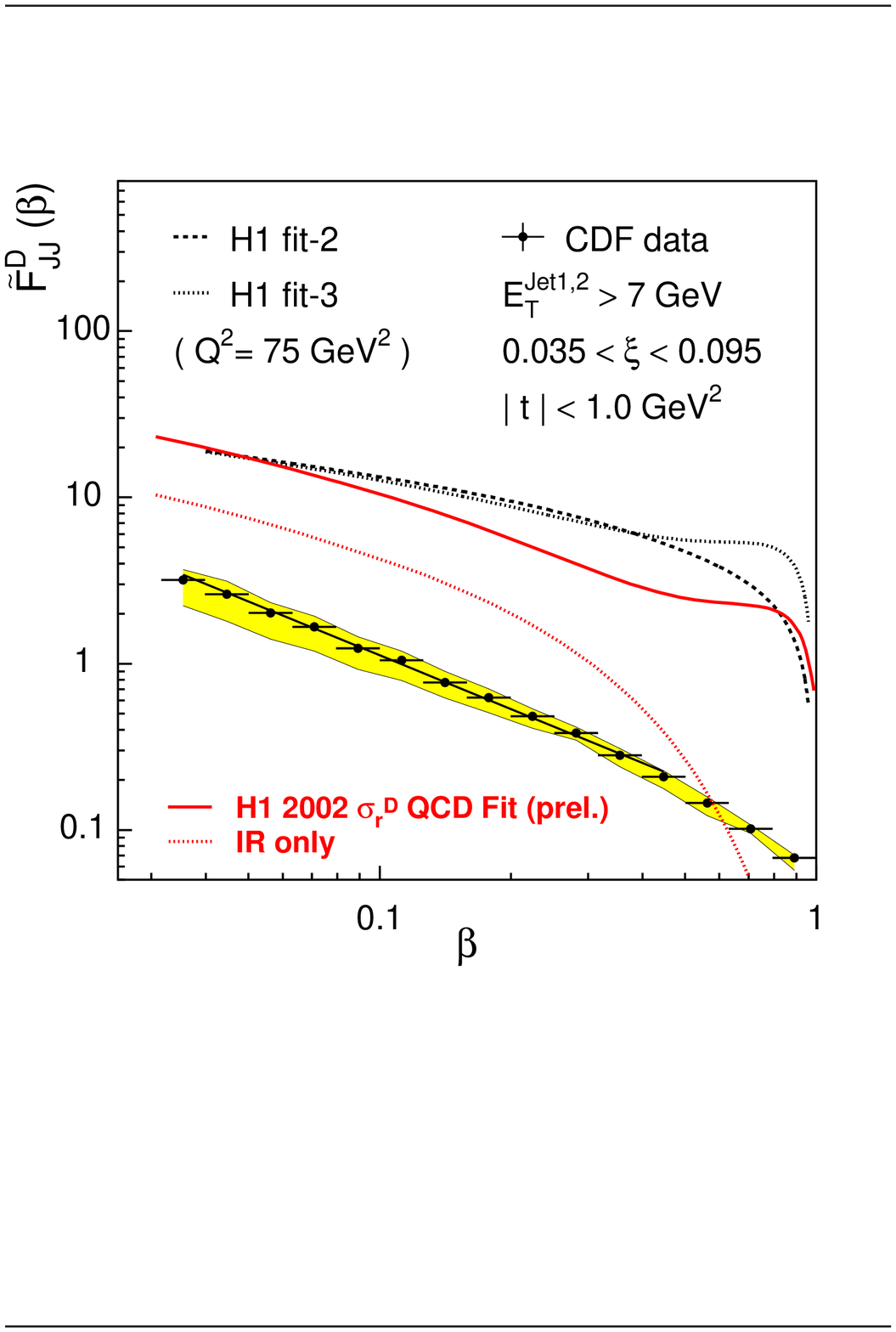,width=9cm,clip=true}
\vspace{8cm}
\caption{Comparison between the CDF measurement of diffractive structure
function (black points) with the expectation of the H1 QCD fits (red full line).}
\label{fig4}
\end{center}
\end{figure}

The CDF collaboration measured diffractive events at the Tevatron and their
characteristics. Diffractive events show as expected less QCD
radiation: as an example, dijet events are more back-to-back or the difference in
azimuthal angles between both jets is more peaked towards $\pi$. We first check 
whether factorisation holds within CDF data
alone, or in other words if the $\beta$ and $Q^2$ dependence can be factorised out
from the $\xi$ one. In Fig.~\ref{fig3}, we notice that the percentage of diffractive 
events shows the same $x$-dependence for diferent $\xi$ bins within
systematic and statistical uncertainties, which supports the fact that CDF data 
are consistent with 
factorisation~\cite{cdfdiff}. The $x$ dependence for
different $Q^2$ bins also leads to the same conclusion. 
These results show that the additional soft interactions or the multiple
interactions are compatible with a weak dependence on the hard scattering which is somewhat natural
since they occur at a
much longer time scale.

The first step of the study of factorisation breaking between Tevatron and
HERA is just confirmed by counting the percentage of diffractive events observed at 
both accelerators: 10\% at HERA and about 1\% of
single diffractive events at the Tevatron. 
The second step is to determine how factorisation is broken between Tevatron and
HERA data. It is possible to measure indirectly the diffractive structure
function at the Tevatron. The CDF collaboration measured the ratio of dijet
events in single diffractive and non diffractive events directly
proportional to the ratio of the diffractive to the ``standard" proton structure
functions $F_2$:
\begin{eqnarray}
R(x) = \frac{Rate^{SD}_{jj} (x)}{Rate^{ND}_{jj} (x)} \sim
\frac{F^{SD}_{jj} (x)}{F^{ND}_{jj} (x)}
\end{eqnarray}
The ``standard" proton structure function in this kinematic region
is known from the usual PDFs using for instance the
CTEQ or MRST parametrisations. The comparison between the CDF measurement 
(black points, with systematics errors as shaded area) and the
expectation from the H1 QCD fits in full line is shown in 
Fig.~\ref{fig4}~\cite{cdffact}. 
We notice a difference by a factor 8 to 10 between the data and the predictions from
the QCD fit assuming factorisation or a survival probability equal to 1. The
breaking of factorisation is thus confirmed and the value of the survival
probability is of the order of 0.1.
Fig.~\ref{fig4} also shows that the difference
is compatible with a constant within systematic and statistical uncertainties
on a large part of the kinematical plane in
$\beta$, which means that the survival probability is compatible with
a constant independent of $\beta$. It will be interesting
to make these studies again in a wider kinematical domain both at the Tevatron and at
the LHC. The understanding of the survival probability and its dependence on the
kinematic variables is important to make precise predictions on inclusive diffraction
at the LHC.

\begin{figure}
\begin{center}
\epsfig{file=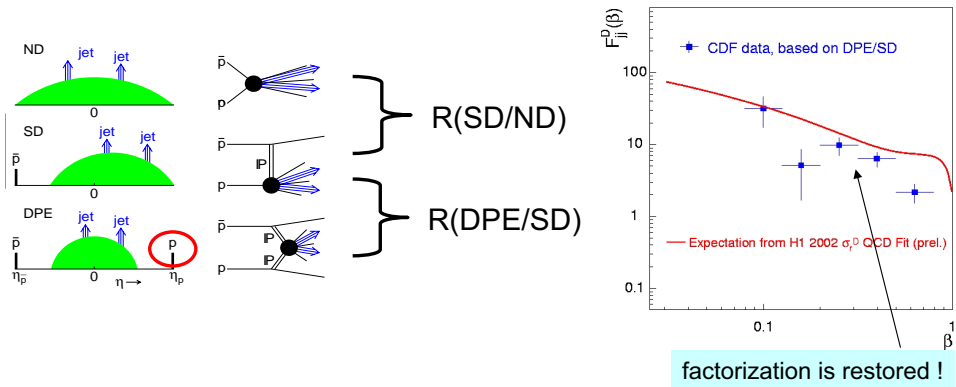,width=6cm,angle=270}
\caption{Restoration of factorisation for the ratio of double pomeron exchange
to single diffractive events (CDF Coll.).}
\label{fig5}
\end{center}
\end{figure}

The other interesting test of factorisation which can be also performed 
at the Tevatron is
to check if factorisation holds between single diffraction and double pomeron
exchange. The results from the CDF collaboration are shown in 
Fig.~\ref{fig5}~\cite{cdffact}.
The left plot shows the definition of both ratios while the right figure
shows the comparison between the ratio of double pomeron exchange to single
diffraction and the QCD predictions using HERA data in full line. 
Factorisation holds for the ratio of double pomeron exchange to single
diffraction. In other words, the price to pay for one gap is the same as the
price to pay for two gaps. The survival probability needs to be applied only 
once to require
the existence of a diffractive event, but should not be applied again for double
pomeron exchange.

\section{From soft to hard physics: vector meson production at HERA}

\subsection{Exclusive vector mesons at HERA}
\epsfigure[width=0.6\hsize]{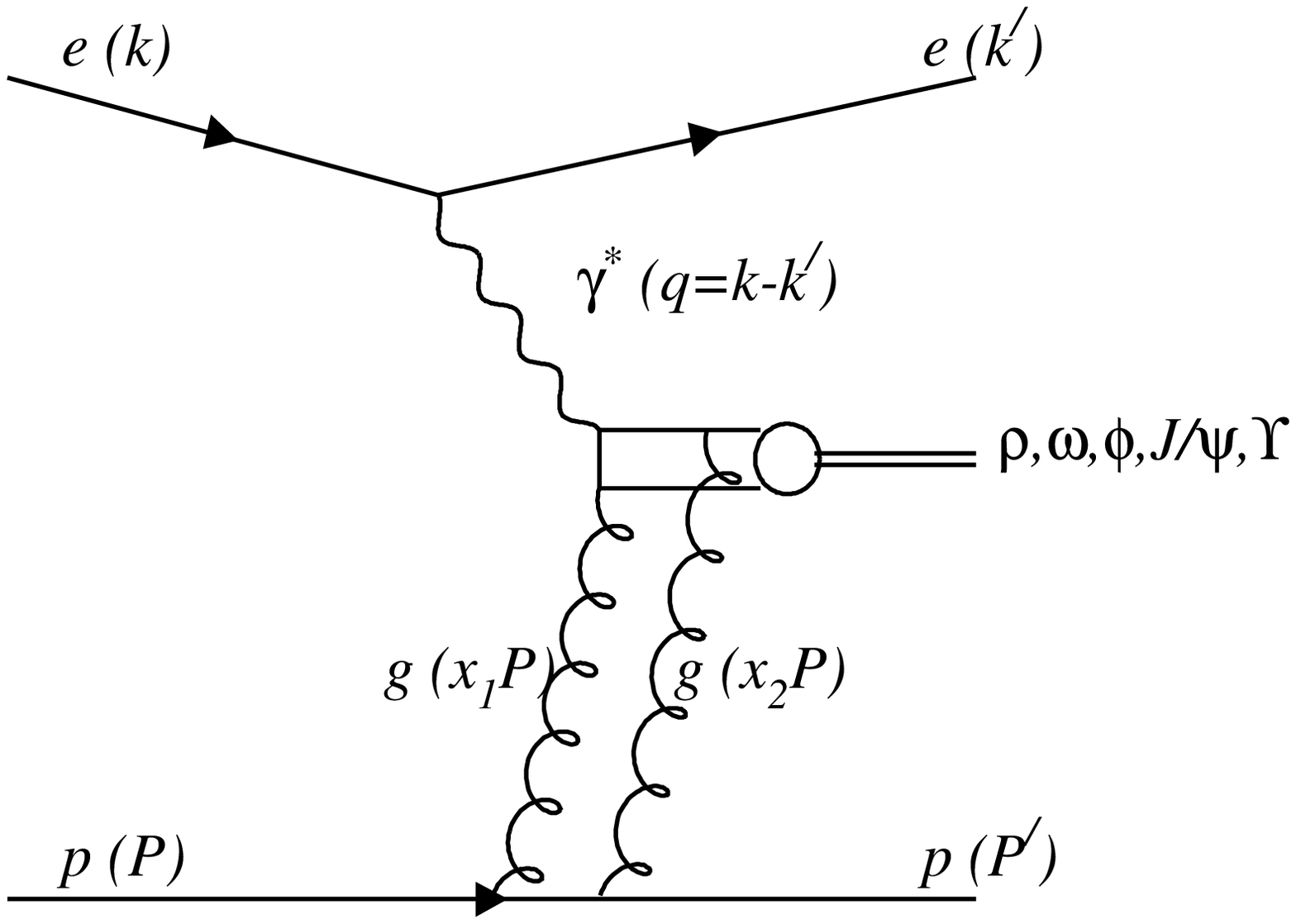} {Exclusive vector
  meson production in QCD based models.  The photon is viewed as
  fluctuating into a quark-antiquark pair, which couples to the
  proton via the exchange of two gluons (with momentum fractions
  $x_1,x_2$).  The vector meson is formed after the scattering has
  occurred.} {qcdvm}

For all processes we have discussed till now, using the point of view of PDFs,
at HERA, Tevatron or LHC, the very basic assumption is of course that
it makes sense to use PDFs to describe thoses reactions.
In other words, we have always assumed that those processes are driven by partons, which is
what we call a hard process. Most of the time, the assumption is implcit when 
there is a hard scale in the problem: for example, the $W$ or $Z$ mass in case
of events at the LHC. However, in general, we do not know the frontier between 
a hard process (parton driven) and a soft reaction, where the description
in terms of partons is yet unknown.
In fact,
together with the high parton density physics of the small-$x$ regime
of $ep$ scattering at HERA came the realization that the hard physics
studied till now is the result of an interplay of hard and soft
phenomena. In case of deep inelastic scattering the unknown soft
physics is hidden in the initial parton distributions which are
parameterized at a relatively small scale $Q_0^2 \sim 1 \gevtwo$.
The lack of a dynamical picture of the proton structure
leads to a large uncertainty about the region of phase space which has
not been probed as yet.  This uncertainty propagates itself in QCD
predictions for high energy hard scattering at future colliders.
The ability to separate clearly the regimes dominated by soft or by
hard processes is essential for exploring QCD both at a quantitative
and qualitative level.  A typical example of a process dominated by
soft phenomena is the interaction of two large size partonic
configurations such as two hadrons. A process which would lend itself
to a fully perturbative calculation, and therefore hard, is the
scattering of two small size heavy onium-states each consisting of a
pair of heavy $q\bar{q}$ pair.
In deep inelastic scattering the partonic fluctuations of the virtual
photon can lead to configurations of different sizes. The size of the
configuration will depend on the relative transverse momentum $k_T$ of
the $q\bar{q}$ pair. Small size configurations (large $k_T\sim Q/2$)
are favored by phase space considerations (the phase space volume
available is proportional to $k_T^2/Q^2$). In the quark parton model (QPM), 
in order to
preserve scaling, it was necessary to suppress their presence by
making them sterile. In QCD there is a simple explanation for this
suppression - the effective color charge of a small size $q\bar{q}$
pair is small due to the screening of one parton by the other and
therefore the interaction cross section will be
small. This phenomenon is known under the name of
color transparency.

At this point comes exclusive vector meson production, for which 
we can differentiate soft and hard components of the 
interaction processes.
In the following, we show that these reactions allows to study
the transition between hard and soft physics.  
At HERA, it was found that the cross
sections for exclusive vector meson production rise strongly with
energy when compared to fixed target experiments, if a hard scale is
present in the process. In the case of $J/\psi$
production, the strong rise of the cross section is indeed measured directly.  
The theoretical
calculations indicate that the cross sections depend on the square of
the gluon density in the proton.  If higher order calculations become
available, the measurement of the energy dependence of the vector
meson cross section may be one of the ideal methods to measure the gluon
density in the proton.
In pQCD models, the scattering (${\gamma p \rightarrow V p}$) is
viewed in the proton rest frame as a sequence of events very well
separated in time.  The process is depicted in
Fig.~\ref{fig:qcdvm}.
A first approximation of the cross section can be written as
\begin{equation}
  \left. {d\sigma \over dt}\right|_{t=0}
     (\gamma^*N\rightarrow VN) 
   = {4\pi^3\Gamma_V m_V \alpha_s^2(Q) 
   \eta_V^2\left(xg(x, Q^2)\right)^2 \over 
    3\alpha_{\rm em} Q^6} \ , 
\end{equation}
where the dependence on the meson structure 
is in the parameter
\begin{equation}
\eta_V = {1\over 2} \int {dz\over z(1-z)} \phi^V(z) /
      \int dz \phi^V(z) 
\end{equation}
and $\phi^V(z)$ is the light-cone 
wave function. 
Exclusive electroproduction of light vector mesons is a particularly
good process to study the transition from the soft to the hard
regime of strong interactions, the former being well described within
the Regge phenomenology while the latter  by perturbative QCD
(pQCD). Among the most striking expectations in this
transition is the change of the logarithmic derivative $\delta$ of the
cross section $\sigma$ with respect to the $\gamma^* p$ center-of-mass
energy $W$, from a value of about 0.2 in the soft regime 
to 0.8 in the hard one (represented by a two-gluon exchange diagram in
Fig.~\ref{fig:qcdvm}), and the decrease of the exponential slope $b$ of
the differential cross section with respect to the
squared four momentum transfer $t$, from a value of about 10
GeV$^{-2}$ to an asymptotic value of about 5 GeV$^{-2}$ when the
virtuality $Q^2$ of the photon increases \cite{rho-other,phi,jpsi}.

\begin{figure}[h]
\centerline{\includegraphics[width=0.7\columnwidth]{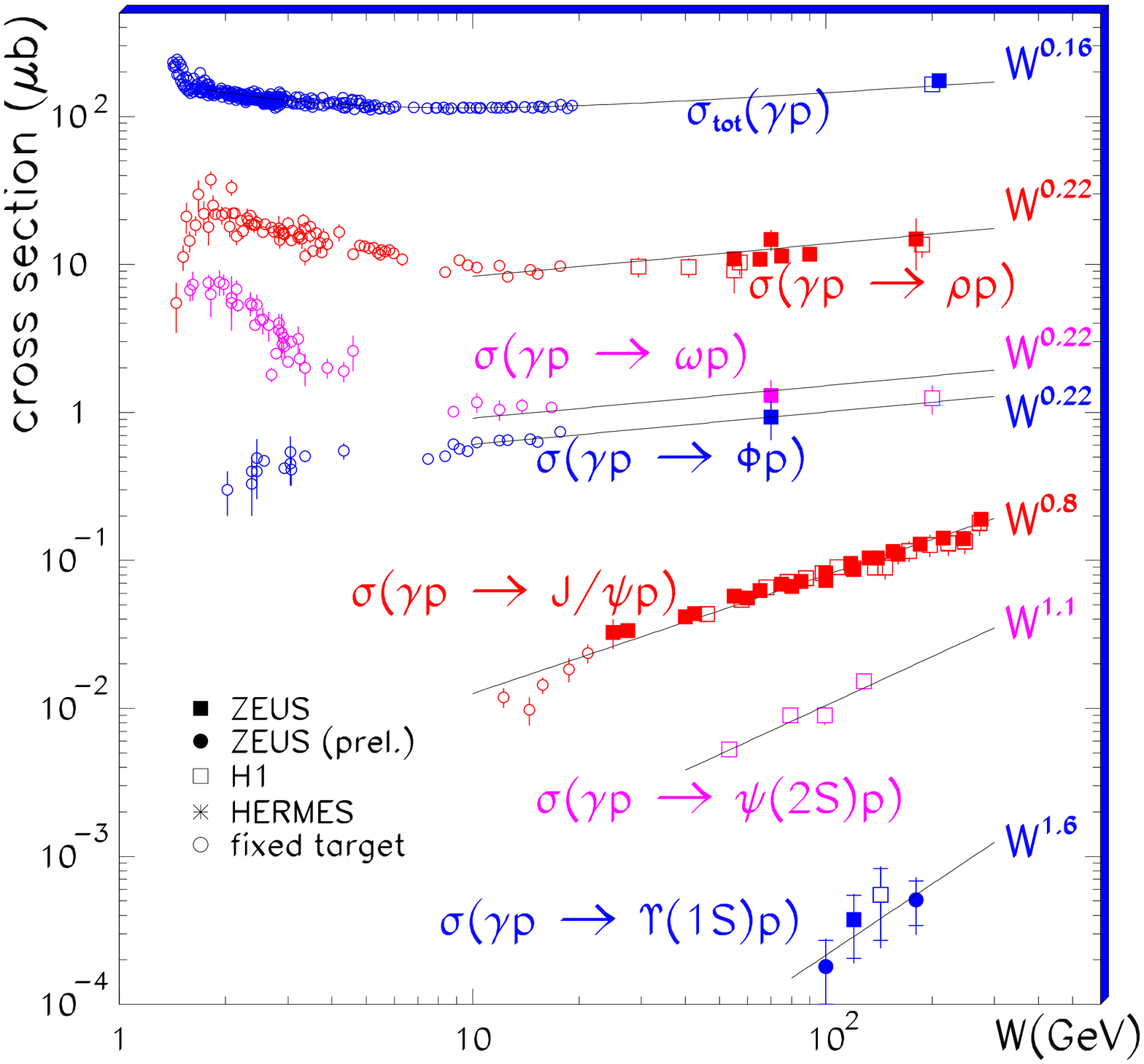}}
\caption{
$W$ dependence of the exclusive vector meson cross section in
photoproduction, $\sigma(\gamma p \to V p)$. The total photoproduction
cross section is also shown. The lines are the fit result of the
form $ W^\delta$ to the high energy part of the data.}
\label{fig:sigvm}
\end{figure}

\begin{figure}[h]
\centerline{\includegraphics[width=0.6\columnwidth]{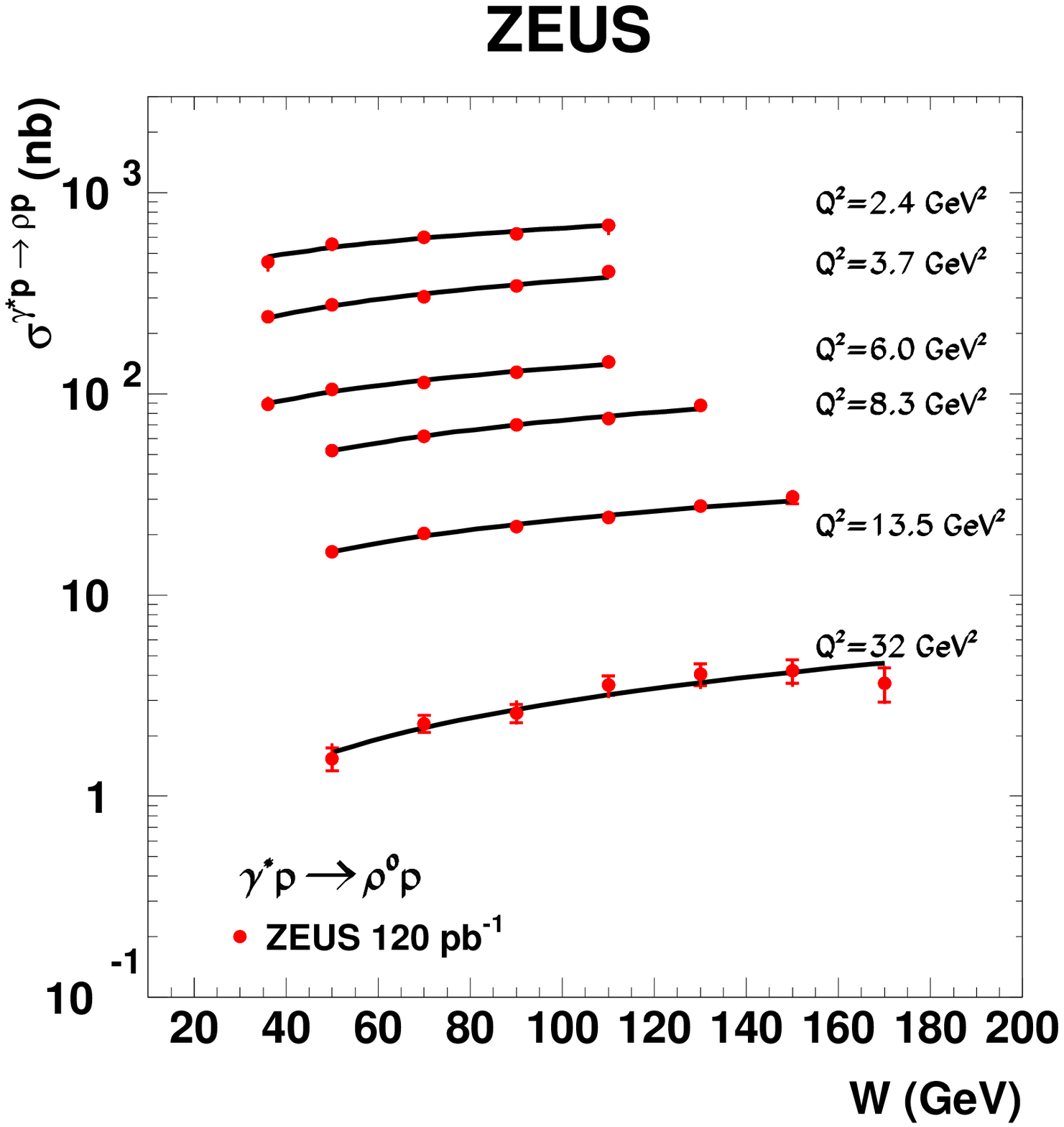}}
\caption{
$W$ dependence of the cross section for exclusive $\rho^0$
electroproduction, for different $Q^2$ values, as indicated in the
figure.  The lines are the fit results of the form $ W^\delta$
to data.}
\label{fig:w}
\end{figure}

The soft to hard transition can be seen by studying the $W$ dependence
of the cross section for exclusive vector meson photoproduction, from
the lightest one, $\rho^0$, to the heavier ones, up to the
$\Upsilon$. The scale in this case is the mass of the vector meson, as
in photoproduction $Q^2$ = 0. Fig.~\ref{fig:sigvm} shows
$\sigma(\gamma p \to V p)$ as a function of $W$ for light and heavy
vector mesons. For comparison, the total photoproduction cross
section, $\sigma_{tot}(\gamma p)$, is also shown. The data at high $W$
can be parameterised as $W^\delta$, and the value of $\delta$ is
displayed in the figure for each reaction. One sees clearly the
transition from a shallow $W$ dependence for low scales (soft) to a steeper
one as the scale increases (hard).

\begin{figure}
\begin{center}
\includegraphics[width=0.6\textwidth]{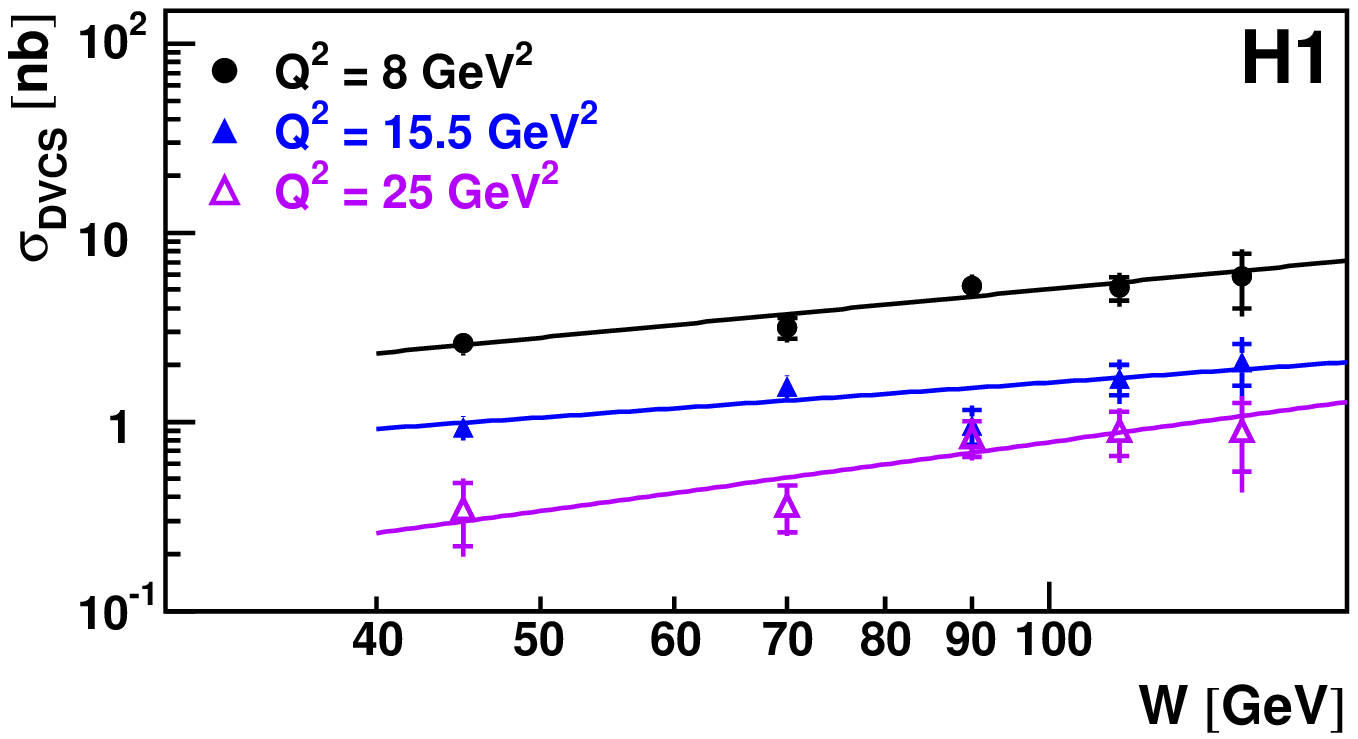}
\caption{
The DVCS cross section as a function of $W$ at three values of
$Q^2$. The solid lines represent the fit results of the form
$W^\delta$.}
\label{fig:dvcs}
\end{center}
\end{figure}

\begin{figure}
\centerline{\includegraphics[width=0.75\columnwidth]{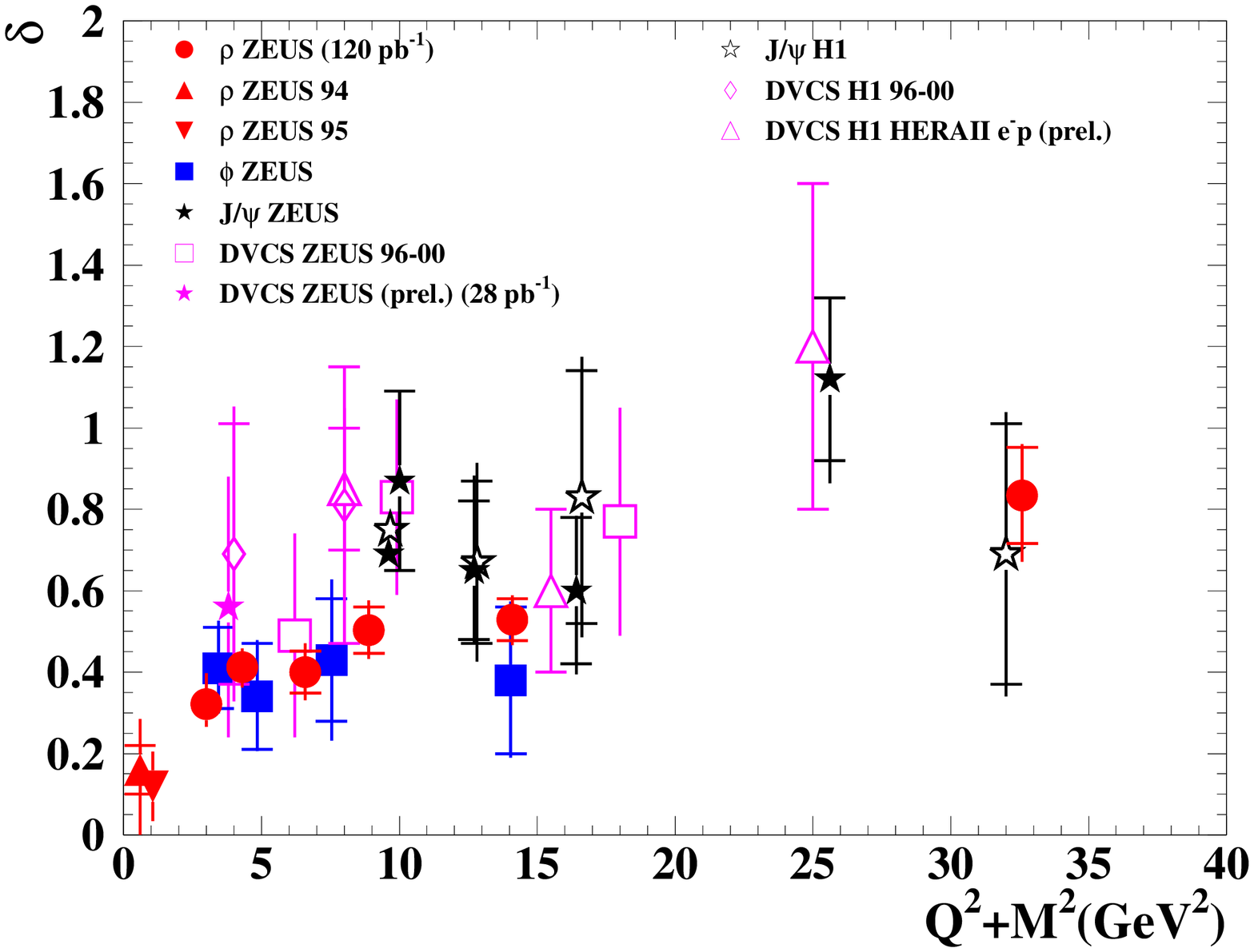}}
\caption{
A compilation of values of $\delta$ from fits of the form
$W^\delta$ for exclusive VM electroproduction, as a function
of $Q^2+M^2$. It includes also the DVCS results.}
\label{fig:del07}
\end{figure}

\begin{figure}
\centerline{
\includegraphics[width=0.5\columnwidth]{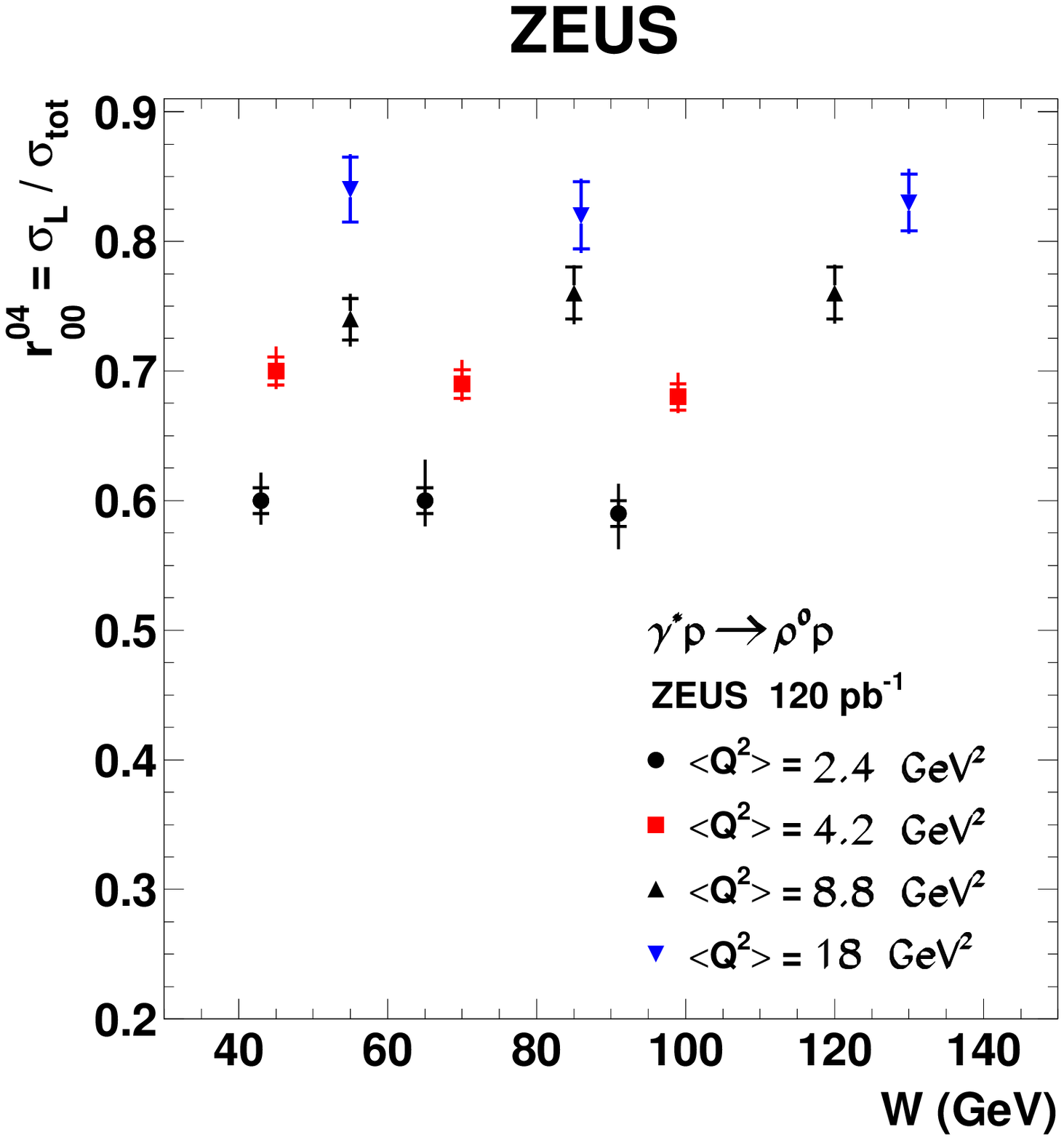}
}
\caption{
The ratio $r^{04}_{00}=\sigma_L/\sigma_{tot}$ 
as a function of $W$ for different values of $Q^2$.}
\label{fig:ratio-rho}
\end{figure}

One can also check this transition by varying $Q^2$ for a given vector
meson.  The cross section $\sigma (\gamma^* p \to \rho^0 p)$ is
presented in Fig.~\ref{fig:w} as a function of $W$, for different
values of $Q^2$. The cross section rises with $W$ in all $Q^2$ bins.
The same conclusion holds for the deeply virtual Compton scattering
(DVCS)  (Fig. \ref{fig:dvcs}),
in which a real photon is produced instead of a VM $ep \rightarrow ep \gamma$.
In order to quantify this rise, the logarithmic derivative $\delta$ of
cross section with respect to $W$ is obtained by fitting the data to the
expression $\sigma \sim W^\delta$ in  $Q^2$ intervals.  
The
resulting values of $\delta$ from recent  measurements are
compiled in Fig~\ref{fig:del07}.  
Also included in this figure are
values of $\delta$ from other measurements~\cite{rho-other} for the
$\rho^0$ as well as those for $\phi$~\cite{phi}, $J/\psi$~\cite{jpsi}
and $\gamma$~\cite{h1-dvcs,h1-dvcsb,h1-dvcsc,zeus-dvcs}. Results are plotted as a function of
$Q^2+M^2$, where $M$ is the mass of the vector meson.  One sees a
universal behaviour, showing an increase of $\delta$ as the scale
becomes larger. The value of $\delta$ at low scale is the one expected
from the soft pomeron intercept, while the one at large
scale is in accordance with twice the logarithmic derivative of the
gluon density with respect to $W$.
A comment is in order concerning the $W$ dependence of DVCS.
It reaches the same value of $\delta$ as in the hard
process of $J/\psi$ electroproduction. Given the fact that the final
state photon is real, and thus transversely polarized, the DVCS
process is produced by transversely polarized virtual photons,
assuming $s$-channel helicity conservation. 
The steep energy dependence
thus indicates that the large configurations of the virtual transverse
photon are suppressed
and the reaction is dominated by small $q {\bar q}$ configurations,
leading to the observed perturbative hard behavior.
A similar effect is observed for  $\rho^0$ production,
as displayed in Fig. \ref{fig:ratio-rho}. The ratio $\sigma_L/\sigma_{tot}$ 
is shown to be constant with $W$, which means that the $W$ 
dependence for $\sigma_L$ and $\sigma_{T}$ are about the same \cite{usscaling}.

%%%%%%%% t-dependence
\subsection{$t$-dependence of the vector meson production cross section}
\begin{figure}
\centerline{\includegraphics[width=0.75\columnwidth]{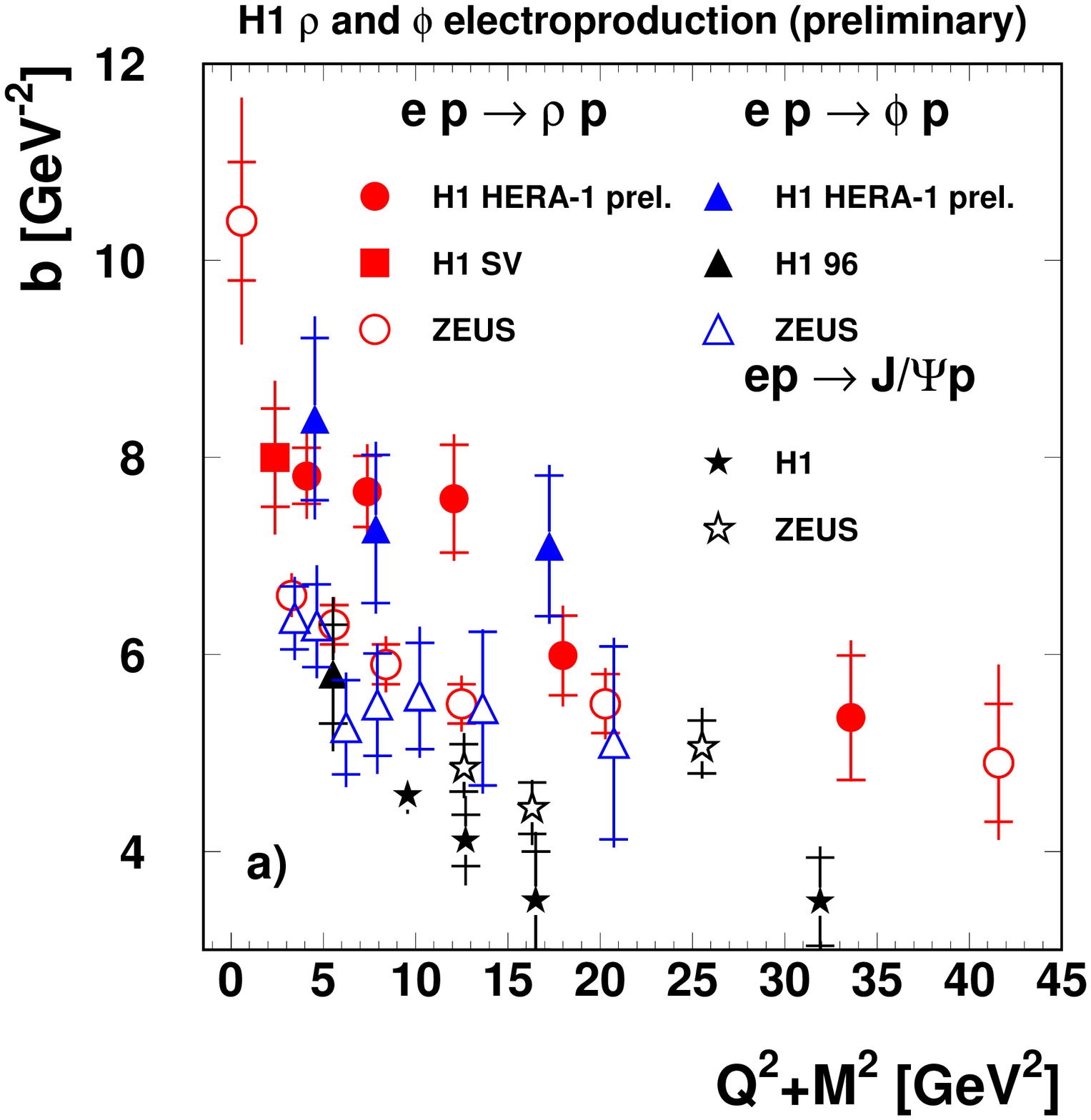}}
\caption{
A compilation of the $b$ slope values from fits of the form
$d\sigma/d|t| \propto
e^{-b|t|}$ for exclusive vector-meson electroproduction, as a function
of $Q^2+M^2$.}
\label{fig:b07}
\end{figure}

One of the key measurement in exclusive processes is the dependence 
of the cross section in $t$, 
where $t=(p-p')^2$ is the square of the momentum transfer at the
proton vertex.
The differential cross section, d$\sigma$/d$t$, is parameterised
by an exponential function $e^{-b|t|}$ (at small $t$) and fitted to the data of
exclusive vector meson electroproduction and also to DVCS. The
resulting values of $b$ as a function of the scale $Q^2+M^2$ are
plotted in Fig.~\ref{fig:b07}. As expected, $b$ decreases to a
universal value of about 5 GeV$^{-2}$ as the scale increases.

A Fourier transform from momentum
to impact parameter space shows that the $t$-slope $b$ is related to the
typical transverse distance between the colliding objects
which data allow us to measure experimentally.
At high scale, the $q\bar{q}$ dipole is almost
point-like, and the $t$ dependence of the cross section is given by the transverse extension 
of the gluons (or sea quarks) in the  proton for a given $x$ range
\cite{bernard,buk,diehl,strik,muller,compass}.
This is an important issue in modern lepton nucleon scattering that we can
call proton tomography. Detailed reviews can be found in
\cite{bernard,buk,diehl,strik}. Applications at LHC energies of the 
parton transverse profiles, derived from impact parameter analysis,
are of fundamental interests, but not yet at a practical level.

\begin{figure}
\centerline{\includegraphics[width=0.8\columnwidth]{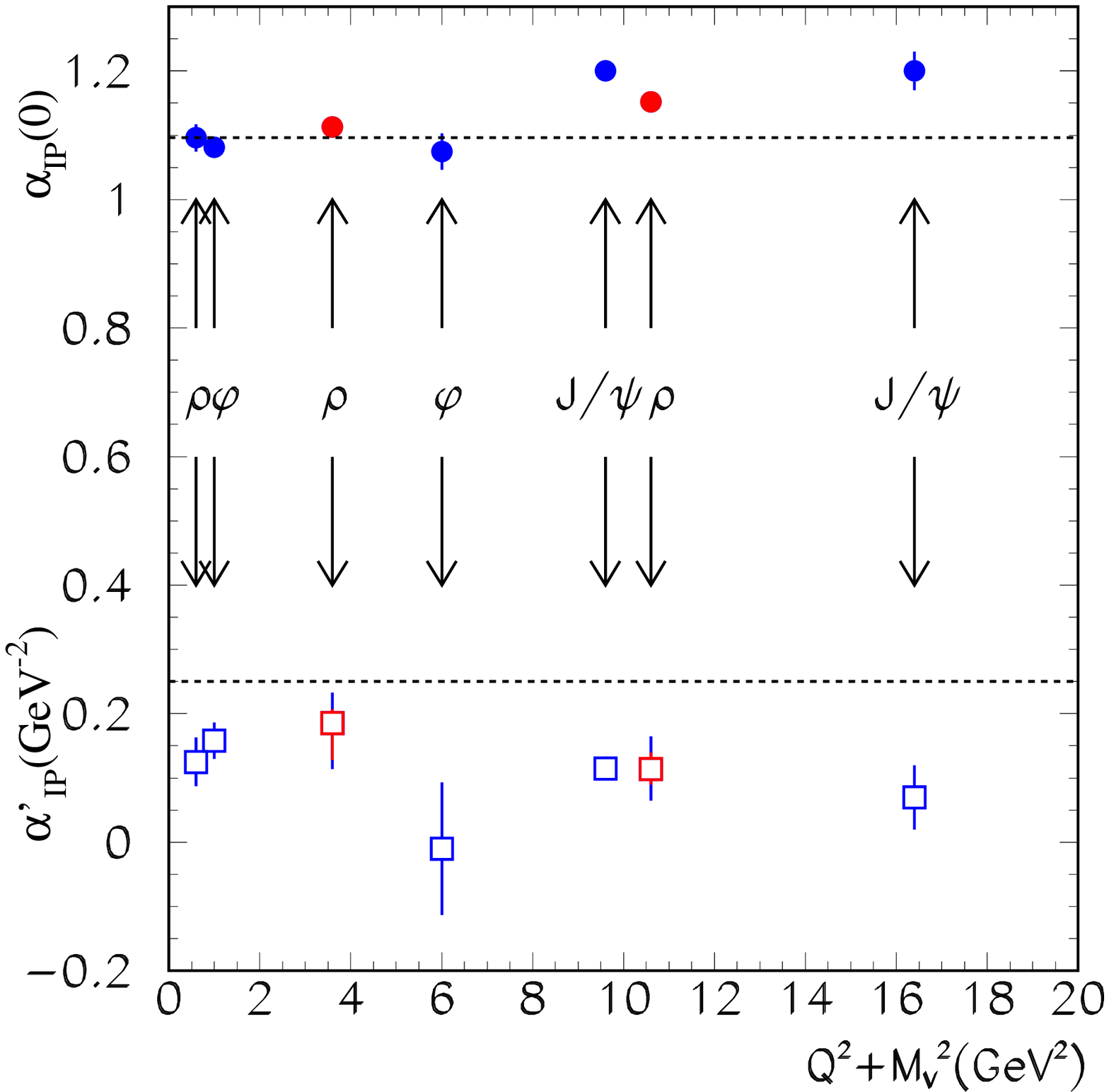}}
\caption{
Values of the intercept and slope of the effective Pomeron trajectory
as a function of $Q^2+M^2$, as obtained from measurements of exclusive VM
electroproduction. The dashed lines show the results from soft
elastic scattering.}
\label{fig:ap-apr-pom}
\end{figure}

\begin{figure}
\begin{center}
\includegraphics[width=7cm,height=5cm]{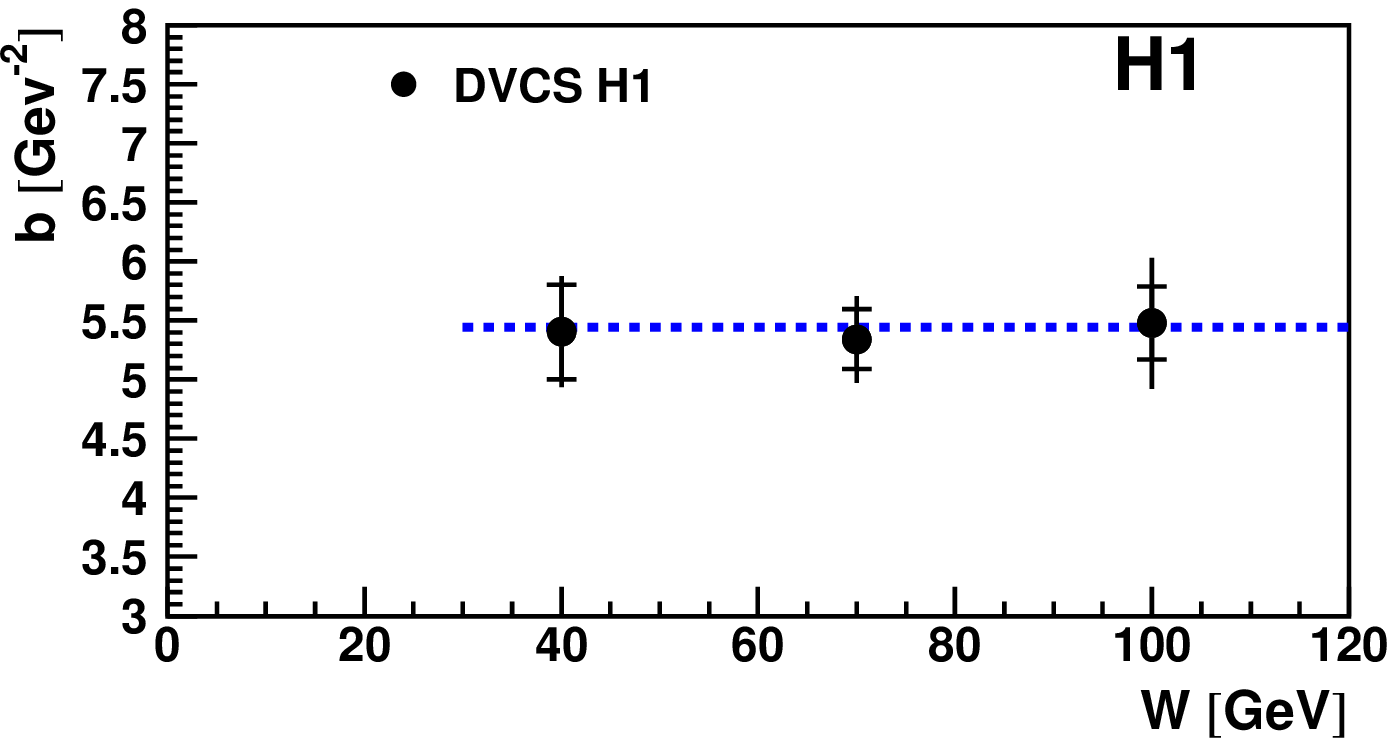}
\caption{The logarithmic slope of the $t$ dependence
  for DVCS as a function of $W$.}
\label{bslopes}
\end{center}
\end{figure}

\begin{figure}
\begin{center}
  \includegraphics [bb= 105 247 487 600,clip,width=0.5\hsize,totalheight=5cm]{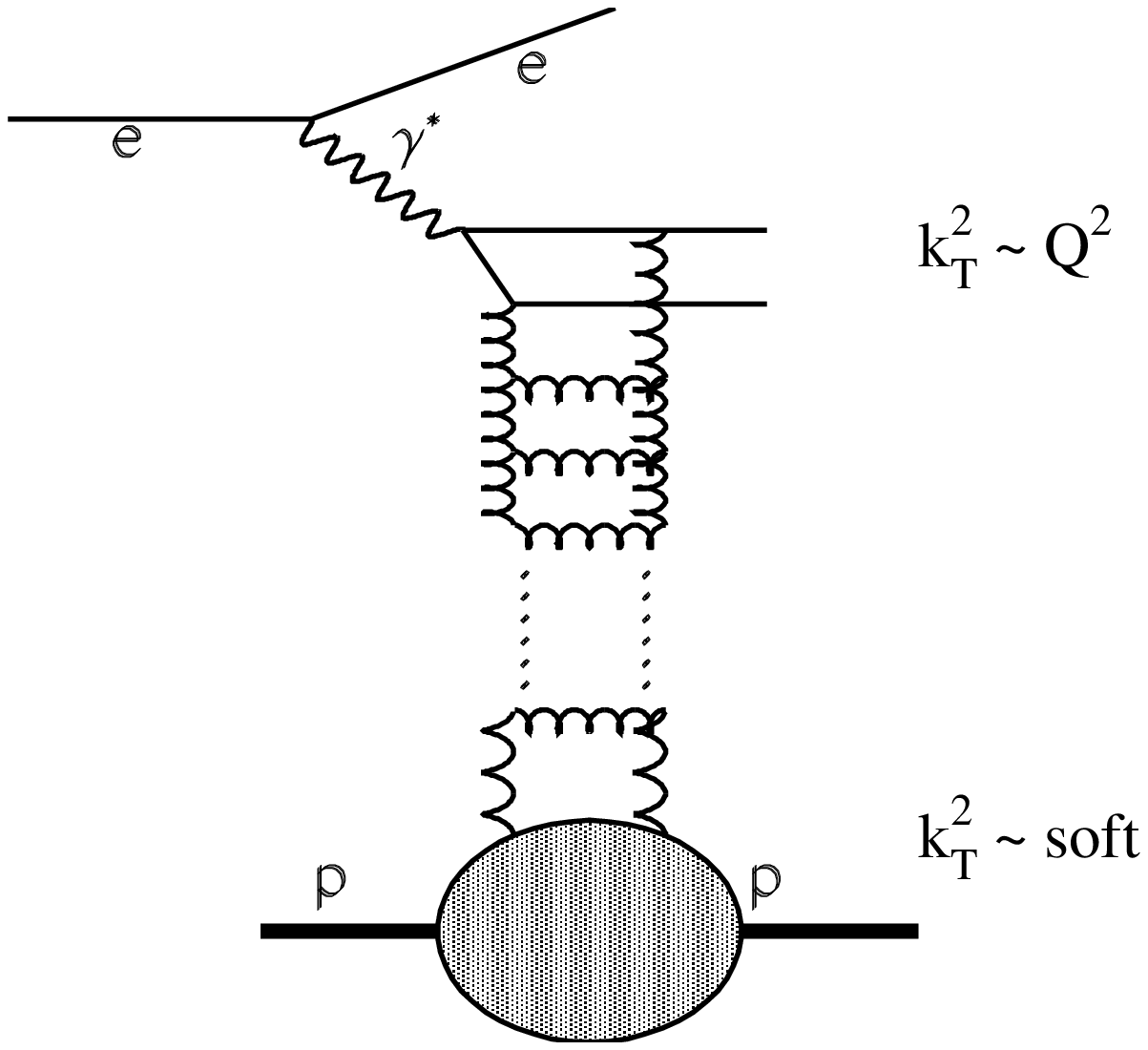}
\caption {
A diagram describing a gluon ladder in a diffractive process.}
\label{fig:ladderb}
\end{center}
\end{figure}

More generally,
one can study the $W$ dependence of d$\sigma$/d$t$ for fixed $t$
values and extract the effective pomeron trajectory
$\alpha_{IP}(t)$ for all VMs. This was done in the case of $\rho^0$ production for two $Q^2$
values and the trajectory was fitted to a linear form to obtain the
intercept $\alpha_{IP}(0)$ and the slope $\alpha_{IP}^\prime$. These
values are presented in a compilation of the effective Pomeron
intercept and slope in Fig.~\ref{fig:ap-apr-pom}. Values are plotted as a function of $Q^2+M^2$.  
We observe that the
value of $\alpha_{P}(0)$ increases with $Q^2$ while the value of
$\alpha_{P}^\prime$ tends to decrease with $Q^2$.
Results found for DVCS are consistent with with  $\alpha_{P}^\prime=0$,
as for the $J/\Psi$ \cite{h1-dvcs} (see Fig. \ref{bslopes}).
The resulting value of $\alpha^\prime \approx$ 0 is an evidence of
no shrinkage of $d\sigma/dt$ in the process $\gamma^*
p \to J/\Psi p$ or $\gamma^*
p \to \gamma p$. 
This gives an important imput for the parametrisations of pomeron flux at the
LHC, when producing  $J/\Psi$ diffractively.

\subsection{Generalised parton distributions}

Let us mention briefly at this level 
one of the newest possibilities of %modern 
lepton-nucleon scattering experiments. 
Since a few years, they can give access to the
spatial distribution of quarks and gluons in the proton
at femtometer scale. 
The general framework which describes these measurements 
is encoded in the 
generalised parton distributions (GPDs) \cite{gpdsreview}.
In fact, the reconstruction of spatial images from scattering
experiments by way of Fourier transforms of the observed 
scattering pattern is a technique widely used in physics,
for example in X-rays scattering from crystals.
Recently, it has been discovered how to extend this technique
to the spatial distribution of quarks and gluons
within the proton, using processes that probe the proton at a tiny resolution scale.
Mapping out the GPDs is an ambitious program that requires
a large amount of experimental information and future programs at JLab
and CERN (COMPASS),
in the continuation of HERA measurements (see
\cite{h1-dvcs,h1-dvcsb,h1-dvcsc,zeus-dvcs} and 
\cite{jlaball,hermes,bcah1,dhose}).
This domain of research follows a great expansion, both on the experimental or theoretical sides.
%The concepts behind GPDs will in many ways revolutionized how scientists think about the
%structure of the nucleon. 
First, as mentioned above via spatial imaging of the nucleon. Second, GPDs
will allow us to quantify how the orbital motion of quarks in the nucleon contributes to the 
nucleon spin, a question of crucial importance for our understanding of the mechanics
underlying nucleon structure.
Third, the spatial view of the nucleon enabled by the GPDs provides us with new
ways to test dynamical models of the nucleon structure. This will be relevant
to understand the proton structure in detail such as the spatial and energy 
partonic structure. However, 
in this review, we focus on results (mainly based
on standard PDFs) that impact directly LHC measurements and searches and the impact of
this new knowledge is rather weak and 
we do not to enter into details of this huge physics topic which
deserves a dedicated review.

\section{Soft and hard diffraction at the LHC}
The LHC with a center-of-mass energy of 14 TeV will allow us to access a completely
new kinematical domain in diffraction. So far, three experiments, namely ATLAS and
CMS-TOTEM have shown interest in diffractive measurements.
The diffractive event selection at the LHC will be the same as at the Tevatron.
However, the rapidity gap selection will no longer be possible at high
luminosity since up to 35 interactions per bunch crossing  are expected to occur
and soft pile-up events will kill the gaps produced by the hard interaction.
Proton tagging will thus be the only possibility to detect diffractive events at
high luminosity. Let us note that this is not straightforward:
we need to make sure that the diffracted protons come from the hard
interaction and not from the soft pile up events. The idea we  develop in
the following is to measure precisely the time of arrival of the diffracted
protons in the forward detectors, and thus know if the protons come from the
vertex of the hard interaction.       

Measurements of total cross section and luminosity are foreseen in the 
ATLAS-ALFA~\cite{atlaslumi} and TOTEM~\cite{totem} experiments, and roman pots are
installed at 147 and 220 m in TOTEM and 240 m in ATLAS. These measurements will
require a special injection lattice of the LHC at low luminosity since they require the
roman pot detectors to be moved very close to the beam.

\begin{figure}
\begin{center}
\epsfig{file=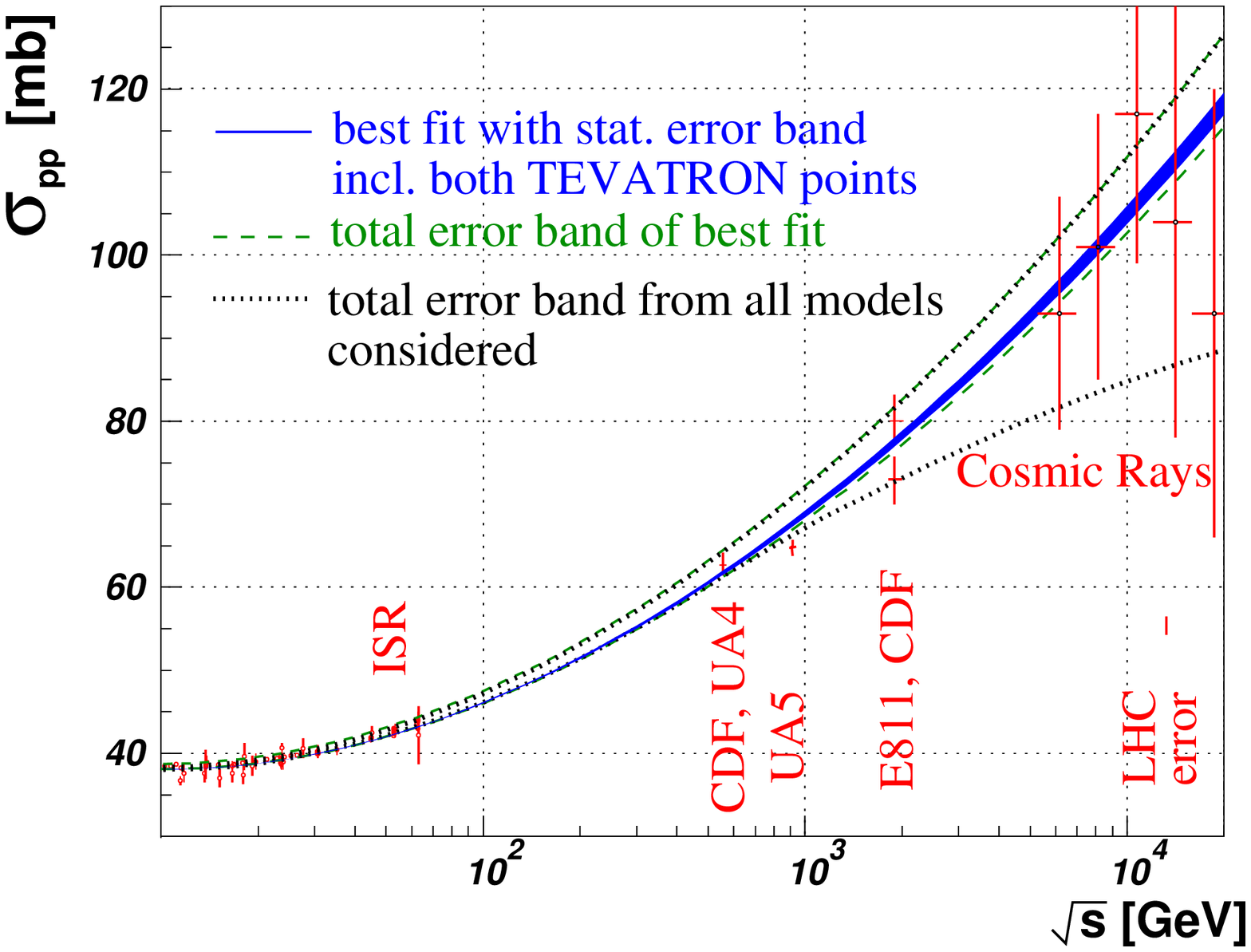,width=10cm}
\caption{Measurement of the total cross section.}
\label{fig18}
\end{center}
\end{figure}

The measurement of the total cross section to be performed by the TOTEM
collaboration~\cite{totem} is shown in Fig.~\ref{fig18}.
We notice that there is a large uncertainty on predictions of the total cross
section at the LHC energy in particular due to the discrepancy between the two
Tevatron measurements. The inelastic $p \bar{p}$ cross section was measured at
a center-of-mass energy of 1.8 TeV at the Tevatron by the E710, E811 and CDF
collaborations which lead to the following respective results: $56.6 \pm 2.2$ mb,
$56.5 \pm 1.2$ mb and $61.7 \pm 1.4$ mb~\cite{tevtotal}. While the E710 and E811 experiments agree (E811 is
basically the follow up of E710), the E811 and CDF measurements disagree by
9.2\%, and the reason is unclear~\cite{tevtotal}.
The measurement of TOTEM will be of special
interest to solve that ambiguity as well.

The ATLAS collaboration prefers to measure the elastic scattering in the Coulomb
region~\cite{atlaslumi}, typically at very low $t$ ($|t| \sim 6.5~10^{-4}$ GeV$^2$). When $t$ is
close to 0, the $t$ dependence of the elastic cross section reads:
\begin{eqnarray}
\frac{dN}{dt} (t \rightarrow 0) = L \pi \left( \frac{-2 \alpha}{|t|} +
\frac{\sigma_{tot}}{4 \pi} (i+\rho)e^{-b|t|/2} \right)^2.
\end{eqnarray}
From a fit to the data in the Coulomb region, it is possible to determine
directly the total cross section $\sigma_{tot}$, the $\rho$ and $b$ parameters
as well as the absolute luminosity $L$. This measurement requires to go down to
$t \sim 6.5~10^{-4}$ GeV$^2$, or $\theta \sim 3.5~\mu$rad (to reach the
kinematical domain where the strong
amplitude equals the electromagnetic one). 
This measurement requires a special high $\beta^*$ lattice, the detectors to be
installed 1.5 mm from the LHC beam, a spatial resolution of these detectors well
below 100 $\mu$m and no significant dead edge on the detector (less than 100
$\mu$m). 
The solution to perform this measurement is to install two sets of 
roman pot detectors on each side of ATLAS located at about 240 m from the
interaction point, which can go close to the beam when the beam is stable.

\begin{figure}
\begin{center}
\begin{tabular}{cc}
\hspace{-1cm}
\epsfig{file=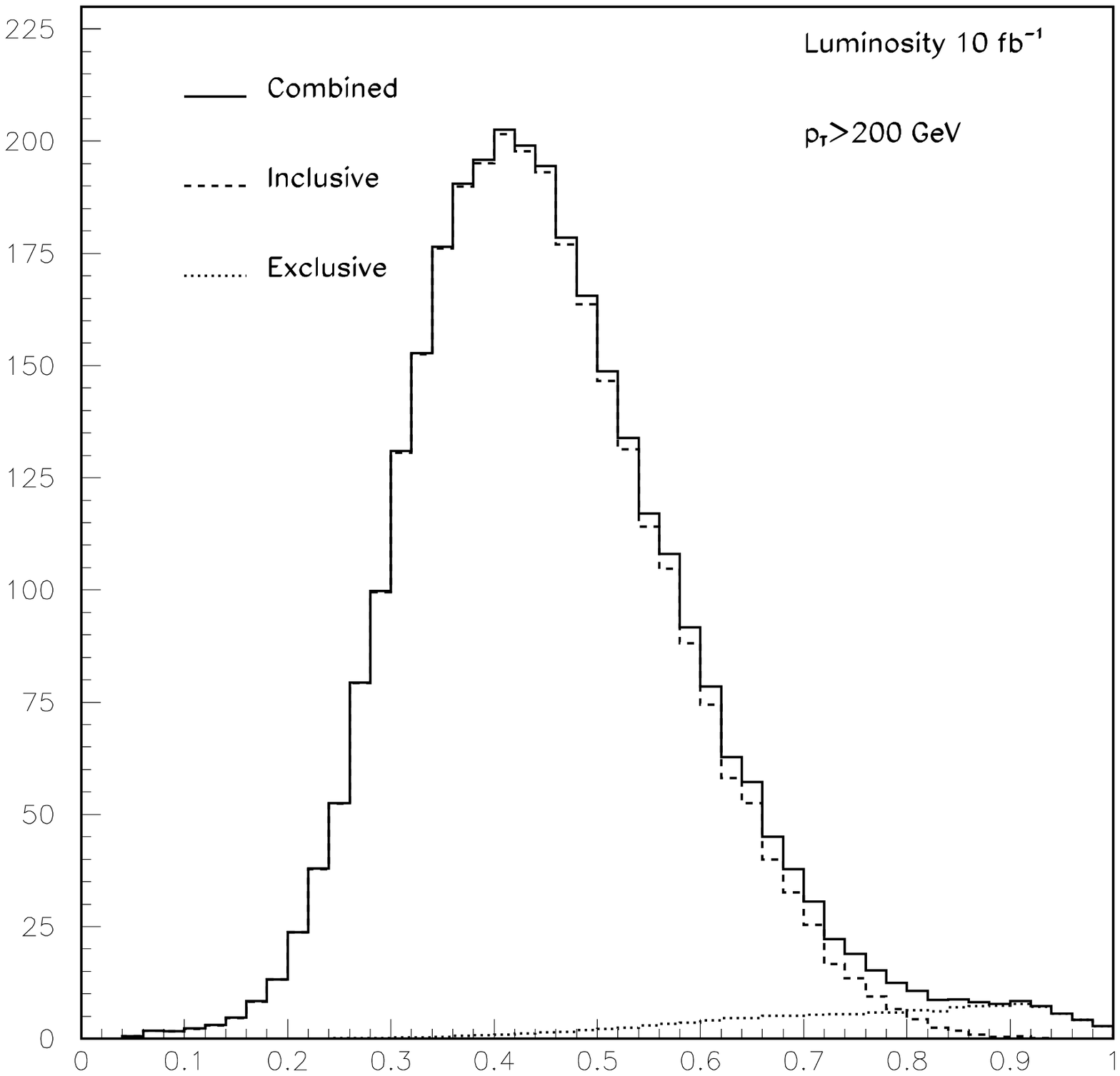,width=6.5cm} &
\epsfig{file=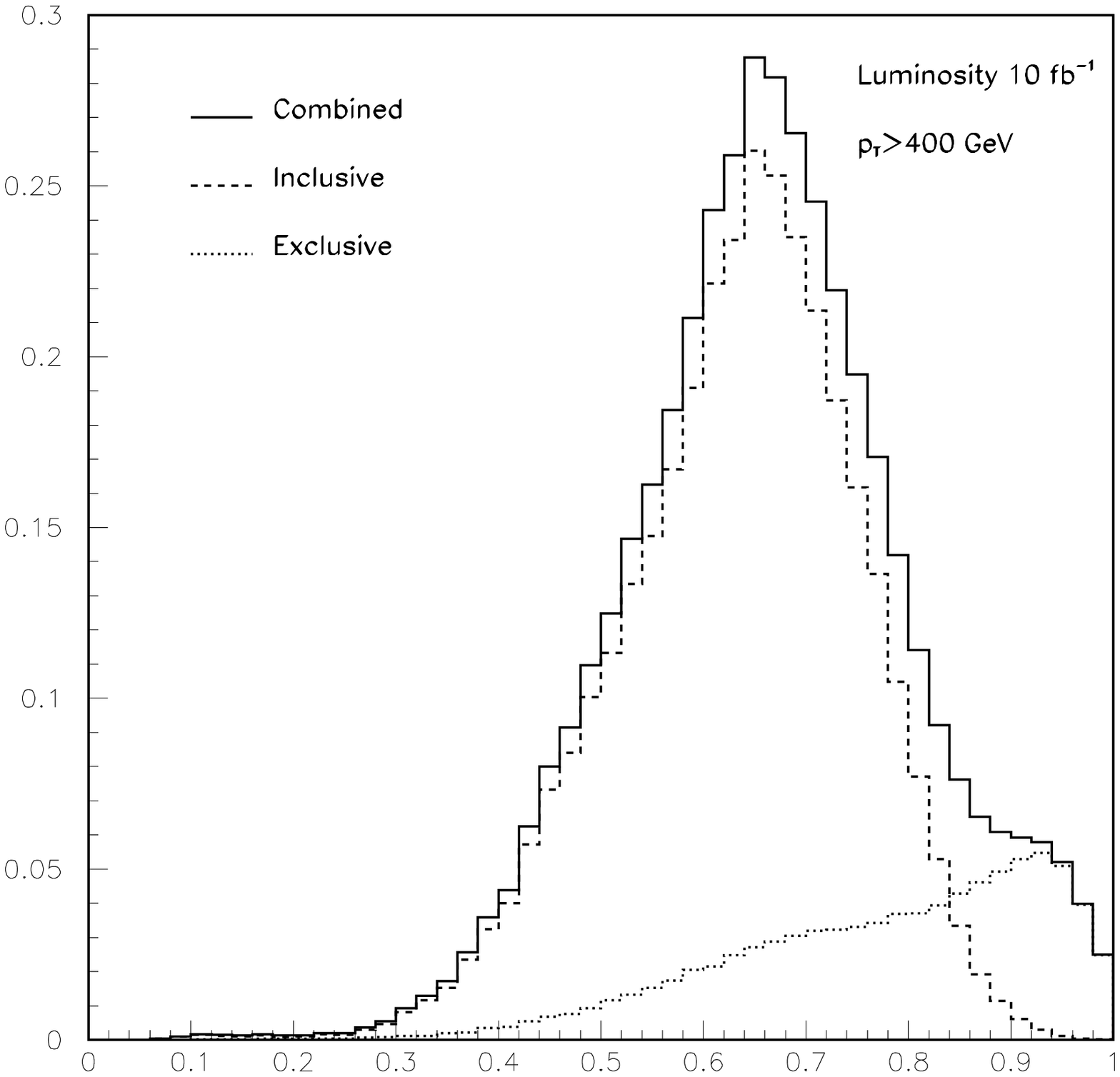,width=6.5cm} 
\end{tabular}
\caption{Dijet mass fraction at the LHC for jets $p_T>200\,\mathrm{GeV}$ and 
$p_T>400\,\mathrm{GeV}$ showing the contribution of both inclusive and exclusive
diffraction.}
\label{FigDMFexcLHC}
\end{center}
\end{figure} 

The second measurement to be performed at the LHC relies on the diffractive
dijet cross section.
The dijet mass fraction as a function of different jet $p_T$ is visible in 
Fig.~\ref{FigDMFexcLHC} after a simulation of the ATLAS/CMS detectors inclusing
inclusive diffraction and the exclusive one which we  discuss in the enxt
section. The
exclusive contribution manifests itself as an increase in the tail of the 
distribution which can be seen for $200\,\mathrm{GeV}$ 
jets (left) and $400\,\mathrm{GeV}$ jets (right) respectively~\cite{oldab},
as we will discuss in the following. 

\section{Exclusive diffractive events at the Tevatron and the LHC}

\subsection{Interest of exclusive events}
\begin{figure}
\begin{center}
\epsfig{file=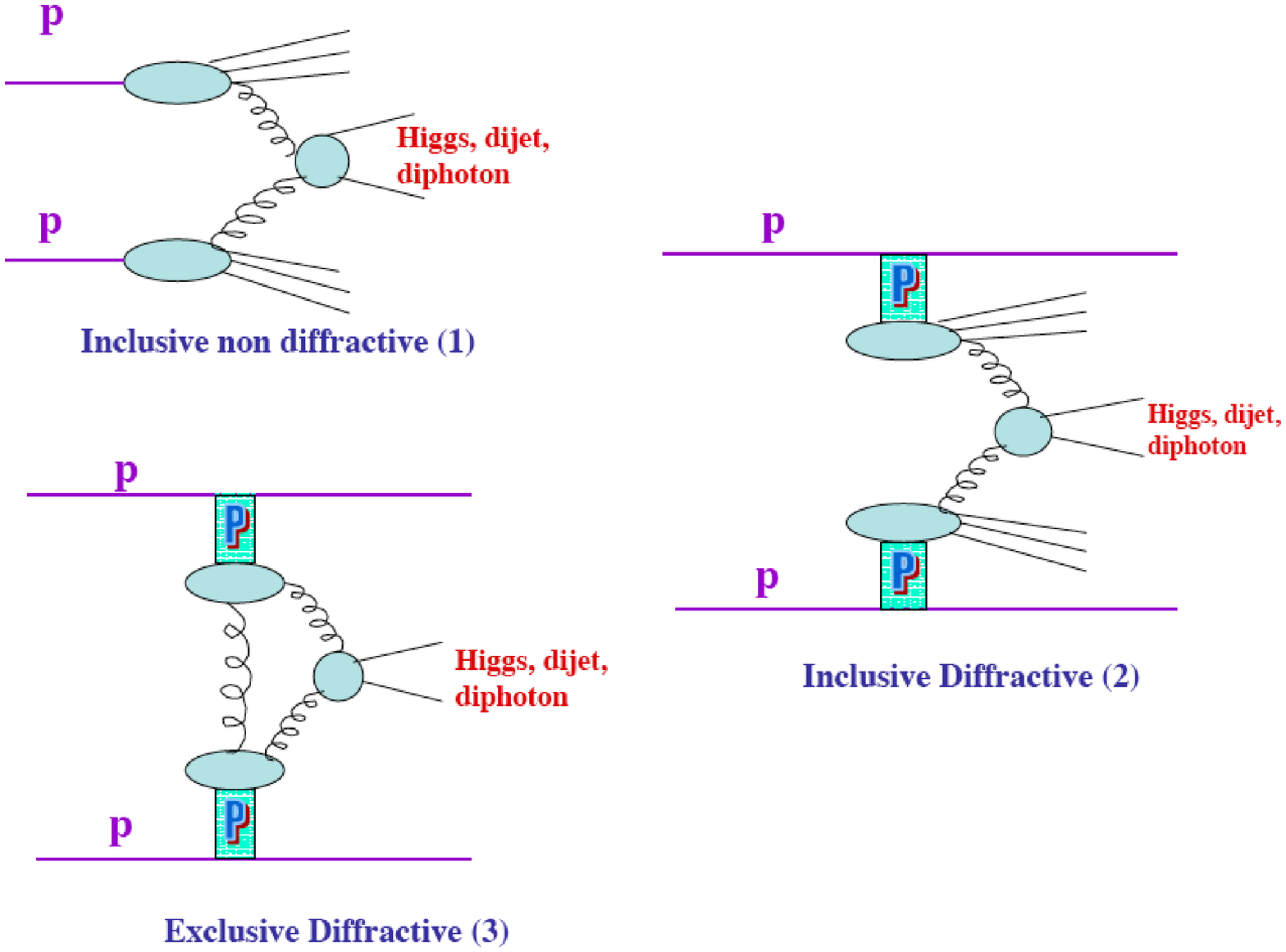,width=12cm}
\caption{Scheme of non diffractive, inclusive double pomeron exchange,
exclusive diffractive events at the Tevatron or the LHC.}
\label{fig7}
\end{center}
\end{figure}

A schematic view of non diffractive, inclusive double pomeron exchange,
exclusive diffractive events at the Tevatron or the LHC is displayed in
Fig.~\ref{fig7}.
The upper left plot (1) shows the ``standard" non diffractive events
where the Higgs boson, the dijet or diphotons are produced directly by a
coupling to the proton and shows proton remnants. The right plot (2) displays 
the standard diffractive double
pomeron exchange where the protons remain intact after interaction and the total
available energy is used to produce the heavy object (Higgs boson, dijets,
diphotons...) and the pomeron remnants. We have so far only discussed
this kind of events and their diffractive production using the
parton densities measured at HERA. There may be a third class of processes
displayed in the lower left figure (3), namely the exclusive diffractive
production. In this kind of events, the full energy is used to produce the heavy
object (Higgs boson, dijets, diphotons...) and no energy is lost in pomeron
remnants. 

There is an important consequence for the diffractive exclusive events: 
the mass of the
produced object can be computed using roman pot detectors and tagged
protons~\footnote{The formula is more complicated for low mass objects when the
proton mass cannot be neglected~\cite{chic}.}
\begin{eqnarray}
M = \sqrt{\xi_1 \xi_2 S}
\end{eqnarray} 
where $\sqrt{S}$ is the center-of-mass energy and $\xi$ is the fraction of the
proton momentum carried away by the Pomeron (called $\xpom$ at HERA).
The advantage of those processes is obvious: we can benefit from the
good forward detector resolution on $\xi$ to get a good mass resolution, and to
measure precisely 
the kinematical properties of the produced
object. It is thus important to know if this kind of events
exists or not. We  now describe in detail the search for exclusive events in the
different channels which is performed by the CDF and D0 collaborations at the Tevatron.
In the next section, we  also discuss the impact of the exclusive events on the
LHC physics potential.

\subsection{Search for exclusive events in $\chi_c$ production}
One way to look for exclusive events at the Tevatron is to search 
for the diffractive exclusive production of 
light particles like the $\chi$ mesons. This leads to high enough 
cross sections -- contrary to the diffractive exclusive production of heavy
mass objects such as Higgs bosons --- to check the dynamical 
mechanisms and the existence of exclusive events. 
The 
CDF collaboration~\cite{chic} put an upper limit for the $\chi$ production
cross section 
of $\sigma_{exc}(p\bar{p} \rightarrow p+J/\psi
 + \gamma+\bar{p}) \sim 49 \pm 18 (stat)
\pm 39 (sys)\  $ pb where the $\chi_c$ decays into $J/\Psi$ and $\gamma$, the
$J/\Psi$ decaying itself into two muons. The experimental signature is very
clean, two
muons and an isolated photon.
The cosmics contamination is difficult to compute and this is
why the CDF collaboration only quotes an upper limit.
To know if the production is expected to be a hint of
exclusive events, it is important to study the tail of inclusive diffraction 
which is a
direct contamination of the exclusive signal. The tail of inclusive diffraction
corresponds to events which show very little energy in the forward direction, or in
other words where the pomeron remants carry very little energy. 
In Ref.~\cite{chic}, we found that the contamination of
inclusive events into the signal region depends strongly on the
assumptions on the gluon distribution in the pomeron at high $\beta$, which
is very badly known. 
Therefore, this channel is unfortunately not
conclusive concerning the existence of exclusive events.

\subsection{Search for exclusive events in the diphoton channel}
The CDF collaboration also looked for the exclusive production of dilepton and
diphoton~\cite{cdfgamma}. 
Contrary to diphotons, dileptons cannot be produced exclusively via pomeron exchanges since
$g g \rightarrow \gamma \gamma$ is possible, but $g g \rightarrow l^+ l^-$ 
directly is impossible. Dileptons are produced via QED processes, and
the CDF dilepton measurement is $\sigma = 1.6
^{+0.5}_{-0.3} (stat) \pm 0.3 (syst)$ pb which is found to be in good agreement
with QED predictions. 3 exclusive diphoton events have been observed by the CDF
collaboration leading to a cross section of
$\sigma = 0.14
^{+0.14}_{-0.04} (stat) \pm 0.03 (syst)$ pb compatible with the expectations
for exclusive diphoton production at the Tevatron. Unfortunately, the number of events
is very small and the conclusion concerning the
existence of exclusive events is uncertain. An update by the CDF collaboration with higher luminosity
is however expected very soon. This channel will be however very
important at the LHC where the expected exclusive cross section is much higher.

\subsection{Search for exclusive events using the dijet mass fraction at the Tevatron}

The CDF collaboration measured the so-called dijet mass fraction
in dijet events --- the ratio of the mass carried by the two jets produced in the event 
divided by the
total diffractive mass --- when the antiproton is tagged in the roman pot
detectors and when there is a rapidity gap on the proton side to ensure that the
event corresponds to a double pomeron exchange~\cite{cdfrjj}. 
We compare this measurement
to the expectation obtained from the pomeron structure in quarks and gluons as
measured at HERA~\cite{us,lolopic}.
The factorisation breaking between
HERA and the Tevatron is assumed to be constant and to
come only through the gap survival probability
(0.1 at the Tevatron).

\begin{figure}
\begin{tabular}{cc}
\hspace{-1cm}
\epsfig{figure=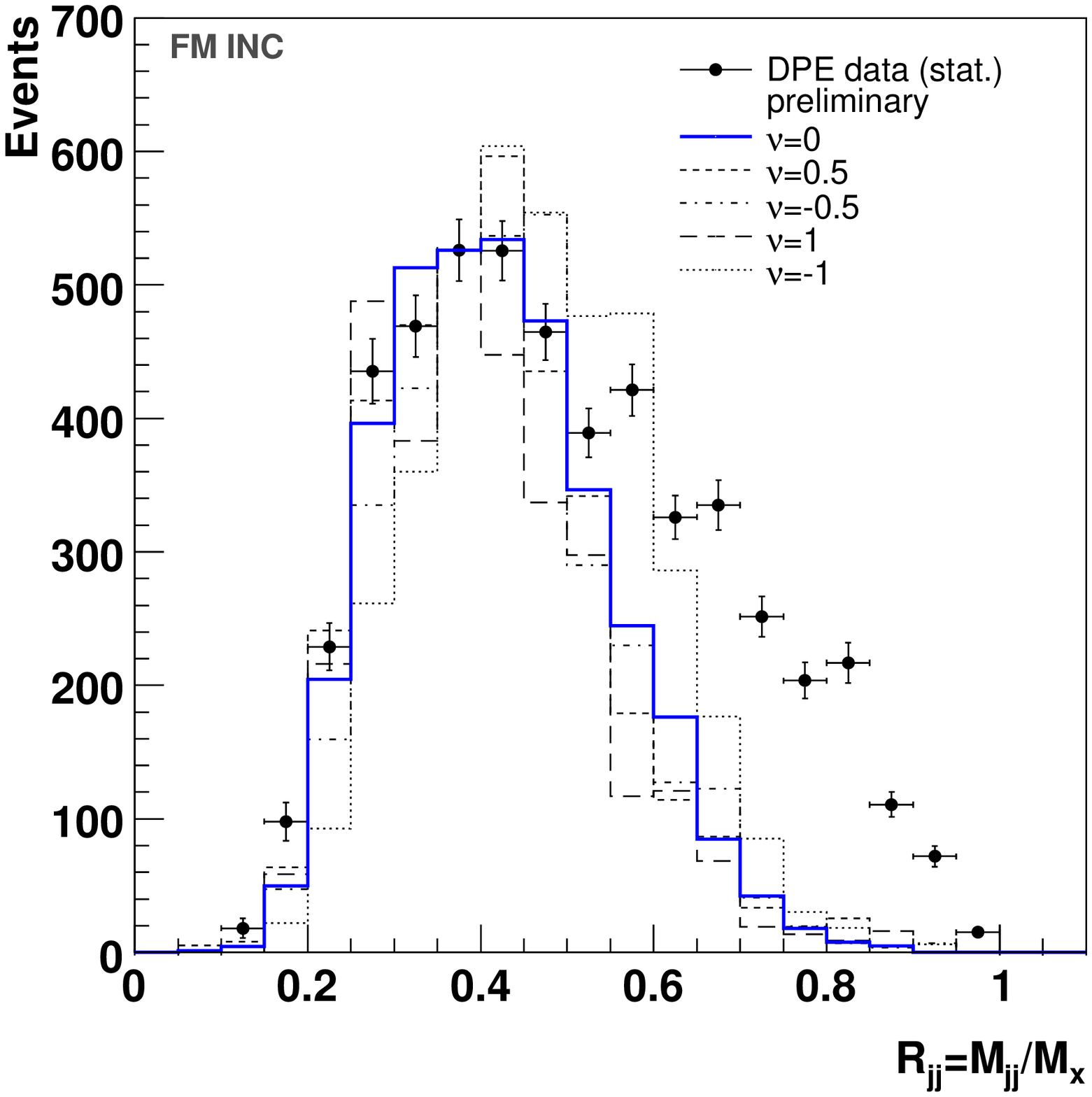,height=2.5in}  &
\epsfig{figure=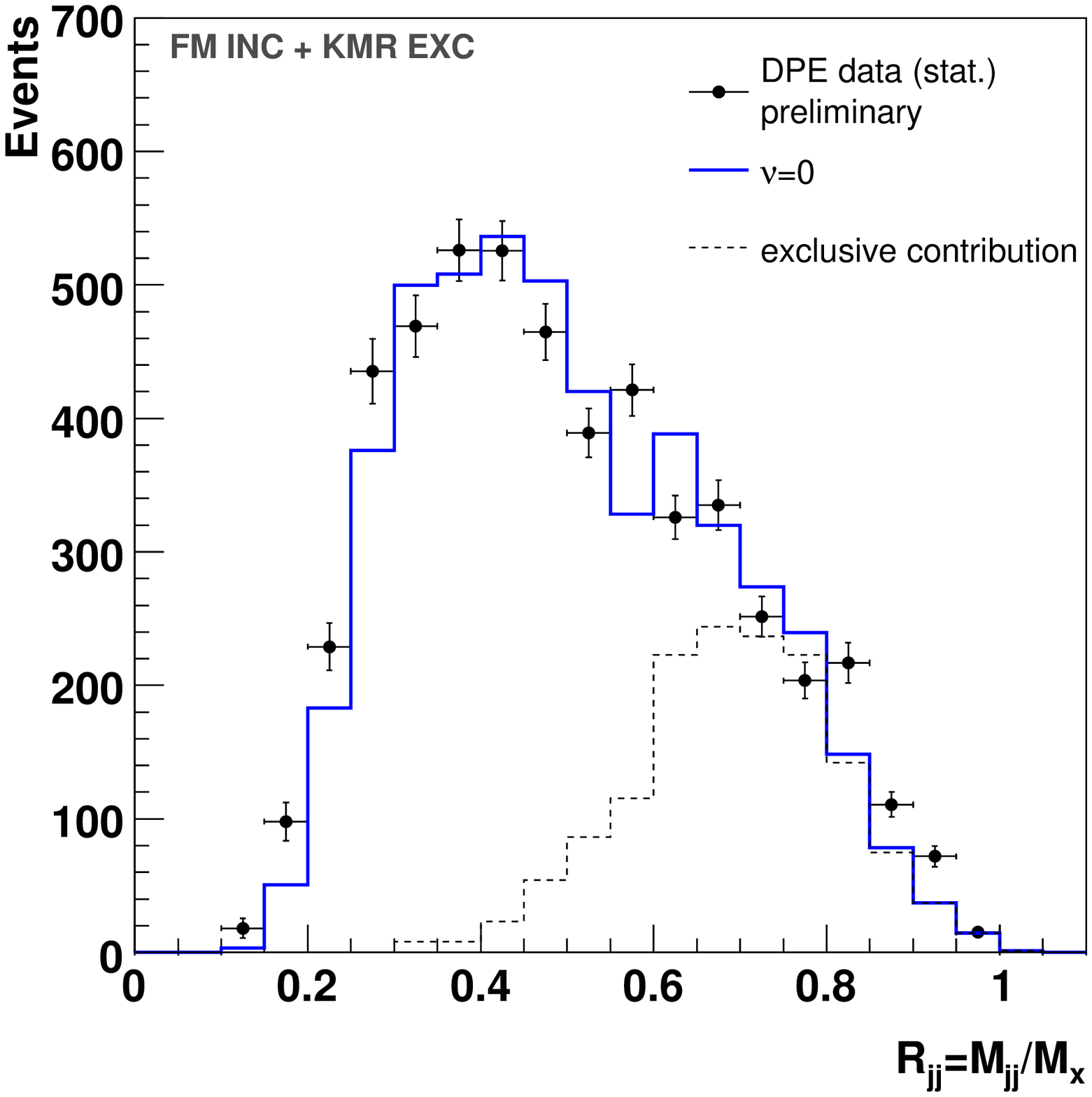,height=2.5in} \\
\end{tabular}
\caption{Left: Dijet mass fraction measured by the CDF collaboration compared to the
prediction from inclusive diffraction based on the parton densities
in the pomeron measured at HERA. The gluon
density in the pomeron at high $\beta$ was modified by varying the parameter
$\nu$.
Right: Dijet mass fraction measured by the CDF collaboration compared to the
prediction adding the contributions from inclusive and exclusive diffraction.}
\label{compare2}
\end{figure}

The comparison between the CDF data for a jet $p_T$ cut of 10 GeV as an
example and the predictions from inclusive diffraction is given in 
Fig.~\ref{compare2}, left, together with
the effects of changing the gluon density at high $\beta$ by
changing the value of the $\nu$ parameter introduced to vary
the gluon density in the pomeron at high $\beta$. Namely, to study the 
uncertainty on the gluon density at high $\beta$, 
we multiply the gluon distribution by the
factor $(1 - \beta)^{\nu}$. The $\nu$ parameter
varies between -1 and 1 (for $\nu=-$1 (resp. $+$1), the gluon density in the pomeron is
enhanced (resp. damped) at high $\beta$). 
QCD fits to the H1 data lead to 
an uncertainty on the $\nu$ parameter of 0.5~\cite{us}. Inclusive
diffraction alone is not able to describe the CDF data at high dijet mass fraction,
where exclusive events are expected to appear~\cite{oldab}. The conclusion
remains unchanged when jets with $p_T>25$ GeV are considered~\cite{oldab}.

Adding exclusive events to the distribution of the dijet mass fraction leads to
a good description of data~\cite{oldab} as shown in Fig.~\ref{compare2}, right. This 
does not prove that exclusive events exist but shows that some 
additional component with respect to
inclusive diffraction compatible with exclusive events is needed to explain CDF data. 
To be sure of the existence
of exclusive events, the observation will have to be done in different channels
and the different cross sections to be compared with theoretical expectations.
 
\begin{figure}
\centerline{\includegraphics[width=0.7\columnwidth]{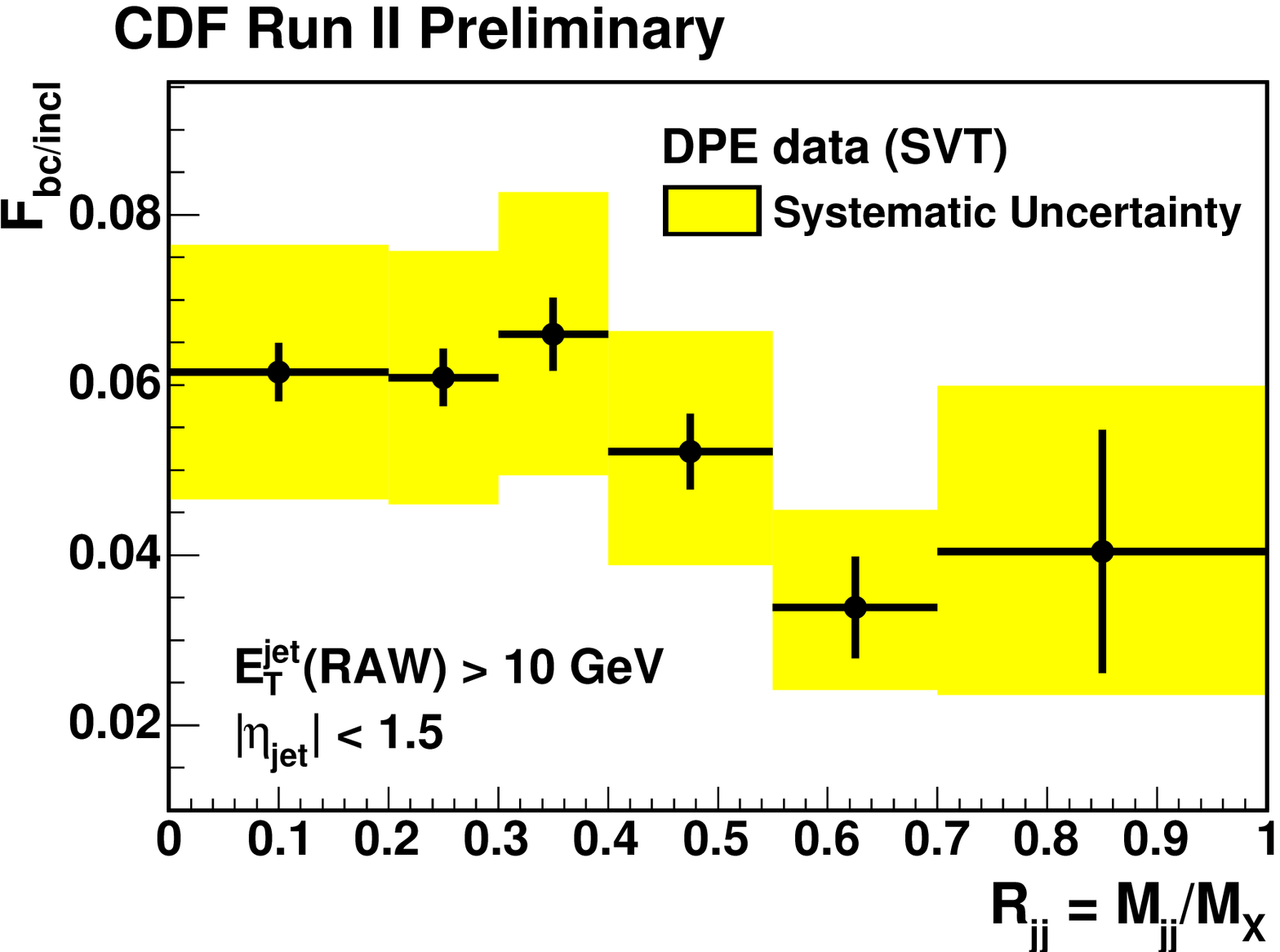}}
\caption{Ratio of $b/c$ jets to inclusive jets in double pomeron exchange events
as a function of the dijet mass fraction.}
\label{bjet}
\end{figure}

Another interesting observable in the dijet channel is to look at the rate
of $b$ jets as a function of the dijet mass fraction. In exclusive events, the
$b$ jets are suppressed because of the $J_Z=0$ selection rule~\cite{ushiggs},
and as
expected, the fraction of $b$ jets in the diffractive dijet sample
diminishes as a function of the dijet mass fraction (see Fig.~\ref{bjet} from
Ref.~\cite{cdfrjj}). 

Another method to search for exclusive events is to study the
correlation between the gap size measured in both $p$ and $\bar{p}$ directions
and the value of $\log 1/\xi$ measured using roman pot detectors~\cite{tev4lhc}. 
The gap size between the
pomeron remnant and the protons detected in roman pot detector 
is of the order of 
$\log 1/\xi$ for usual diffractive events while
exclusive events show a larger rapidity gap since the gap
occurs between the jets and the proton detected in roman
pot detectors (in other words, there is no pomeron remnant). 

As we mentioned in a previous section, we also compared the CDF measurements 
with the expectations from the soft colour interaction (SCI) model. The SCI
model is the only model which explains the different normalisation between HERA
and Tevatron diffractive data without the need of the survival probability
(since it assumes that diffraction is purely due to a soft color rearrangement
in the final state). The CDF dijet data were compared with the predictions from
the SCI models~\cite{oldab}, and the proportion of needed exclusive events to
describe the dijet mass fraction is found to be smaller. However, the SCI model
can not describe properly the jet rapidity distributions which are predicted to
be asymmetric around 0 by the SCI model (the CDF requires one tagged proton on
one side and a rapidity gap on the other side, and within the SCI model, it is
difficult to obtain an intact proton in the final state). For these reasons, the
SCI model is disfavoured but it would be probably useful to revisit the ideas
and the implementation of such a model in Pythia.

\subsection{Search for exclusive events at the LHC}
The search for exclusive events at the LHC can be performed in the same channels
as the ones used at the Tevatron. Additional possibilities
benefitting from the higher luminosity and cross sections at the LHC appear. 
One of the cleanest ways
to show the existence of exclusive events would be to measure the dilepton and
diphoton cross section ratios as a function of the 
dilepton/diphoton mass~\cite{ushiggs,tev4lhc}. If
exclusive events exist, this distribution should show a bump towards high values
of the dilepton/diphoton mass since it is possible to produce exclusively
diphotons but not dileptons at leading order. 

The search for exclusive events at the LHC will also require a precise analysis
and measurement of inclusive diffractive cross sections and in particular the
tails at high $\beta$ since it is a direct
background to exclusive event production. It will be also useful to measure directly
the exclusive jet production cross section as a function of jet $p_T$ as an example
and compare the evolution to the models. This will allow to know precisely the
background especially to Higgs searches which we  discuss in the following.

\subsection{Exclusive Higgs production at the LHC}

One special interest of
diffractive events at the LHC is related to the existence of exclusive events
and the search for Higgs bosons at low mass in the diffractive mode.
So far, two projects are being discussed at the LHC: the installation of 
forward detectors at 220 and 420 m in ATLAS and CMS~\cite{afp} which we 
describe briefly at the end of this review.

\begin{figure}
\begin{center}
\centerline{\psfig{figure=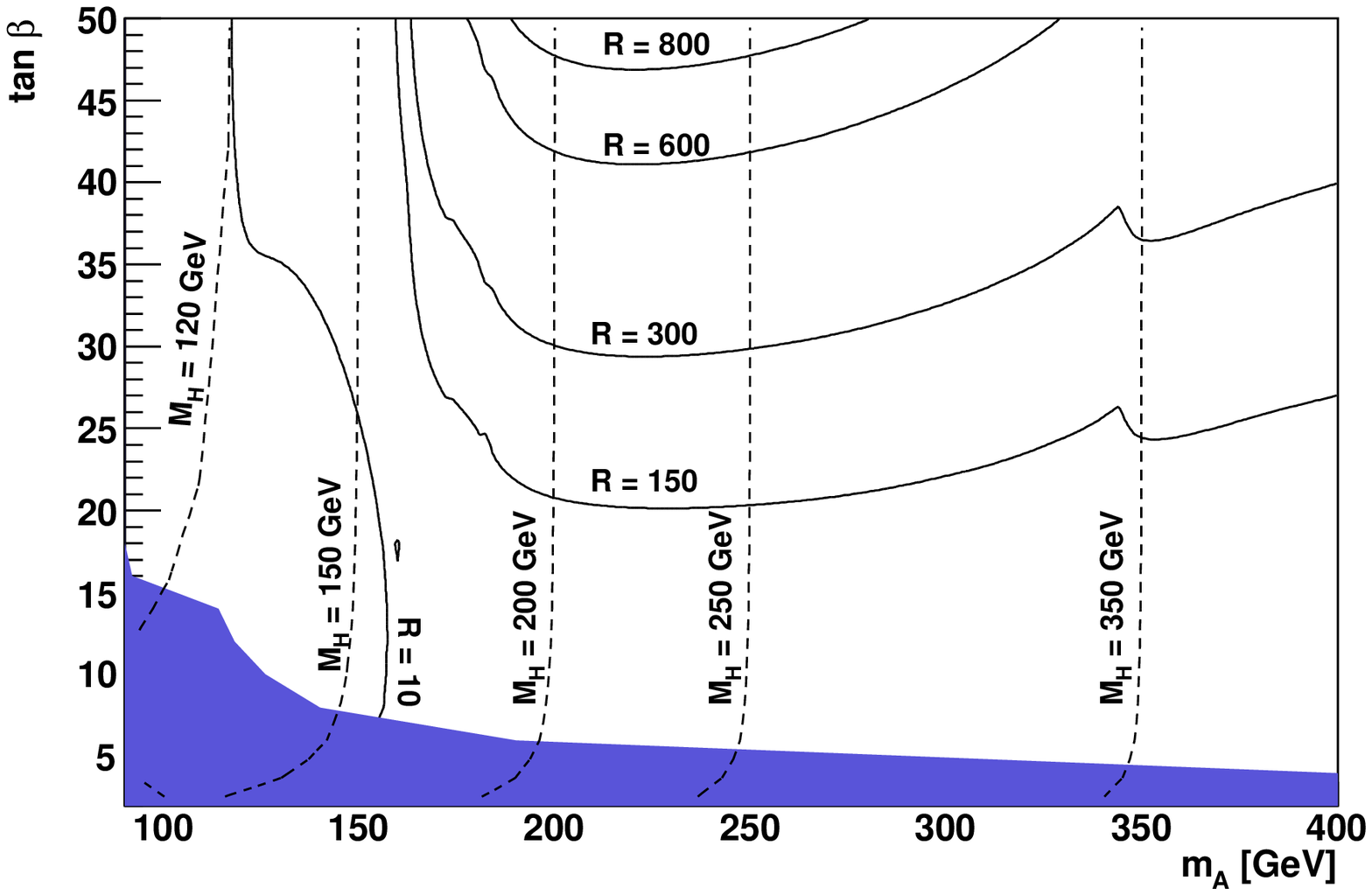,height=3.1in}}
\caption{Ratio $R$ at generator level between the number of diffractive Higgs
events in MSSM to SM in the ($\tan \beta$, $M_A$) plane for heavy
CP-even Higgs bosons. The lines of
constant Higgs boson mass are also indicated in dashed line.}
\label{marek}
\end{center}
\end{figure}

\begin{figure}
\begin{center}
\begin{tabular}{cc}
\hspace{-1cm}
\epsfig{figure=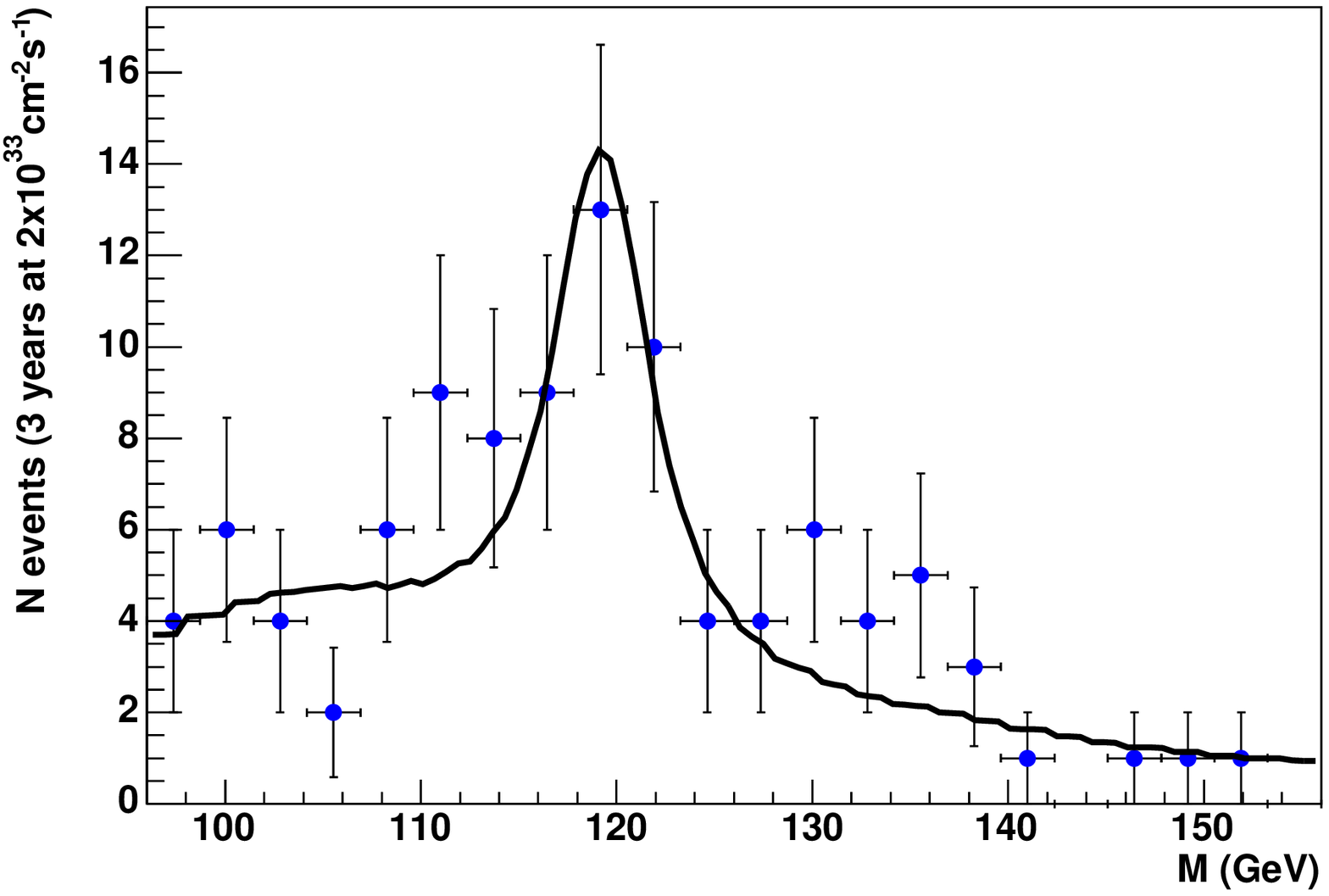,height=1.8in} &
\epsfig{figure=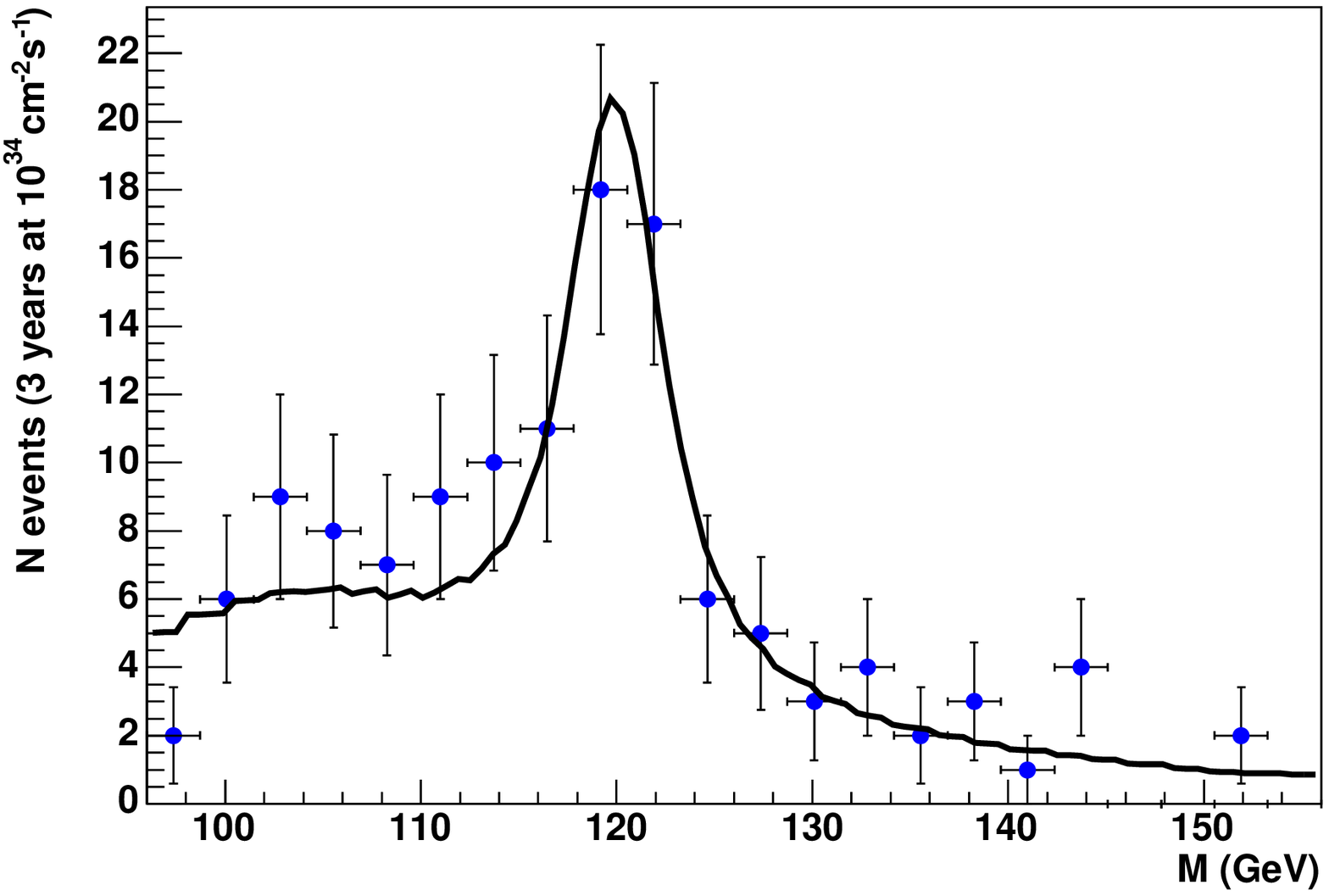,height=1.8in} \\
\end{tabular}
\caption{Higgs signal and background obtained for MSSM Higgs production
for neutral light CP-even Higgs bosons. 
The signal significance is larger than 3.5 $\sigma$ for 60 fb$^{-1}$
(left plot) and larger than 5 $\sigma$ after three years of data taking at high
luminosity at the LHC and using timing detectors with a resolution of 2 ps
(right plot).}
\label{signif}
\end{center}
\end{figure}

Many studies (including pile up effects and all background sources
for the most recent ones) 
were performed
recently~\cite{ushiggs,lavignac,higgsnew} to study in detail the signal 
over background for
MSSM Higgs production in particular. The ratio $R$ of the number of diffractive Higgs bosons in
MSSM to SM is given in Fig.~\ref{marek}. Typically if $R>$10, the number of
events is high enough to allow a discovery with 30 fb$^{-1}$ per experiment
using the diffractive mode.
We notice that almost the full plane
in ($\tan \beta$, $M_A$) can be covered even at low luminosity.
In Fig.~\ref{signif}, we give the number of  background and MSSM Higgs signal events
for a Higgs mass of 120 GeV for $\tan \beta \sim$40.
The signal significance is larger than 3.5 $\sigma$ for 60 fb$^{-1}$
(see Fig.~\ref{signif} left) and larger than 5 $\sigma$ after three years of data taking at high
luminosity at the LHC and using timing detectors with a good timing resolution
(see Fig.~\ref{signif} right).

In some scenario such as NMSSM where the Higgs boson decays in $h \rightarrow
aa \rightarrow \tau \tau \tau \tau$ where $a$ is the lighter of the two
pseudo-scalar Higgs bosons, the discovery might come only from
exclusive diffractive Higgs production~\cite{higgsnew} ($m_a<2 m_b$ is natural in NMSSM with $m_a>2
m_{\tau}$ somewhat preferred).

\subsection{Photon induced processes at the LHC}

In this section, we discuss  particularly
a new possible test of the Standard Model (SM) predictions
using photon induced processes at the LHC, and especially
$WW$ production~\cite{jochen,olda}. The cross sections of these
processes are computed with high precision using Quantum Electrodynamics
(QED) calculations, and an experimental observation leading to differences with
expectations would be a signal due to beyond standard model effects. The
experimental signature of such processes is the decay products of the $W$ in the
main central detectors from the ATLAS and CMS experiments and the presence of
two intact scattered protons in the final state.

The experimental signature of diboson events is very clear. Depending on the decay
of the $W$ there is zero, one, or two leptons in the final state. When both $W$s decay
purely hadronically 
four jets are produced in the final state. This topology can be easily  mimicked 
in the high luminosity environment
with pile-up interactions and also suffer from a high QCD
background. Therefore we always require
that at least one of the $W$ decays leptonically.
In addition, the interpretation of the signal is simple contrary to e.g. $e^{+}e^{-}\rightarrow WW$ production
at LEP where such production could be due to $\gamma$ or $Z$ exchange and one could 
not clearly separate the $\wwgamma$ and $WWZ$ couplings. In our case, only the $\gamma$
exchange is possible since there is no $Z\gamma\gamma$ vertex in the SM.
 
In summary, we require the following constraints at particle level to select $WW$ events: 
\begin{itemize}
   \item both protons are tagged in the forward detectors in the acceptance $0.0015<\xi<0.15$
   \item at least one electron or muon is detected with $p_T>30\units{GeV/c}$ and 
    $|\eta|<2.5$ in the main detector
\end{itemize}

The main source of background is the $W$ pair production
in Double Pomeron Exchange (DPE), i.e. $pp\rightarrow p+WW+Y+p$ through 
$\pomeron\pomeron\rightarrow WW+Y$
where $Y$ denotes the pomeron remnant system.
To remove most of the DPE background, it is possible to cut on the $\xi$ of the protons
measured in the proton taggers.
Indeed, two-photon
events populate the low $\xi$ phase space whereas DPE events show a flat $\xi$ distribution. 
The $pp\rightarrow pWWp$ cross section can be measured precisely with a \lumi=$1\,\invfb$ 
with a statistical
significance higher than 20$\sigma$ depending on the active $\xi$ range. Using
the full $\xi$ acceptance $0.0015<\xi<0.15$, one expects about 30
tagged $WW$ events. As $\xi_{max}$, the upper cut on $\xi$, decreases, one obtains a cleaner
signal, but the number of observed events drops.
Already with a low integrated luminosity of \lumi=$200\,\invpb$ it is 
possible to observe 5.6 $W$ pair two-photon events for a background of DPE lower than 0.4,
leading to a signal 
above 8 $\sigma$ for $WW$ production via photon induced processes.

New physics with a characteristic scale (i.e. the typical mass of new particles) 
well above  what can be probed
experimentally at the LHC can manifest itself as a  modification of gauge
boson couplings due to the exchange of new heavy particles. The conventional
way to investigate the sensitivity to the potential new physics is
to introduce an effective
Lagrangian with additional higher
dimensional terms parametrized with anomalous parameters.  
We consider the modification of the $\wwgamma$ triple gauge boson vertex with additional terms conserving $C-$ and $P-$
parity separately, that are parametrized with two anomalous parameters $\dkap$, $\lam$. The effective Lagrangian reads
\begin{eqnarray}
   \mathcal{L}/ig_{WW\gamma}&=&(W^{\dagger}_{\mu\nu}W^{\mu}A^{\nu}-W_{\mu\nu}W^{\dagger\mu}A^{\nu})
   +(1+{\Delta\kappa^{\gamma}})W_{\mu}^{\dagger}W_{\nu}A^{\mu\nu}+ \nonumber \\
   &~&\frac{{\lambda^{\gamma}}}{M_W^2}W^{\dagger}_{\rho\mu}
   W^{\mu}_{\phantom{\mu}\nu}A^{\nu\rho},
\end{eqnarray}
where $g_{WW\gamma}=-e$ is the $\wwgamma$ coupling in the SM and the double-indexed terms are
$ V_{\mu\nu}\equiv\partial_{\mu} V_{\nu}-\partial_{\nu}V_{\mu}\nonumber$, for $V^{\mu}=W^{\mu},A^{\mu}$. In the 
SM, the anomalous parameters are $\dkap=\lam=0$. The sensitivity to anomalous coupling can 
be derived by counting the number
of observed events and comparing it with the SM expectation.
In order to obtain the best $S/\sqrt{B}$ ratio, the
$\xi$ acceptance was further optimized for the $\lam$ parameter.
The event is accepted if $\xi_i>0.05$.  
In case of $\dkap$, the full acceptance of the forward detectors is used since 
the difference between the enhanced and SM cross section
is almost flat around relevant values of the coupling $|\dkap|$. For 30 fb$^{-1}$,
the reach on  $\dkap$ and $\lam$ is respectively 0.043 and 0.034, improving the direct limits
from hadronic colliders by factors of 12 and 4 respectively (with respect to the LEP indirect 
limits, the improvement is only about a factor 2).
Uisng a luminosity of 200 fb$^{-1}$, present sensitivities
coming from the hadronic colliders
can be improved by about a factor 30, while the LEP sensitivity can be improved
by a factor 5. 

\begin{figure}
\begin{center}   
\includegraphics[width=\picwidth]{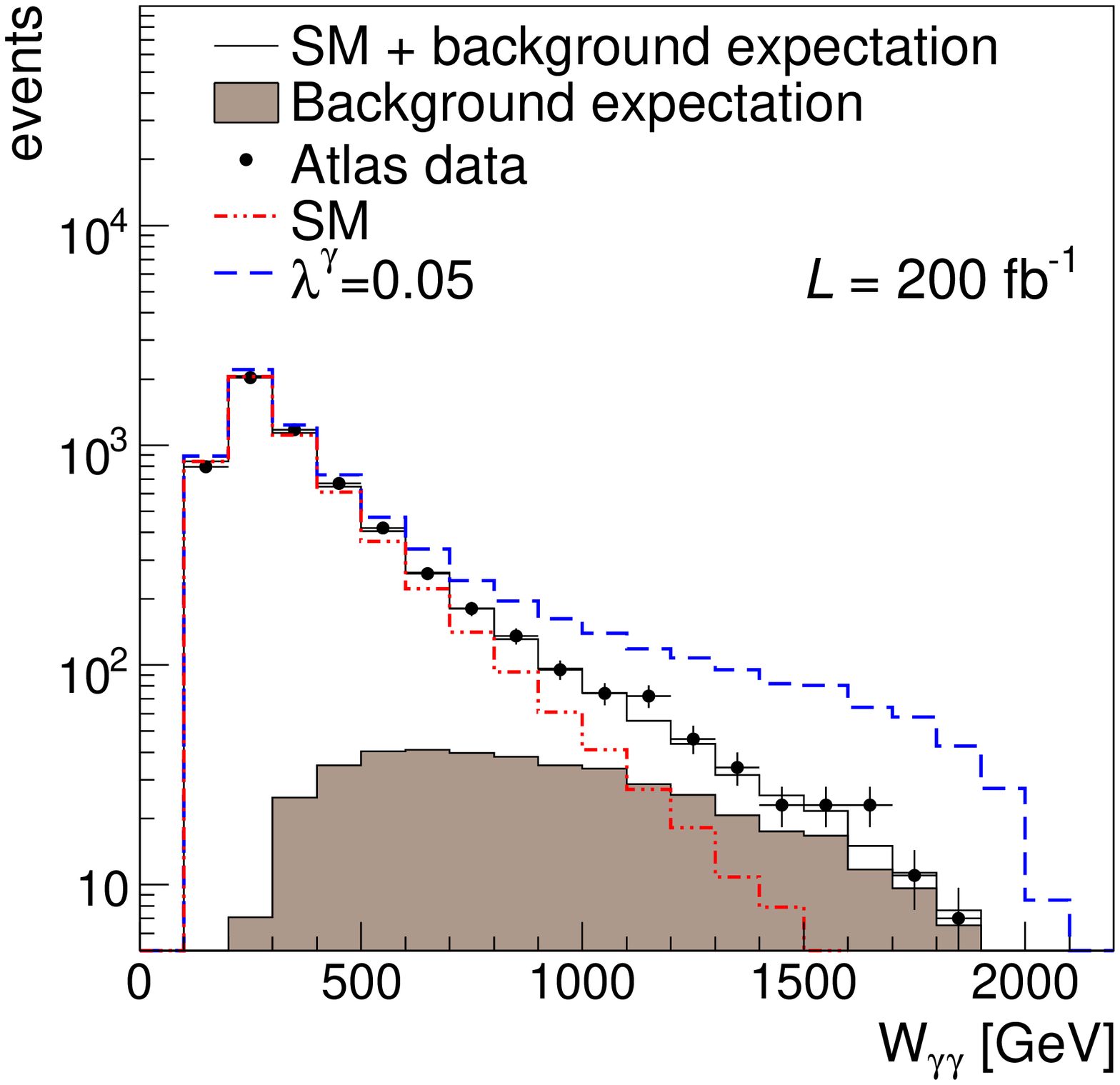}
\caption{Distributions of the $\gamma\gamma$ photon invariant mass $W_{\gamma\gamma}$ 
measured with the forward detectors using $W_{\gamma\gamma}=\sqrt{s\xi_1\xi_2}$. The 
effect of the $\lam$ anomalous parameter appears at high $\gamma\gamma$ 
invariant mass (dashed line).  The SM background is indicated in dot-dashed line, 
the DPE background as a shaded area and their combination
in full line. The black points show the ATLAS data smeared according to a Poisson distribution. }
\label{fig:Wspectrum200}
\end{center}
\end{figure}

It is worth noticing that many observed events are expected in the region $W_{\gamma\gamma}>1$ TeV where 
beyond standard model effects, such as SUSY, new strong dynamics at the TeV
scale, anomalous coupling, etc., are expected (see Fig.~\ref{fig:Wspectrum200}). 
It is expected that the LHC
experiments will collect 400 such events predicted by QED with $W>$1 TeV for a luminosity of 200
fb$^{-1}$ which will allow to probe further the SM expectations. In the same way
that we studied the $WW\gamma$ coupling, it is also possible to study the
$ZZ\gamma$ one. The SM prediction for the $ZZ\gamma$ coupling is 0, and any
observation of this process is directly sensitive to anomalous coupling (the main
SM production of exclusive $ZZ$ event will be due to exclusive Higgs boson
production decaying into two $Z$ bosons).

Many other studies can be performed using $\gamma$ induced processes.
The $WW$ cross section measurements are also sensitive to anomalous quartic
couplings~\cite{QuarticCoupling}, and recent studies showed that the sensitivity on quartic
couoling is 10000 times
better than at LEP with only a luminosity of 10 fb$^{-1}$. In addition, it is
possible to produce new physics beyond the Standard Model. 
Two photon production of SUSY leptons as an example has been 
investigated and the cross section for 
$\gamma \gamma \rightarrow \tilde{l}^+ \tilde{l}^-$ can be larger than 1~fb. 

\subsection{Projects to install forward detectors at the LHC}
In this section, we describe briefly the project to install forward detectors at
220 and 420 m in the ATLAS and CMS collaborations~\cite{afp} allowing to accomplish the physics
program we just described, namely a better understanding of diffractive events,
the search for exclusive events and the Higgs boson in this mode, and the search for
photon anomalous coupling.
To obtain a good acceptance in mass (above 50\% for masses between
115 and 650 GeV), both detectors at 220 and 420 m are needed, since many events
even at low masses show an intact proton in the 220 m detector on one side and
in the 420 m one on the other side. Two kinds of detectors namely the 3D Si and
the timing detectors, will be hosted in movable beam pipes located at 220 and
420 m.

The idea of movable beam pipes is quite simple and was used already in the ZEUS
collaboration at HERA: when the beam is injected, the movable beam pipe is in
its ``home" position so that the detectors can be far away from the beam and its
halo, and when the beam is stable, the movable beam pipe moves so that the
detectors go closer to the beam. A typical movement is of the order of 2 cm.
Beam Pipe Monitors will be located in the fixed and movable beam pipe areas to
measure how close the detectors are located with respect to the beam. The needed
precision is of the order of 10-15 $\mu$m. 4 horizontal pockets containing the 3D Silicon
and timing detectors will be located within the movable beam pipe structure. 
The protons are emitted diffractively in
the horizontal plane and this is why only horizontal detectors and pockets are needed.

It is assumed that it will be possible to go as close to the beam as 15 $\sigma$
at 220 and 420 m. To get a full coverage for the diffracted protons, 
detectors of 2 cm $\times$ 2 cm and of 0.6 cm $\times$ 2 cm are needed at 220
and 420 m respectively. The position of the protons have to be measured with a
precision of 10-15 $\mu$m in $x$-direction in a radiation environment and this is why the
solution of 3D Silicon pixels has been chosen. The size of the pixels is of 50
$\mu$m
$\times$ 400 $\mu$m and a supermodule shows an active area of 7.2 mm $\times$ 8 mm.
There will be 10 layers of such supermodules staggered in $x$ and $y$ directions
perpendicular to the beam. This will allow to obtain the needed resolution.
To achieve the full coverage, 3 and 6 supermodules per layer are needed at 420 and 220 m
respectively. The alignment of these detectors will be achieved using exclusive
dimuon events --- this can be performed at a store-by-store basis at 420 m and
only on a  week basis at 220 m since the 220 m detectors are sensitive to higher
dimuon mass events --- and also possibly using elastics events which would imply
the installation of vertical detectors as well.

The 3D Si detector at 220 m can also provide a L1 trigger allowing to cut 
on the proton momentum loss (the 420 m detector is too far away to make it to
the L1 trigger). Two kinds of trigger are considered. The first one triggers on
events when both protons are detected at 220 m. The second (more difficult)
triggers on events when only one proton is present at 220 m. In that case, the
idea is to cut on the acceptance at 220 m corresponding to the possibility of a
tag at 420 m. A typical jet trigger will be: two jets with a transverse momentum
above 40 GeV, one proton at 220 m with a momentum loss smaller than 0.05
compatible with the presence of a proton at 420 m on the other side, and the
exclusiveness of the process (most of the energy is carried by the two hard
jets). With these cuts, the L1 rate is expected to be smaller than 1 kHz for a
luminosity smaller than 2.10$^{33}$ cm$^{-2}$s$^{-1}$. The expected output at L2
is assumed to be only a few Hz only since the timing and the 420 m detector
informations will be available at this stage. 

The other kind of detector which is crucial for this project is the timing
detectors. At the LHC, up to 35 interactions occur at the same bunch crossing
and we need to know if the observed protons in the final state originate from the
main interaction producing for instance the Higgs boson or the $WW$ event, or
from a secondary one which is not related to the hard interaction. To achieve
this goal, it is needed to know if the protons are coming from the main vertex
of the event, and for this sake, to measure precisely the time of flight of the
protons with a precision of the order of 10-15 ps or better. 
Two kinds of detectors have been proposed. The GASTOF measures
the Cerenkov light emitted by the protons and collected in a multichannel
photomultiplier. This detector has a very good intrinsic resolution measured in
beam tests of about 10-15 ps, but present the inconvenient of showing no lateral
space resolution which is needed in the case of multiple protons are produced in
one bunch crossing. The other device, QUARTIC, uses quartz radiator bars to emit
the photons and leads to a resolution per detector of 30-40 ps. Several
detectors are thus needed to achieve the wanted resolution. The advantage of
such detectors is that it can have a couple of mm space resolution and the
inconvenient is the smaller number of photoelectrons produced. The idea would be
thus to combine the advantages of the GASTOF and QUARTIC detectors which is
under study now in a world-wide collaboration regrouping the institutes of
Louvain, Chicago. Fermilab, Argonne National Lab, Brookhaven National Lab, Stony
Brook, Alberta, Texas Arlington, Saclay and Orsay. This kind of detectors is
also specially interesting for medical applications since it would allow to
improve the present resolution of the PET imaging detectors by one order of
magnitude.

\clearpage

%\input{exclusive}

%%%%%%%%%%%%%%%%%%%%%%%%%%%%%%%

\section{Conclusion}

The aim of this review is to describe the present vision we have of
the proton structure at high energy from the HERA and Tevatron data
and to discuss the potential improvements brought by the LHC.  

Parton momentum density distributions are important ingredients in the
calculation of high energy hadron-hadron and lepton-hadron scattering
cross sections. In these calculations the cross sections are written
as a convolution of the parton densities and the elementary cross
sections for parton-parton or lepton-parton scattering. Whereas the
latter can be perturbatively calculated in the framework of the
Standard Model, the parton densities are non-perturbative quantities
and are obtained from DGLAP based fits to measured cross sections at
various experiments. 

After showing the present status of PDF determination at
the Tevatron, we discussed the possible LHC measurements at LHC that
will increase our knowledge of the PDFs, as well as the dependence of
the LHC measurements and discovery potential on the present PDF
uncertainties. We stressed that the current PDF uncertainty on some LHC
observables is underestimated as a consequence of hidden
assumptions in current PDF sets. The usual sea quark symmetry
assumptions, relating the $\bar{u}$, $\bar{d}$ and $\bar{s}$
densities, are prominent examples that affect the study of electroweak
boson production. We review a number of processes that can be measured
at the LHC and allow to test the validity of these assumptions.
In some cases, cross section ratios can be defined that are less
sensitive to PDF uncertainties while preserving the sensitivity to new
physics effects.

In a second part of the paper, we study another striking kind of
events as they appear at HERA or Tevatron, namely diffractive events,
where in most cases, the proton is found intact after the
interaction. Diffractive events provide another way to examine the
structure of matter under specific conditions, at large gluon
densities at HERA for example. We first describe inclusive and
exclusive diffraction at HERA, which is the starting point to study
diffraction at Tevatron and then LHC. Some evidence of a new kind of
diffractive events, namely exclusive events, where all available energy
is used to diffractively produce high mass objects, was
shown at Tevatron. These events are particularly interesting at LHC
where diffractive Higgs events might occur through this
mechanism. Tagging the intact final state protons will also
allow to measure photon induced processes, and to study
the $\gamma W$ coupling in $W$ pair production.  The link of the
diffractive data with saturation effects which we also discussed  is
an interesting and promising issue for the future and could lead to a
global unified view of the proton.   

To summarize, the vision of the proton we have at present was
definitely improved with the recent data from HERA and Tevatron but
still suffers from large uncertainties at low or high $x$, with
significant impact on LHC plysics. No doubt that the understanding of
the proton structure will be further improved at LHC, and new
observables less sensitive to PDF uncertainties can be used to
disantangle in a better way the PDF effects from the ones due to new 
physics. Diffractive events still deserve to be explained via a
unified understanding of the proton structure and it might be that
saturation physics will help us achieving this goal. Finally, the rich
PDF and diffractive program at LHC will definitely lead to new
unforeseen insights of the proton.

\section*{Acknowledgments}

We thank R. Peschanski for a careful reading of the manuscript.

\vfill
\clearpage
\newpage

%=========================================================================

\end{document}